\documentclass[preprint]{aastex}
\usepackage{graphicx}
\usepackage{natbib}
\begin{document}

\title{Of Horseshoes and Heliotropes: Dynamics of Dust in the Encke Gap}
\author{M.M. Hedman$^{a,*}$, J.A. Burns$^{a,b}$, D.P. Hamilton$^c$, M.R. Showalter$^d$}
\affil{\it $^a$ Department of Astronomy, Cornell University, Ithaca NY 14853 USA \\
$^b$ Department of Mechanical Engineering, Cornell University, Ithaca NY 14853 USA \\
$^c$ Department of Astronomy, University of Maryland, College Park MD 20742 USA \\
$^d$ SETI Institute, Mountain View CA 94043 USA \\}

\maketitle

\noindent {\bf ABSTRACT:} The Encke Gap is a 320-km-wide opening in Saturn's outer A ring that contains the orbit of the small moon Pan and an array of dusty features composed of particles less than 100 microns across. In particular, there are three narrow ringlets in this region that are not longitudinally homogeneous, but instead contain series of bright clumps. Using images obtained by the Cassini spacecraft, we track the motions of these clumps and demonstrate that they do not follow the predicted trajectories of isolated ring particles moving under the influence of Saturn's and Pan's gravitational fields. We also examine the orbital properties of these ringlets by comparing images taken at different longitudes and times. We find evidence that the orbits of these particles have forced eccentricities induced by solar radiation pressure. In addition, the mean radial positions of the particles in these ringlets appear to vary with local co-rotating longitude, perhaps due to the combined action of drag forces, gravitational perturbations from Pan, and collisions among the ring particles. The dynamics of the dust within this gap therefore appears to be much more complex than previously appreciated.


\pagebreak

\section{Introduction}

The Encke Gap is a 320-km-wide opening in the outer part of Saturn's A ring centered on the orbit of the small moon Pan. In addition to Pan itself, this gap contains several faint ringlets with spectral and photometric properties that indicate they are composed primarily of dust-sized grains less than 100 microns wide. These ringlets attracted interest when they were first observed by the Voyager spacecraft because they contained prominent ``clumps'' of bright material associated with distinct ``kinks'' in the ringlets' radial position \citep{Smith82, FB97}. However, it was difficult to investigate the structure and dynamics of these longitudinally-confined features due to the restricted  amount of data obtained by the Voyager missions. 

Now, thanks to the Cassini spacecraft,  a much more extensive data set is available for investigations of the Encke Gap ringlets. In particular, the Encke Gap has now been imaged multiple times since Cassini arrived at Saturn in 2004,  allowing the evolution and motion of this material to be tracked  over timescales from weeks to years. Cassini data also provide information about other dusty ringlets in Saturn's rings \citep{Porco05, Horanyi09},  which can help clarify the dynamical processes operating in the Encke Gap.  For example, a ringlet located within the Cassini Division's Laplace Gap demonstrates ``heliotropic'' behavior: its geometric center is displaced away from Saturn's center towards the Sun \citep{Hedman10}. This happens because the particles in this ringlet are sufficiently small that solar radiation pressure can induce significant orbital eccentricities.  Since the spectral and photometric properties of the Encke gap ringlets indicate that they are also composed primarily of dust-sized particles \citep{Hedman11},  their structure should also be affected by such non-gravitational forces.

After a brief introduction to the Encke Gap's architecture  (Section 2), this report will describe the Cassini imaging observations of the Encke Gap obtained between 2004 and 2011 that provide the best information about the structure and evolution of material in this region (Section 3). Section 4 documents the distribution and motion of bright clumps in the denser ringlets. This study reveals that the bright clumps do not follow the expected trajectories of test particles under the influence of the combined gravitational fields of Saturn and Pan. Section 5 discusses structures produced by Pan's perturbations on the nearby dusty material. Section 6 examines the orbital properties of the particles in the ringlets and demonstrates that non-gravitational forces like solar radiation pressure are indeed influencing the structure of these ringlets. Finally, Section 7 discusses  some of the physical processes that could explain the longitudinal variations in the ringlets' orbital properties, the distribution of both the clumps along each ringlet and the radial locations of the ringlets within the gap. Note that these theoretical considerations only represent an initial examination of some of the dynamical phenomena that could be relevant to the Encke Gap ringlets' structure and evolution, and are not meant to provide an exhaustive 
or complete picture of the ringlets' complex dynamics.

\section{Architecture of the Encke Gap}

\begin{figure}
\centerline{\resizebox{4in}{!}{\includegraphics{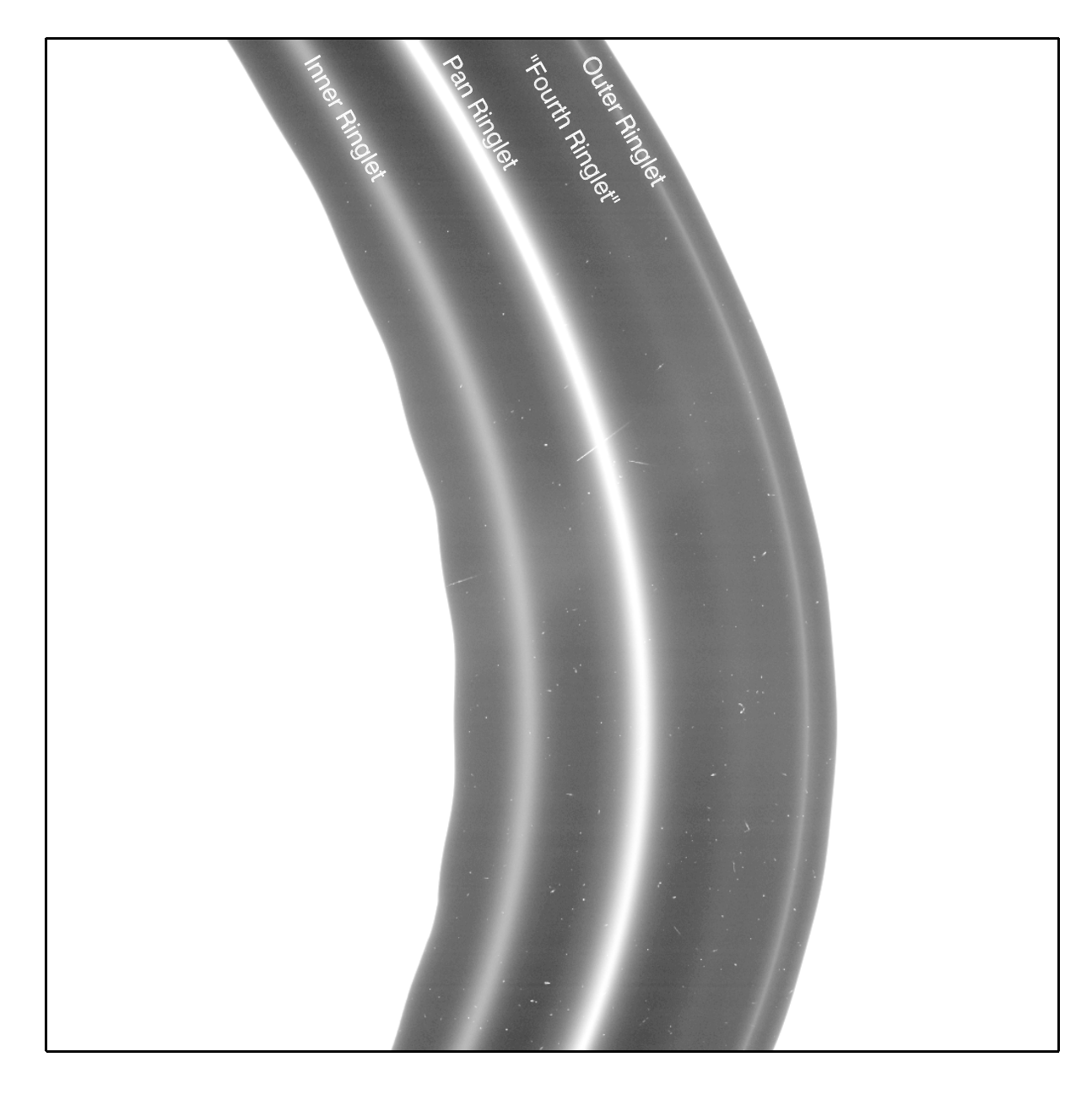}}}
\caption{One of the highest resolution images of the Encke Gap obtained by
the Cassini spacecraft. This observation was made on day 183 of 2004 during Cassini's orbit insertion (N1467351325). The image has been heavily stretched to show the ringlets in the Encke Gap, causing the regions outside the gap to appear saturated. Labels mark the positions of the four ringlets observed in this region. The inner edge of the gap appears scalloped because Pan's gravity has excited radial motions in the nearby ring material \citep{Porco05}.}
\label{egapim}
\end{figure}

\begin{figure}
\centerline{\resizebox{4in}{!}{\includegraphics{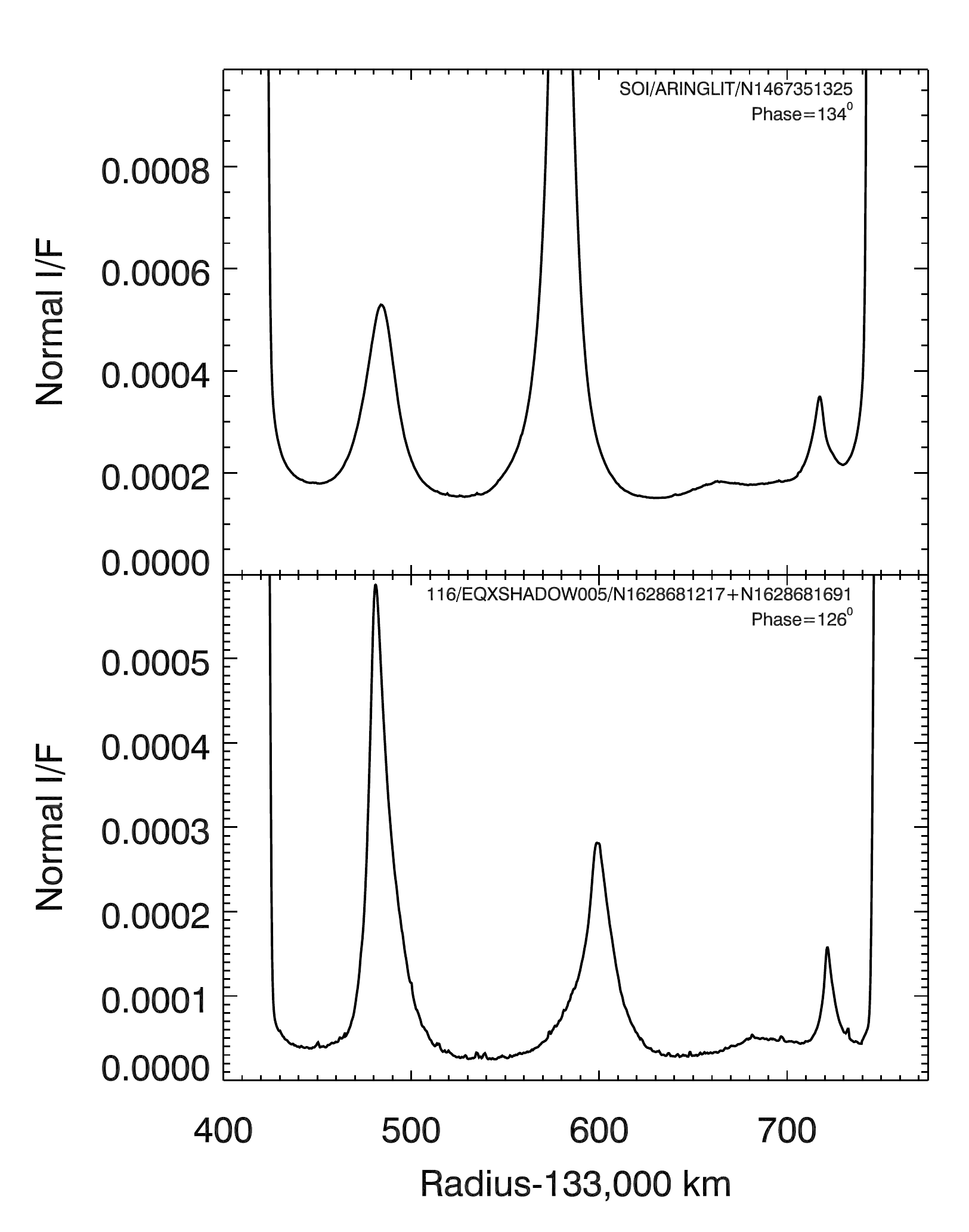}}}
\caption{Profiles of average brightness versus radius through the gap derived
from the two observations of this gap with the best combination of resolution and signal-to-noise. Brightness is measured in terms of normal $I/F$, which is the observed $I/F$ values multiplied by the cosine of the emission angle (see Section 3). The upper profile is derived from the same image shown in Figure~\ref{egapim}, while the lower profile is derived from images taken on day 223 of 2009 during Saturn's equinox. Both profiles show the same basic features, including three narrow ringlets and a broad shelf at 133,680 km (for the names of these features, see Figure~\ref{egapim}). Note the differences in radial positions and relative brightnesses of the three narrow ringlets. These are due to the longitudinal variability of these structures.}
\label{egapprof}
\end{figure}

The basic architecture of the Encke Gap is best illustrated by Figures~\ref{egapim} and~\ref{egapprof}, which provide images and radial brightness profiles  derived from the highest resolution and best signal-to-noise images of the Encke Gap obtained so far by Cassini (cf. Porco {\it et al.} 2005). These images and plots show that most of the faint material in this region is organized into three narrow ringlets and one broader feature. One narrow ringlet lies near the center of the gap, close to Pan's orbit at 133,584 km from Saturn's center. This feature is designated the ``Pan ringlet'' here, although it could just as well be called the ``central ringlet''. The two other narrow ringlets are situated on either side of the Pan ringlet. For want of a better terminology (thus far, no moon has been found within either of these ringlets), we will call the ringlet centered around 133,484 km  the ``inner ringlet'' and the ringlet centered around 133,720 km  the ``outer ringlet''.  Note that the widths, peak brightnesses and locations of all three  ringlets are different for the two profiles shown in Figure~\ref{egapprof}. This is an example of the longitudinal variability exhibited by all three of these ringlets. Closer inspection of these images and profiles reveals a broad shelf of material extending inward from the outer ringlet  to an orbital radius of about 133,680 km. This shelf, which was called the ``fourth ringlet"  by Porco {\it et al.} (2005), is considerably fainter than the other features in the Encke Gap and can only be seen with an appropriate combination of image resolution and viewing geometry. This broad feature also appears to be much more homogeneous than the three narrow ringlets. While wakes can be observed in this feature close to Pan (see Section 5 below), we have never observed anything like the clumps or kinks seen in the other three ringlets. 

These ringlets all exist within a complex dynamical environment that is strongly influenced by the gravity of Saturn's small moon Pan \citep{Showalter91}. Pan travels in a nearly circular orbit (eccentricity $\sim 10^{-5}$) through the center of the gap with a semi-major axis $a_P=133,584$ km and an orbital period of 0.575 days \citep{Jacobson08}. Due to Keplerian shear, material within and surrounding the gap drifts in longitude relative to Pan and therefore periodically encounters the moon. Since the gap is so narrow, these relative motions are very slow and encounters with Pan are correspondingly infrequent. For example, particles at the edges of the gap (at orbital radii of 133,423 km and 133,745 km) will reach conjunction with Pan only once every 543 orbits, or roughly every 315 days. 
Nevertheless, each time a particle has a close encounter with Pan, its orbital parameters will be perturbed by the moon's gravity. Indeed, Pan's influence is clearly visible in both the few-kilometer-high waves on the edges of the gap and the moonlet wakes found in the A-ring material on either side of the gap \citep{CS85, Showalter86, Horn96, Weiss09}.  Based on the amplitudes of the waves Pan generates at the edge of the Encke Gap, the mass ratio of Pan to Saturn ($m_P/M_S$) has been estimated to be about $0.8*10^{-11}$,  which corresponds to a mass $m_P \simeq 5*10^{15}$ kg \citep{Porco07, Weiss09}. 

\begin{figure}
\resizebox{6in}{!}{\includegraphics{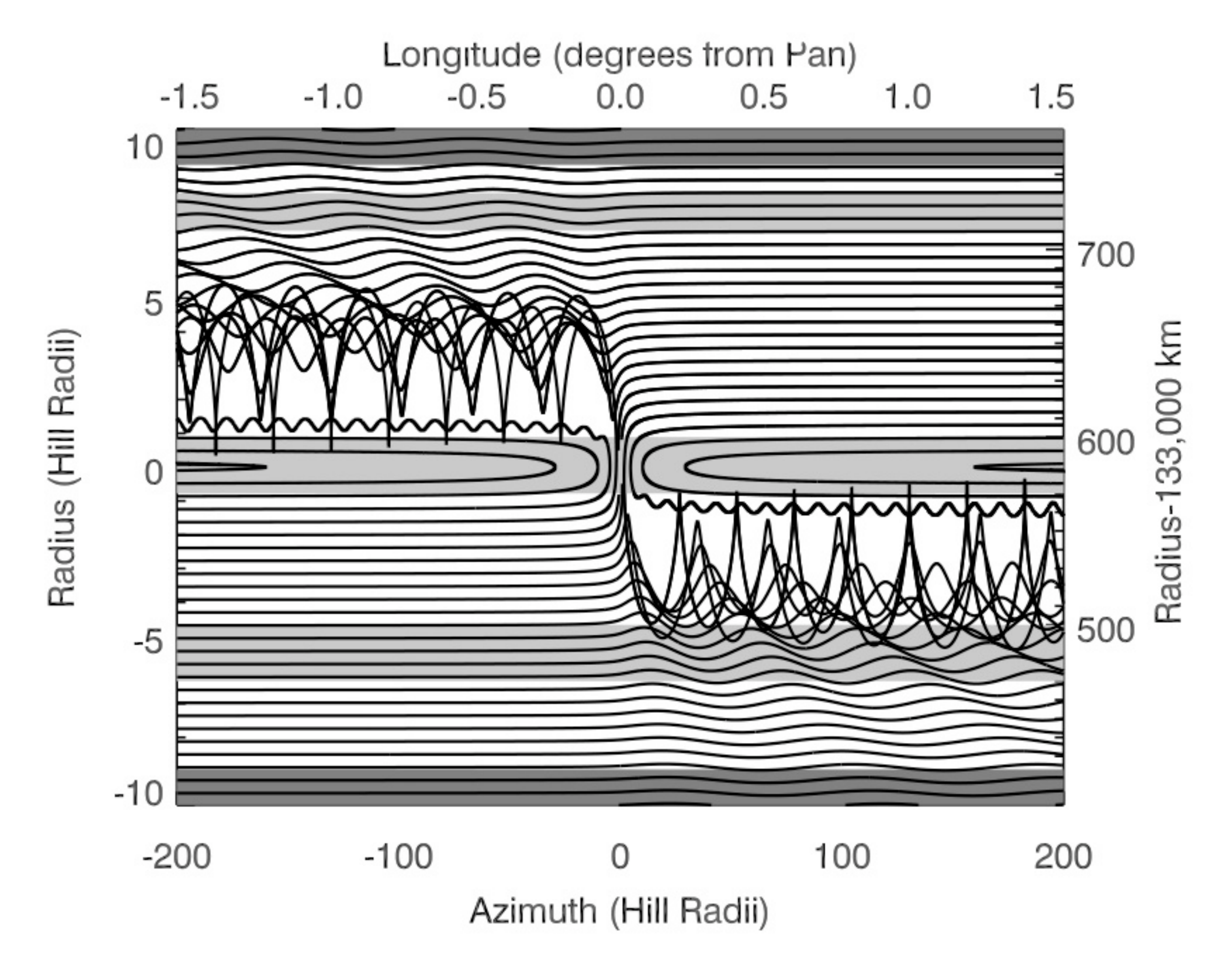}}
\caption{Schematic representation of the expected particle trajectories relative to Pan, computed using Hill's equations \citep{MD99}. Units of Hill radii (indicated along the bottom and left axes) are converted  into physical coordinates (indicated along the top and right axes), assuming Pan's Hill radius is 18 km and that Pan's semi-major axis $a_P=$133,584 km. Note that the trajectories are computed assuming particles approach Pan on initially circular orbits with a range of semi-major axes $a$. The particles approach Pan from the left when $a<a_P$ and from the right when $a>a_P$. Dark shaded bands  at the top and bottom of the plot indicate the edges of the gap, and the lighter shaded bands indicate the locations of the inner, Pan and outer ringlets. }
\label{hillplot}
\end{figure}

Particles orbiting within the Encke Gap are even more strongly affected by Pan's gravity. Figure~\ref{hillplot} illustrates the expected trajectories of small particles within the Encke Gap, assuming that the only forces acting on the particles come from Pan's and Saturn's gravitational fields. These trajectories are computed using Hill's eqautions (cf. Murray and Dermott 1999), and the scale of structures in this diagram is set by Pan's Hill radius $R_H = a_P(m_P/3M_S)^{1/3} \simeq 18$ km. For example, while particles on orbits more than a few Hill radii from Pan's semi-major axis drift past the moon, particles orbiting close to $a_P$ are unable to drift past Pan, but will instead execute horseshoe or tadpole motion around the moon's L3, L4 and L5 Lagrange points (i.e. their orbital longitude relative to Pan will librate instead or circulate). The transition between these two regimes occurs at a critical distance from Pan's semi-major axis  $\Delta a_{crit} \simeq 2.4 a_P(m_P/M_S)^{1/3} \simeq 65$ km \citep{DM81, MD99}. However, orbits with semi-major axes near $a_P\pm\Delta a_{crit}$ are actually highly unstable because they involve extremely close encounters with Pan \citep{DM81}.  Such close encounters produce large changes in the particles' orbital semi-major axes and eccentricities, and cause the orbital parameters to undergo large stochastic variations \citep{DQT89}. Particles in this ``chaotic zone'' are likely to be lost either to collisions with the moon itself or with the gap edges.  Numerical experiments and analytical theory suggest that the orbits of particles drifting past the moon will become chaotic when the semi-major axes are closer to Pan's orbit than $\Delta a_{d} \simeq 1.3a_P(m_P/M_S)^{2/7} \simeq 120$  km \citep{DQT89}. Similarly,  particles on horseshoe orbits will become chaotic when their semi-major axes are greater than  $\Delta a_{h} \simeq f_{h} a_P(m_P/M_S)^{1/3}$ from Pan's orbit, where $f_{h}$ is a numerical constant between 0.5 \citep{WW74, GT82} and  1.3 \citep{Dermott80}. Stable simple horseshoe orbits are therefore only found within 15 or 35 km of Pan's orbit.

The Pan ringlet always lies within $\Delta a_{crit}$ of Pan's orbit, and thus almost certainly consists of material moving in horseshoe and tadpole orbits around the moon's Lagrange points \citep{Showalter91}. By contrast, the inner, outer and fourth ringlets all are more than $\Delta a_{crit}$ from 133,584 km, and thus are likely composed of material that drifts continuously past Pan. The motions of the bright clumps in the inner and outer ringlets, as well as the presence of moonlet wakes in all these structures are consistent with this supposition (see below). However, note that both the inner and fourth ringlets may overlap the  semi-major axis range where particle orbits should be chaotic (i.e., they lie within $\Delta a_d$ of Pan's orbit). This could imply that inter-particle interactions or some other process is affecting these particles' orbits and stabilizing these ringlets. Indeed, one might be tempted to regard the outer edge of the inner ringlet and the inner edge of the fourth ringlet as marking the edges of the chaotic zone.

\section{Observations and data reduction procedures}

This investigation of the Encke Gap structures will rely exclusively on pictures obtained by  the Narrow Angle Camera (NAC) of the Imaging Science Subsystem onboard the Cassini spacecraft \citep{Porco04}. The observations that are most informative about the overall structure and dynamics of the Encke Gap ringlets include:
\begin{itemize}
\item Movie sequences obtained when the camera pointed at one place in the Encke Gap and watched material orbit through the field of view over a significant fraction of an orbital period. These observations provide snapshots of the longitudinal structure of the ringlets at particular times. The thirteen movies used in this analysis, which are the best in terms of longitudinal coverage, are listed in Table~\ref{moslist}.
\item The so-called SATELLORB observations designed to periodically observe various small moons in order to refine and track their orbits. A subset of these images targeted at Pan also capture nearby parts of the Encke Gap. Specifically, Table~\ref{suplist} lists 189 images where the ring opening angle was sufficiently high (more than $1^\circ$), the radial resolution was sufficiently good (better than 20 km/pixel) and a sufficiently broad range of longitudes were observable (at least 1$^\circ$). These images were obtained in between the more extensive movies, and thus provide additional information about the evolution and motion of certain clumps.
\item  The  PANORBIT observation made in 2007-143 during Rev 45. This is a sequence of 158 images (N1558590310- N155861997,  emission angle 68$^\circ$, phase angle 79$^\circ$) targeted at Pan as it moved around the planet. These images also captured the part of the Encke Gap surrounding Pan, enabling us to observe how the structure of the central ringlet changes with true anomaly.
 \end{itemize}
We also presented above some data from  selected high-resolution, high signal-to-noise images of the Encke gap (N1467351325 and N1628681217-N16281691, see Figure~\ref{egapim} and ~\ref{egapprof}). However, this report will not include a thorough analysis of all the highest resolution images of the Encke Gap. While such images can provide very useful data regarding the fine-scale morphology of individual clumps, we will limit our scope here to the region's global behavior. 
 
All the relevant images were calibrated using the standard CISSCAL routines \citep{Porco04} to remove instrumental backgrounds, apply flatfields and convert the raw data numbers to $I/F$, a standardized measure of reflectance that is
unity for a Lambertian surface at normal incidence and emission. The images were geometrically navigated using the appropriate SPICE kernels and this geometry was refined based on the position of sharp ring edges in the field of view. Whenever practical, this navigation used the outer edge of the Keeler Gap as a fiducial, but when the resolution of the images was either insufficient to resolve this gap or so high that the gap was not present in the field of view, the edges of the Encke Gap were used instead. While neither the Keeler Gap's outer edge nor the Encke Gap's edges are perfectly circular,  the variations in the relevant edge positions are sufficiently  small (only a few km) that they do not impact efforts to quantify and track the longitudinal positions of the clumps. However, these imperfections cannot be ignored in detailed studies of the ringlets' radial positions (see below). 

\begin{figure}[tb]
\resizebox{6in}{!}{\includegraphics{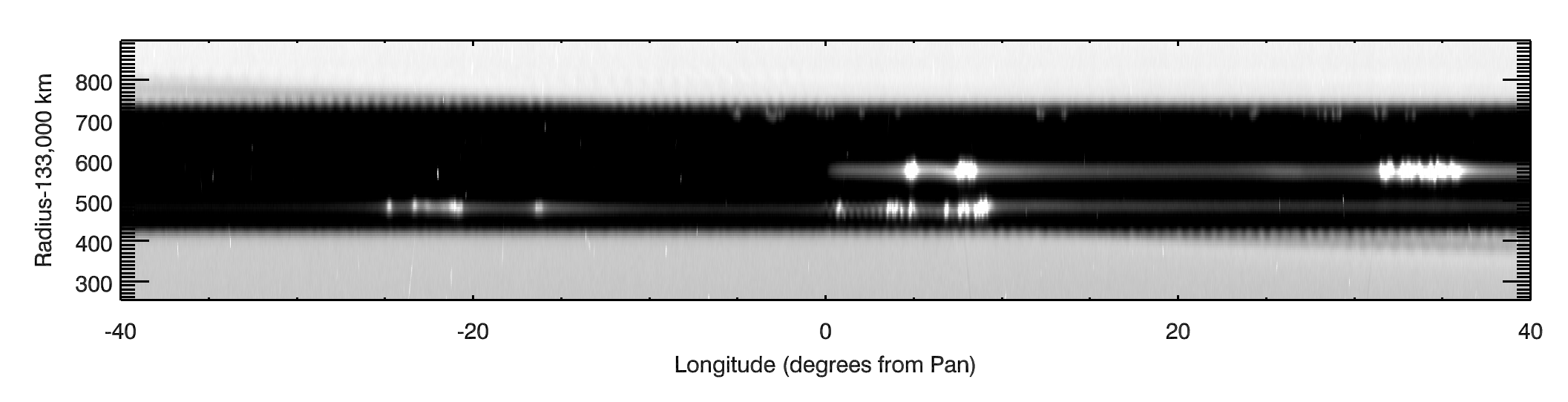}}
\caption{Example of part of a mosaic generated from Rev 030 HIPHAMOVD observation. This mosaic shows the brightness of the rings as a function of radius and longitude, and within this figure one can clearly see clumps in the Pan ringlet at a radius of 133,584 km and the Inner ringlet at 133,484 km. One can even see a few features in the outer ringlet just interior to the Gap's outer edge at 133,745 km.}
\label{mos0}
\end{figure}

For the high-resolution images described above, the rings are sufficiently homogeneous that we can reduce the geometrically-navigated data from each image into a single radial brightness profile by simply averaging over all longitudes. For the other observations, however, a single image can contain multiple clumps or kinks, so reducing the data to a single radial scan is not appropriate. Instead, the brightness measurements from each image are re-projected to produce ``maps'' of the Encke Gap on a uniform grid of radii and longitudes relative to Pan (derived from the appropriate SPICE kernels). For the SATELLORB and PANORBIT observations, these maps provide a useful basis for subsequent data analysis. However,  for the movie sequences listed in Table~\ref{moslist}, which cover a broad range of co-rotating longitudes at a single time, individual images are less useful than the combined data set. Hence the relevant maps derived from individual images are interpolated onto a common radius and longitude scale and then assembled into a single mosaic spanning a large fraction of the Encke Gap (see Figure~\ref{mos0}).  These mosaics can then be processed using the same basic procedures as the individual maps.

Besides re-projecting the data into convenient maps and mosaics, the relevant geometric information is also used to compute the cosine of the emission angle $\mu$. By multiplying the observed brightness values by this quantity, the observed $I/F$  can be converted into an estimate of the ``normal $I/F$'', which for low optical depth features like the Encke Gap ringlets should be independent of emission angle. 

Depending on the resolution and quality of the observation, different procedures were used to quantify the brightness and location of these ringlets. The finite resolution of the images influence both  the peak  brightness and radial width of the ringlets, so the brightness of the ringlet is instead quantified using the radially integrated normal $I/F$ of the ringlet, or ``normal equivalent width'' (abbreviated NEW in Figures~\ref{panb1},~\ref{inb1} and ~\ref{outb1} below). For low optical-depth features like the Encke Gap ringlets, this integrated quantity is independent of the image resolution. Profiles of normal equivalent width versus longitude derived from different observations can therefore be compared to one another relatively easily and reliably. 

Whenever possible, the ringlet's radial brightness profile at each longitude was fit to a Lorentzian in order to obtain estimates of both the ringlets'  radial position and its  equivalent width. The fitting procedure for each ringlet is tuned to minimize contamination from the other ringlets and to cope with variations in the radial position of the ringlet with longitude and time.

For the Pan and inner ringlets, extrema in the derivative of the radial brightness profile are used to make a preliminary estimate of the location of the ringlet and to determine the radial range included in the fit.  For the Pan (inner) ringlet, the point of maximum positive slope between 133,520 and 133,600 km  (133,420 km and 133,500 km) provides an estimate of  the ringlets' inner edge position $r_1$, while the point of largest negative slope between 133,560 and 133,630 km (133,470 and 133,530 km) yields an estimate for the ringlet's outer edge location $r_2$. The average of these two numbers therefore provides an estimate of the center of the ringlet, and a radial region centered on this location with a width that is the larger of 60 km and $2(r_2-r_1)$ is selected and fit to a Lorentzian plus linear background (the lower limit of 60 km ensures that the fitted region is broad enough to contain the entire ringlet, see Figure~\ref{egapprof}).  

The outer ringlet is located closer to the edge of the gap than the other ringlets, and therefore required a somewhat more complex procedure that includes removing the background signal due to the nearby gap edge. This background was estimated by interpolating the brightness profile on either side of the ringlet, which requires a preliminary estimate of the ringlet's position and radial extent.  The center of the ringlet  is estimated as the location of the minimum in the second derivative of the brightness profile between 133,710 and 133,730 km.  Preliminary estimates of the ringlet edge positions were obtained as the maximum of 20 km and 1.5 times the distance to the minimum slope within 20 km of the ringlet center (the lower limit of 20 km ensures that the fitted region is broad enough to contain the entire ringlet, see Figure~\ref{egapprof}). However, in order to obtain a sensible background level, the outer edge of the fit region is constrained to at least two radial bins short of the point of maximum slope on the gap edge. The background level under the ringlet is then obtained by a spline interpolation of the brightness data outside the selected region. The interpolation is actually  applied to the logarithm of the brightness measurements because the abrupt change in slope near the edge of the gap made interpolation of the raw brightness measurements difficult. After removing the background, the remaining data are then fit to a Lorentzian plus constant offset. 

For observations obtained at lower resolutions or at lower phase angles (where the ringlets are comparatively faint), the above fitting routines were not appropriate and so it was not possible to estimate the radial positions of the ringlet. However, the integrated brightness of the ringlet can still be computed. For the Pan ringlet we compute the integrated brightness within 50 km of 133,585 km. A background level based on the average brightness outside this region can be removed from these profiles if required. For the inner and outer ringlets, which lie closer to the edges of the gap, the radial region containing the ringlet and the appropriate background levels are computed using the same basic method as described in the previous paragraph. The edges of the ringlet region are determined based on extrema in the slopes, and the background in this  region is determined by a cubic spline interpolation of the log-transformed data on either side of this region.

Mosaics where the peak-fitting procedures were successful are marked with P or R in Table~\ref{moslist}. By contrast, mosaics where only the integrated brightness could be computed are marked with an I. The data obtained from the SATELLORB observations (Table~\ref{suplist}) are entirely derived from simple integrations, and the PANORBIT observations are all processed with peak-fitting routines. Note that  the different resolutions and processing techniques used on these different data sets could potentially complicate any effort to compare the absolute brightness of the ringlets derived from different observations, and hence we will not attempt such photometric comparisons here. Instead, this paper will focus exclusively on the structure and morphology of these ringlets, which are more robustly determined by these procedures. 

Uncertainties in these relative brightness and position estimates are dominated by systematic errors in the fits and background removal rather than statistical noise, and thus are difficult to quantify {\it a priori}. Based on the lack of obvious long-wavelength drifts outside the clump-rich regions in the brightness profiles for the inner and Pan ringlets, systematic errors in the brightness structure of the clumps in these ringlets are expected to be negligible. The brightness variations outside the clumps are more substantial for the outer ringlet, but even here the morphology of the clumps are very repeatable between observations (see Figure~\ref{outb1} below), so systematic errors in the brightness of these clumps should also be small (probably less than 10\%). Finally, the repeatability of long-wavelength structure in the radial positions for these ringlets (see Section 6) implies that systematic errors in the radial positions of the inner and Pan ringlets are typically less than 1 km. However, these estimates are based on heuristic {\it a posteriori}  arguments and not rigorous quantitative analyses. Hence in order to avoid giving a misleadingly precise impression of the relevant uncertainties, we will not plot error bars on the various longitudinal profiles presented in this paper.  

\section{Brightness variations in the ringlets}

\begin{figure}
{\resizebox{3in}{!}{\includegraphics{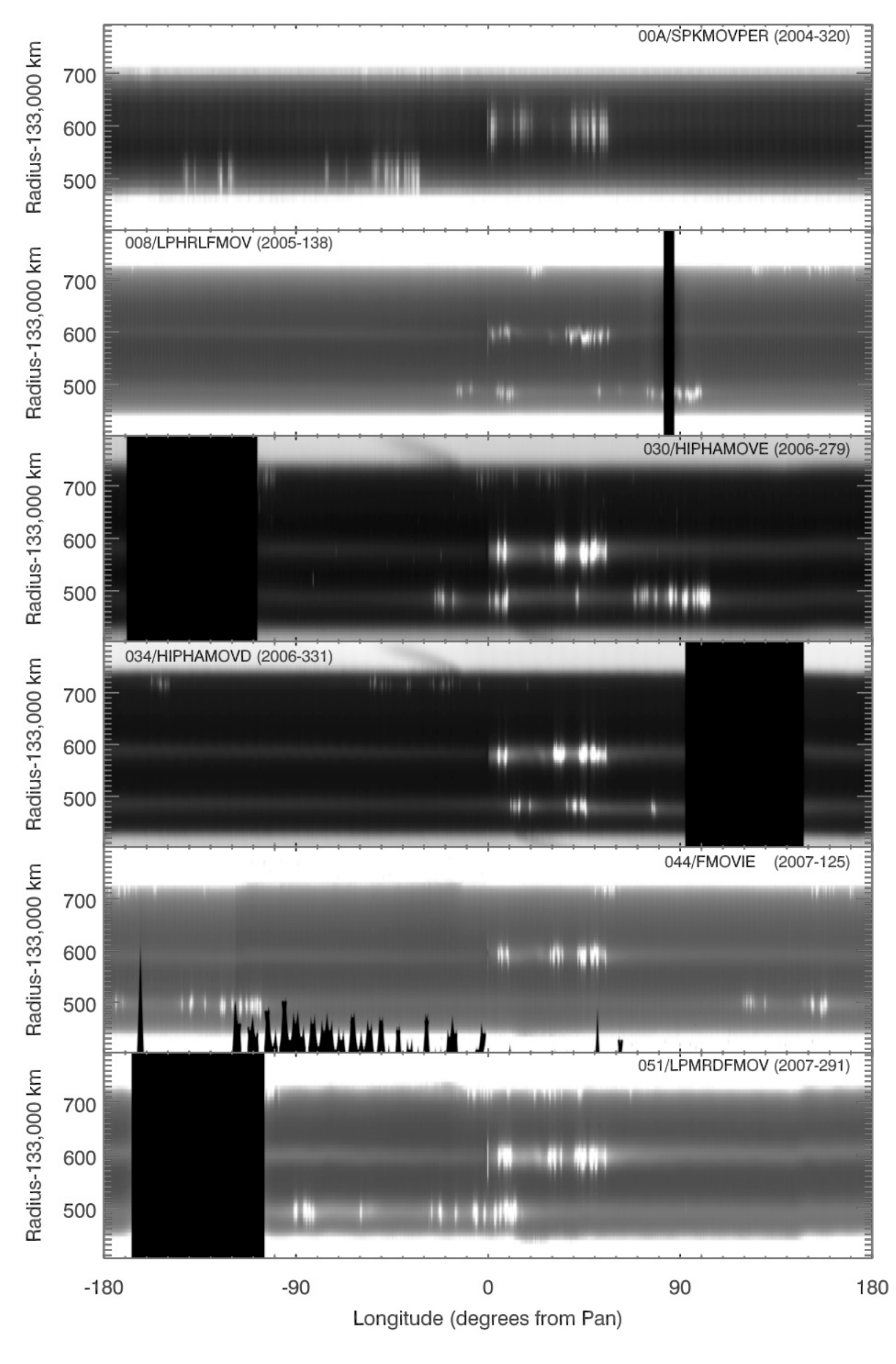}}}
{\resizebox{3in}{!}{\includegraphics{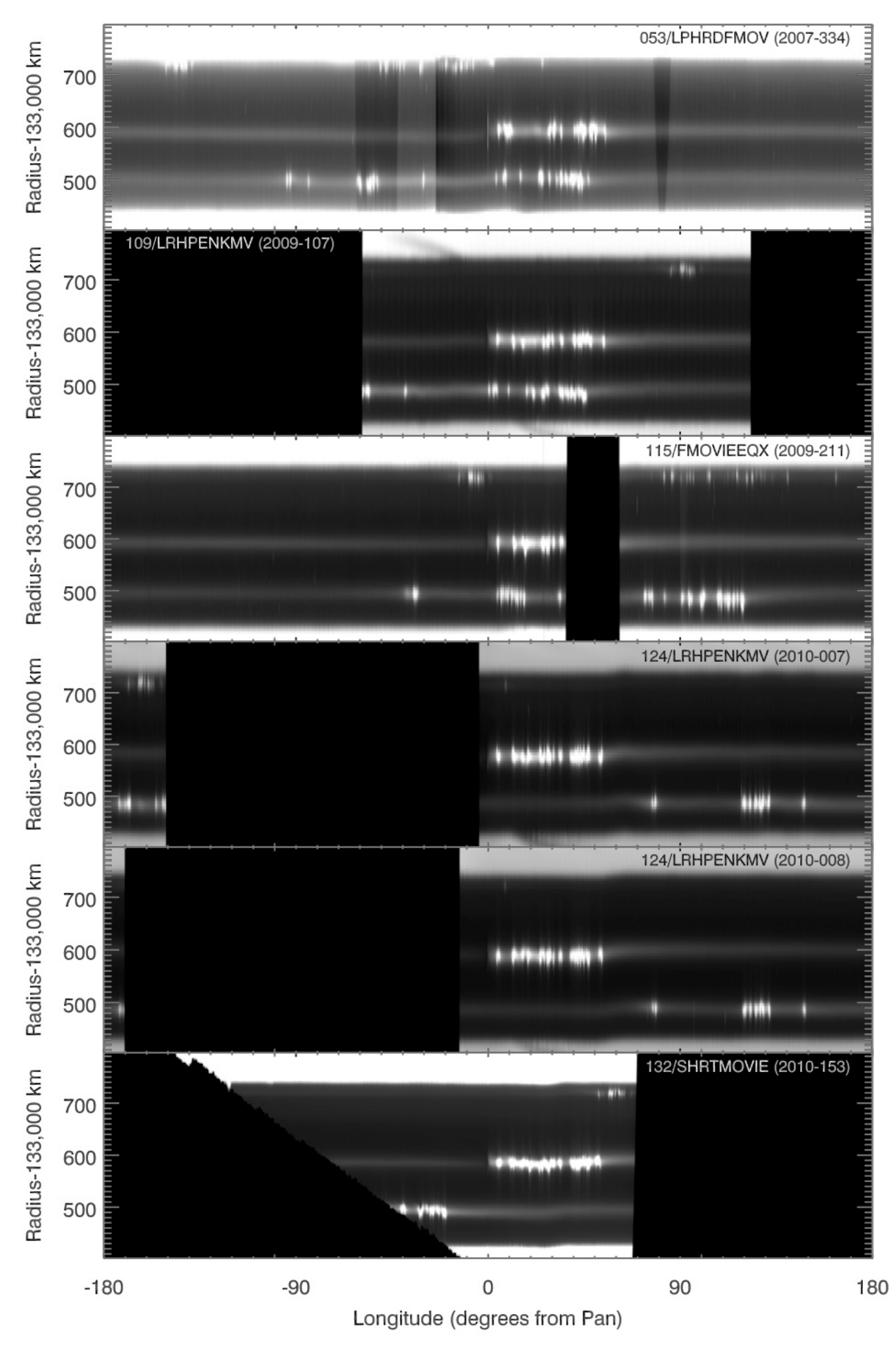}}}
\caption{Images of the Encke Gap mosaics constructed from the observing sequences listed in Table~\ref{moslist}. The data from the SATSRCH observation are not shown here due to their low resolution. Each panel displays the ring brightness as a function of radius and longitude relative to Pan. Each image is individually stretched to best highlight the ringlets in the gap. Black regions in each map correspond to areas that were not observed during the observing sequence. Note the restricted longitude range of the clumps in the central Pan ringlet, and the steady movement of the clumps in the inner and outer ringlets relative to Pan.}
\label{mos1}
\end{figure}

Figure~\ref{mos0} illustrates the brightness variations that can be seen within the inner, outer and Pan ringlets. All three ringlets contain localized regions of enhanced brightness, which we interpret here as concentrations or ``clumps'' of material.\footnote{Alternative interpretations of the brightness variations as the result of vertical structures producing changes in the amount of material along certain lines of sight are much less plausible. If the bright regions were just the result of projection effects, then the distribution of these features would change radically with the observation geometry.
Instead, image sequences taken in very different observing geometries exhibit the same basic pattern of clumps (see Table~\ref{moslist} and Figures~\ref{panb1},~\ref{inb1} and \ref{outb1}), which is much more consistent with simple variations in the local particle density.}  Figure~\ref{mos1}  shows the full mosaics derived from most of the observations listed in Table~\ref{moslist} (the SATSRCH observation is not illustrated due to its lower resolution). These mosaics show that these clumps are not distributed randomly along each ringlet. In particular, the clumps in the Pan ringlet are always found between longitudes of $0^\circ$ and $ +60^\circ$ in a Pan-centered coordinate system, that is,  between Pan and its leading Lagrange point.  Studies of Voyager images of this ringlet taken around 1980 \citep{FB97} showed a similar pattern, indicating that such an asymmetric clump distribution is a persistent feature of this ringlet.

Next, consider the inner and outer ringlets. These features are located outside of Pan's horseshoe zone (see above), so this material should drift  slowly relative to Pan. Indeed, the clumps in the inner ringlet can be observed to slip slowly ahead of Pan, while those in the outer ringlet move slowly backwards, as expected. However, within each ringlet, the distribution of clumps is again remarkably persistent. For the inner ringlet, the clumps cluster in a region between 110$^\circ$ and 160$^\circ$ wide. This is again consistent with the Voyager observations 25 years earlier  \citep{FB97}, implying that something may be preventing these clumps from efficiently dispersing all around the ringlet. The clumps in the outer ringlet, by contrast, seem to be a bit more broadly distributed, with a dense cluster of clumps roughly 20$^\circ$ wide lagging 120$^\circ$ behind a more spread-out array of clumps (see top right panel of Figure~\ref{mos1}). Again, this basic pattern of clumps seems to persist for years. Note that all the clumps in both the inner and outer ringlets drifted past Pan multiple times during the course of these observations, so the distribution of the clumps in these ringlets appears to be moderately robust against perturbations from that moon. 

The evolution of these clumps'  morphology and spatial  distribution between 2004 and 2011 can be more closely examined with the longitudinal brightness profiles shown in  Figures~\ref{panb1},~\ref{inb1} and \ref{outb1}. These plots  show the radially-integrated brightness of the ringlets as a function of longitude derived from the various mosaics listed in Table~\ref{moslist}. Also useful are the plots shown in Figures~\ref{clumptrackpan}-\ref{clumptrackdet2}, ~\ref{clumptrackin} and~\ref{clumptrackout},  which graph the positions of brightness maxima in these profiles as functions of time. In order to facilitate comparisons between observations taken at various times,  a different co-rotating longitude system has been used to plot the data for each ringlet.

Identifying individual clumps and tracking their motions is challenging because clumps are not  always isolated brightness peaks that drift relative to each other.  Instead, regions of enhanced brightness have a range of morphologies, including tightly-packed clusters and looser archipelagos of brightness maxima that can split, merge or even drift as units. 
This complicates any effort to quantify the motion or evolution of these structures, and consequently we will not attempt to generate a comprehensive catalog of these features. However, in all three ringlets, certain regions consisting of one or more bright clumps appear to be remarkably persistent across the various observations.  Hence we can identify and track  these broader-scale features over several years with some degree of confidence (cf. Showalter 2004), although we must admit that even some of these features could form or dissolve between observations taken years apart. In the following sections, we will examine the overall distribution of the brightness maxima and the detailed evolution of a few particular structures in each ringlet. \nocite{Showalter04}

\subsection{Pan ringlet}

\begin{figure}
\centerline{\resizebox{3.2in}{!}{\includegraphics{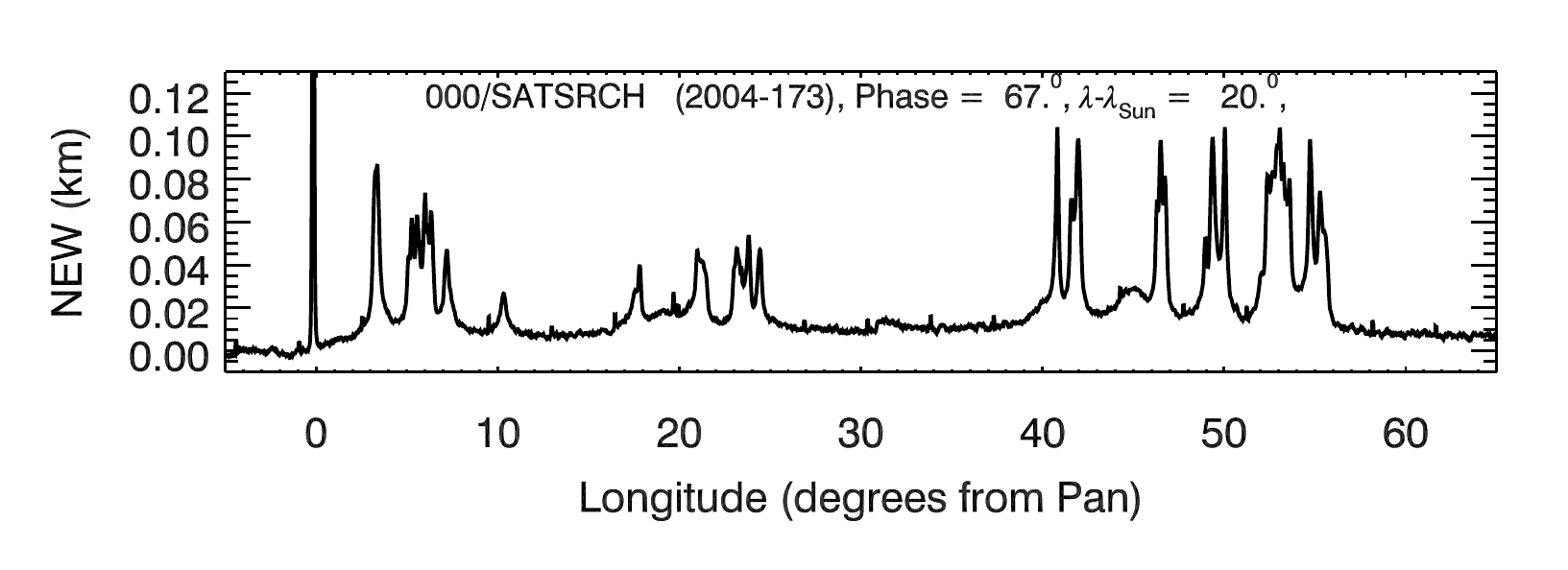}}}
{\resizebox{3in}{!}{\includegraphics{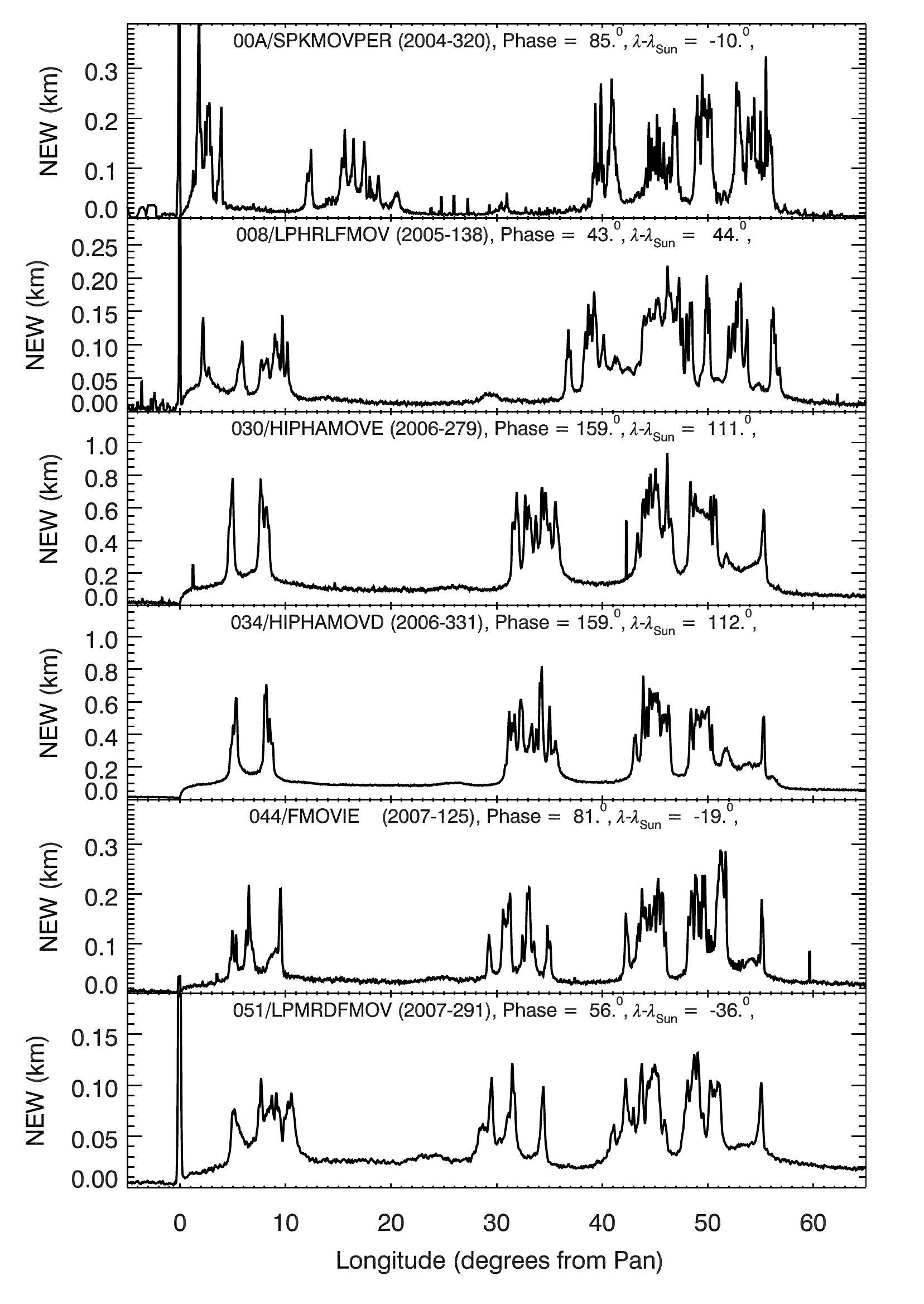}}}
{\resizebox{3in}{!}{\includegraphics{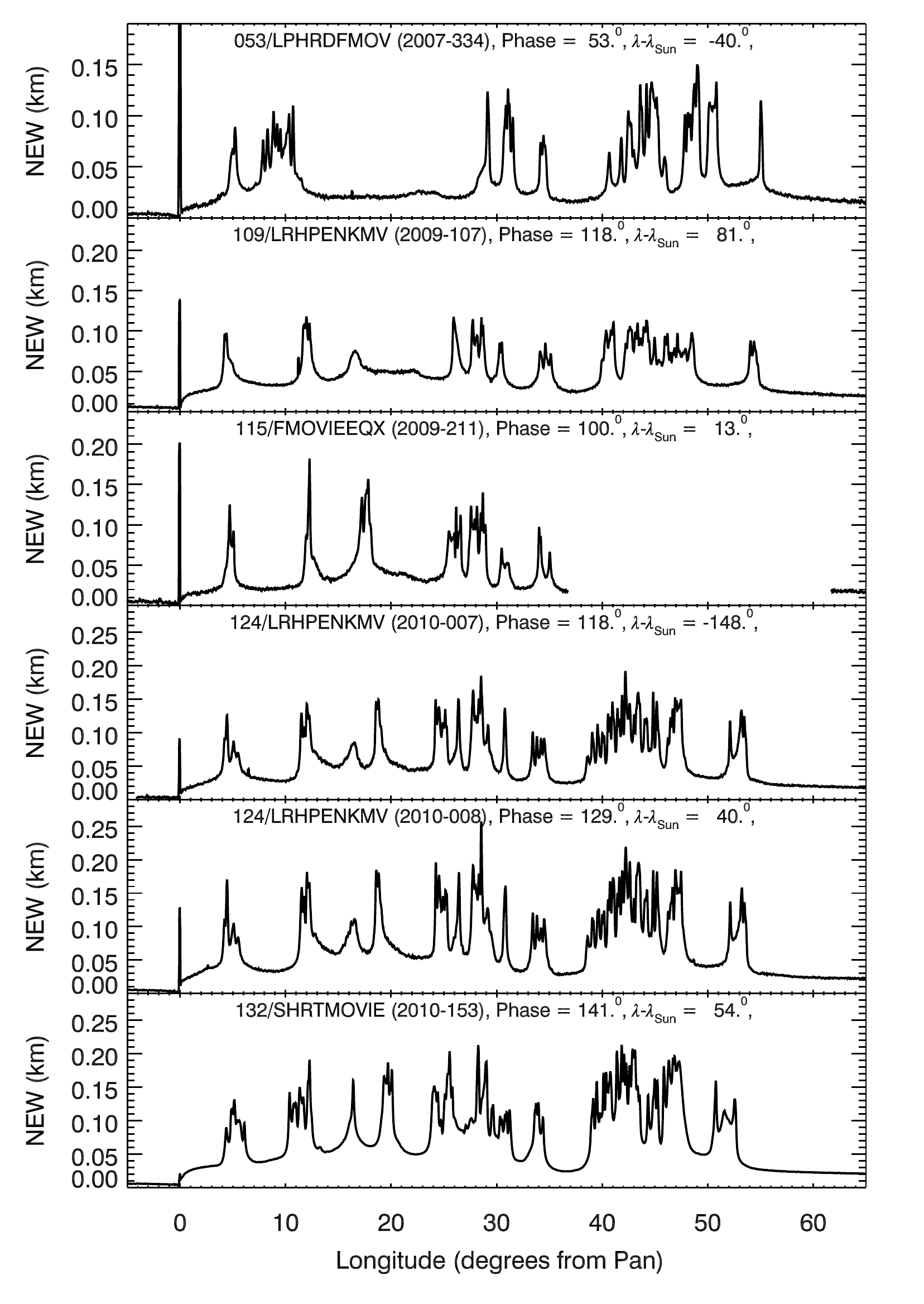}}}
\caption{Plot of the {\bf Pan ringlet's} radially-integrated brightness (normal equivalent width)  versus longitude from Pan based on the data from the observations  listed in Table~\ref{moslist}. The 000/SATSRCH profile comes from radial integration of the brightness profile, while the other  brightness profiles are all derived from Lorentzian fits to the ringlet. Fits with peak radii more than 30 km from 133,585 km are removed and the remaining data smoothed over 5 samples for the sake of clarity. Narrow spikes between 23$^\circ$ and 30$^\circ$ in the 00A/SPKMOVPER profile and around $60^\circ$ in the 044/FMOVIE profile are due to stars and cosmic rays, while the clumps all have a finite longitudinal width. }
\label{panb1}
\end{figure}

\begin{figure}
\centerline{\resizebox{5in}{!}{\includegraphics{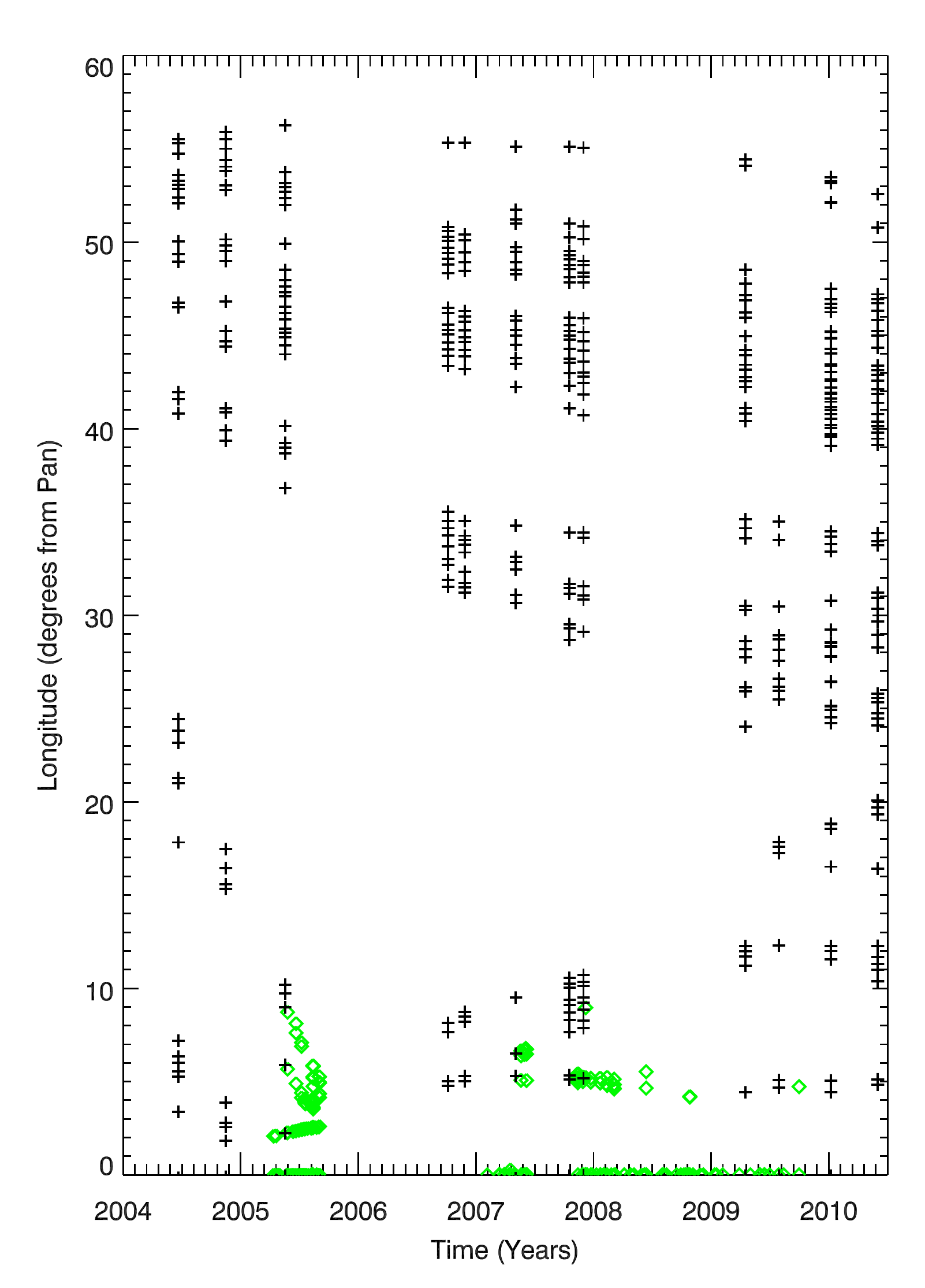}}}
\caption{Plot showing the locations of brightness peaks in the {\bf Pan ringlet} as a function of longitude and time. The black plusses are measurements derived from the largely complete mosaics shown in Figure~\ref{panb1}, while the green diamonds are derived from the SATELLORB images listed in Table~\ref{suplist}. Note that the latter data only cover the region immediately in front of Pan.}
\label{clumptrackpan}
\end{figure}

\begin{figure}
\centerline{\resizebox{5in}{!}{\includegraphics{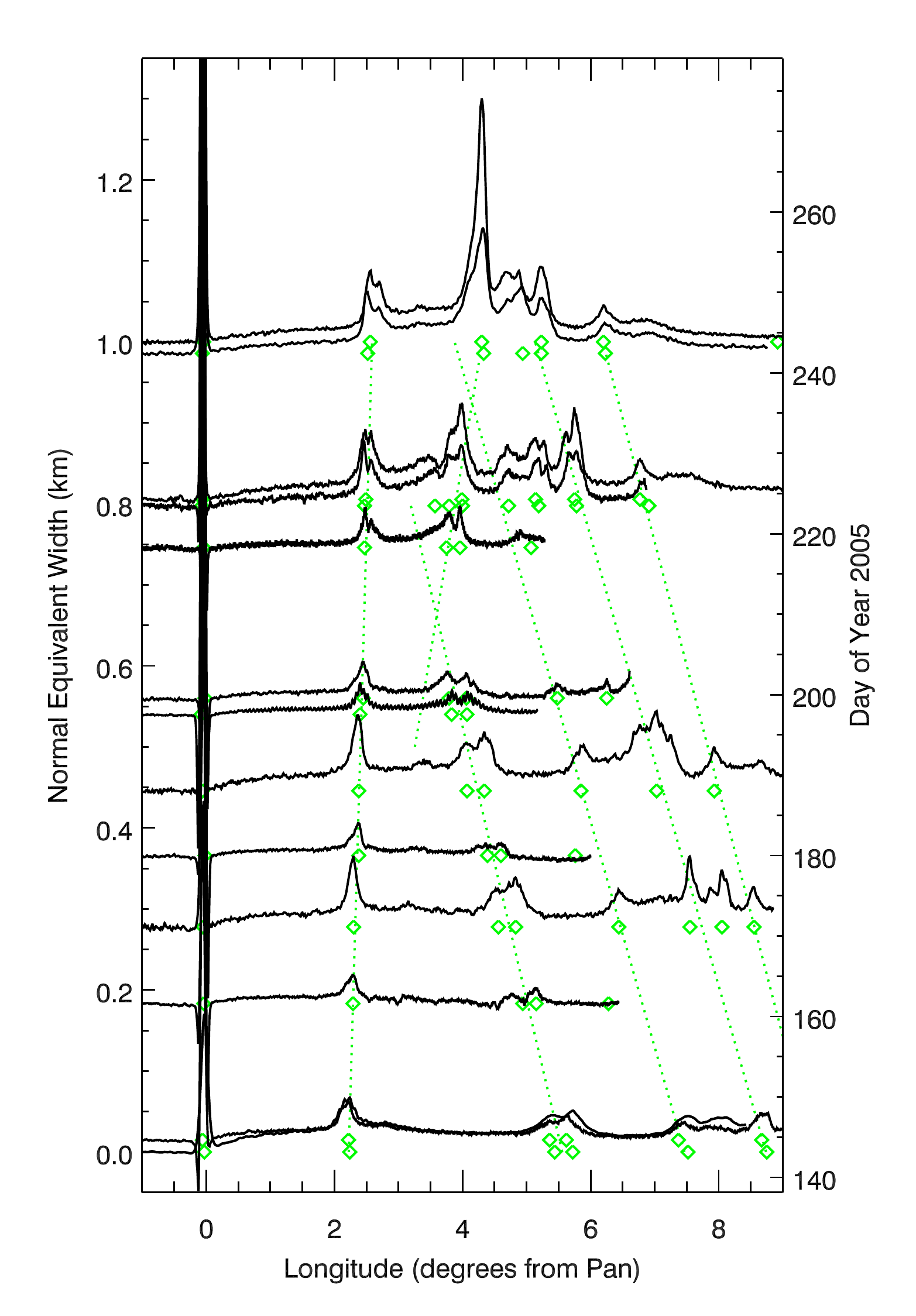}}}
\caption{Longitudinal profiles of the Pan ringlet brightness obtained between days 140 and 250 of 2005. The profiles are stacked vertically with spacings proportional to their time separation, and the green diamonds mark the locations of brightness maxima at the times given on the right-hand vertical axis. Dotted lines tracing the motion of particular clumps are included to guide the eye. Note the clump that starts near 5.5$^\circ$ first drifts towards Pan, but then appears to reverse direction between days 200 and 220,  such that it collides with the clump that had been following it between days 220 and 240.}
\label{clumptrackdet1}
\end{figure}

\begin{figure}
\centerline{\resizebox{5in}{!}{\includegraphics{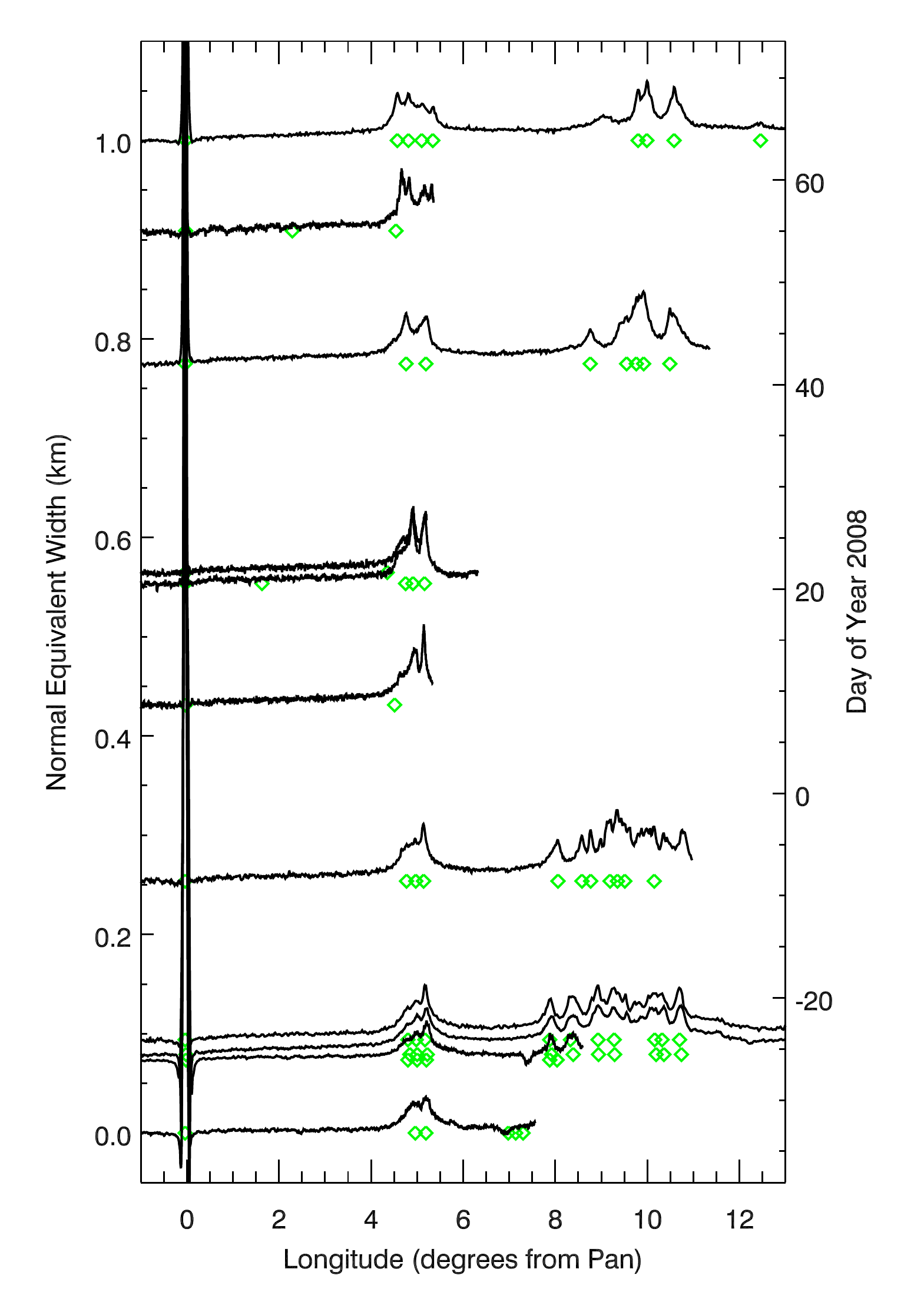}}}
\caption{Longitudinal profiles of the Pan ringlet brightness obtained between days 330 of 2007  and 70 of 2008. The profiles are stacked vertically with spacings proportional to their time separation, and the green diamonds mark the locations of brightness maxima at the times given on the right-hand vertical axis. In this case, the motions of individual clumps are less obvious. However, the morphology of the clump around 5$^\circ$ in front of Pan changes in an interesting way. In 2007, this clump had an asymmetric profile with a single brightness peak. In 2008 a second peak appears and the two peaks begin to separate. Around day 50, each of those two peaks splits to produce a total of four peaks, which again move apart over time.}
\label{clumptrackdet2}
\end{figure}

First, let us  consider the Pan ringlet data shown in Figures~\ref{panb1} and~\ref{clumptrackpan}. Note that the coordinate system used in these plots is simply longitude relative to Pan. When this region was first observed in 2004 the clumps were concentrated in three regions roughly 5$^\circ$, 20$^\circ$ and 50$^\circ$ in front of Pan. Over the next year, the clumps less than 30$^\circ$ in front of Pan seem to rapidly converge  into a region roughly $5^\circ$ in front of Pan, while the clumps around 50$^\circ$ dispersed slightly. When these clumps were again seen in late 2006, the clumps could still be divided into two groups. The smaller group close to Pan appears to have spread over the region between 5$^\circ$ and 10$^\circ$, while the clumps $50^\circ$ in front of Pan had continued to disperse. In fact, this group appears to have split into two clusters, one centered around $35^\circ$  and one remaining around 45$^\circ$. Over the next year and a half, the cluster closest to Pan spread away from Pan, while the cluster around 35$^\circ$ drifted slowly towards Pan. During 2008-2009, one of the clumps appears to stay within 5$^\circ$ of Pan, while the remaining clumps from this region appear to have drifted outward so that they were seen a little beyond 10$^\circ$ in early 2009. At the same time, the clumps around 35$^\circ$ dispersed and the clumps around $45^\circ$ shifted a bit closer to Pan. The motions of these different clumps during the next year were modest, but during this time a new clump cluster seems to have formed roughly 17$^\circ$ in front of Pan. As can be seen in Figure~\ref{panb1}, this feature started as a broad hump in the Rev 109 LRHPENKMV data, then became a stronger peak with two maxima in the Rev 115 FMOVIEEQX data, which then moved apart to become a pair of clumps in subsequent observations. By the middle of 2010, clumps were distributed throughout much of the region between 0$^\circ$ and $60^\circ$ in front of Pan.

The fastest drift rates observed in these data are associated with the clumps that moved from just outside 20$^\circ$ to about 5$^\circ$ between mid-2004 and late 2006. These clumps moved at a rate of between 0.035$^\circ$/day and $0.040^\circ$/day relative to Pan. However, this drift rate appears to be unusual, and most of the other clump features only moved a few degrees per year, or less than 0.01$^\circ$/day relative to Pan. If these drift rates were due to the clump material having slightly different semi-major axes from Pan, then most of these clumps would be within 1.5 km of $a_P$, with the fast-moving clumps being only 5-6 km away. However, the actual trajectories of these clumps are not consistent with those expected for concentrations of material at such semimajor axes (cf Murray and Dermott 1999). Particles at these locations would be expected to execute horseshoe or tadpole motion around Pan's Lagrange points, where the particle approaches Pan at some speed, turns around, then recedes at the same speed until it is somewhere beyond 60$^\circ$ in front of Pan. The clump trajectories shown in Figure~\ref{clumptrackpan} do not match these expectations.  For example, consider the most distant clump from Pan, which is a relatively isolated feature between 2005 and 2010 and thus can be tracked with confidence. It first emerges from the leading side of a large clump complex in 2005, when it is moving slowly away from Pan towards the leading Lagrange point at 60$^\circ$. However, in 2006-2008, this clump seems to have stalled at about $56^\circ$, and in 2009 and 2010 it is clearly moving towards Pan, away from the Lagrange point. This clump therefore accelerated away from Pan's Lagrange point between 2007 and 2009, which is inconsistent with any sort of horseshoe or tadpole orbit. This clump therefore is not moving like a simple test particle in the combined gravitational fields of Pan and Saturn. 

Even more curious are the motions of the clumps found within 10$^\circ$ of Pan, which can be studied in greater detail thanks to the extensive SATELLORB observations of these regions in both 2005 and 2007-2008. Figures~\ref{clumptrackdet1} and~\ref{clumptrackdet2} illustrate how these clumps evolved over the course of these two time periods. During the 2005 observation sequence, the clump closest to Pan steadily drifts outwards at a rate of about 0.004$^\circ$/day, while the other clumps are initially drifting towards Pan at rates between 0.029$^\circ$/day and 0.035$^\circ$/day. (see Figure~\ref{clumptrackdet1}). If these approaching clumps were on horseshoe orbits, their semi-major axes would be $\delta a \sim$4-5 km exterior to Pan's. Such particles should be able to approach Pan until they reach a critical distance $y_{\rm min}$, where they will turn around on their horseshoe orbits. This minimum distance can be calculated from the semi-major axis separation \citep{DM81}:
\begin{equation}
 y_{\rm min} = \frac{8}{3}\frac{m_p}{M_S}\left(\frac{a_P}{\delta a}\right)^2a_P.
\end{equation}
For such clumps, $y_{\rm min}$ corresponds to 1$^\circ$-1.5$^\circ$, but none of these approaching clumps ever gets that close to Pan. Instead, the closest of the approaching clumps seems to stop moving when it gets only 4$^\circ$ in front of Pan, and even starts moving away from Pan a bit before it appears to merge with the clump that had been following it. Looking at the profiles obtained between days 220 and 230 of 2005, it almost appears as if this clump was ``repelled'' by the slowly-moving clump at 2$^\circ$ (Note that additional peaks appeared in both clumps during this time). Yet this same clump then seems to have merged with the clump that had been following it just a few weeks later. Note the two profiles from around day 245 were both obtained at the same phase angle (about 60$^\circ$), so the sudden brightening at 4$^\circ$ could be the result of this merging event. In any case, these data demonstrate the interactions of these clumps can be quite complex.

By contrast, the clumps seen during late 2007-2008 do not appear to move very much (see Figure~\ref{clumptrackdet2}). Instead, we can observe the morphology of the clump around $5^\circ$ slowly change over time. In late 2007, this clump has a single obvious brightness maximum, but in early 2008 a second maximum appears and the two maxima begin to drift apart. Sometime around day 50 of 2008, each of these two maxima splits again to produce a total of four maxima, all separating from each other. This transformation of one clump into multiple clumps is similar to that seen in the region 17$^\circ$ in front of Pan during 2009 described above. But in addition to these morphological changes, what is remarkable is that the clump is not moving at all during this time, which is inconsistent with any of the drift rates seen in Figure~\ref{clumptrackdet1}. Indeed, looking at Figure~\ref{clumptrackpan}, we notice that the clump closest to Pan (if it can be interpreted as a persistent feature) has moved alternately closer and further from Pan between 2004 and 2010.  Again, this indicates that the motions of these clumps cannot be easily described in terms of simple horseshoe motion, and we will re-consider this issue at the end of this report.

\subsection{Inner ringlet}

\begin{figure}
{\resizebox{3in}{!}{\includegraphics{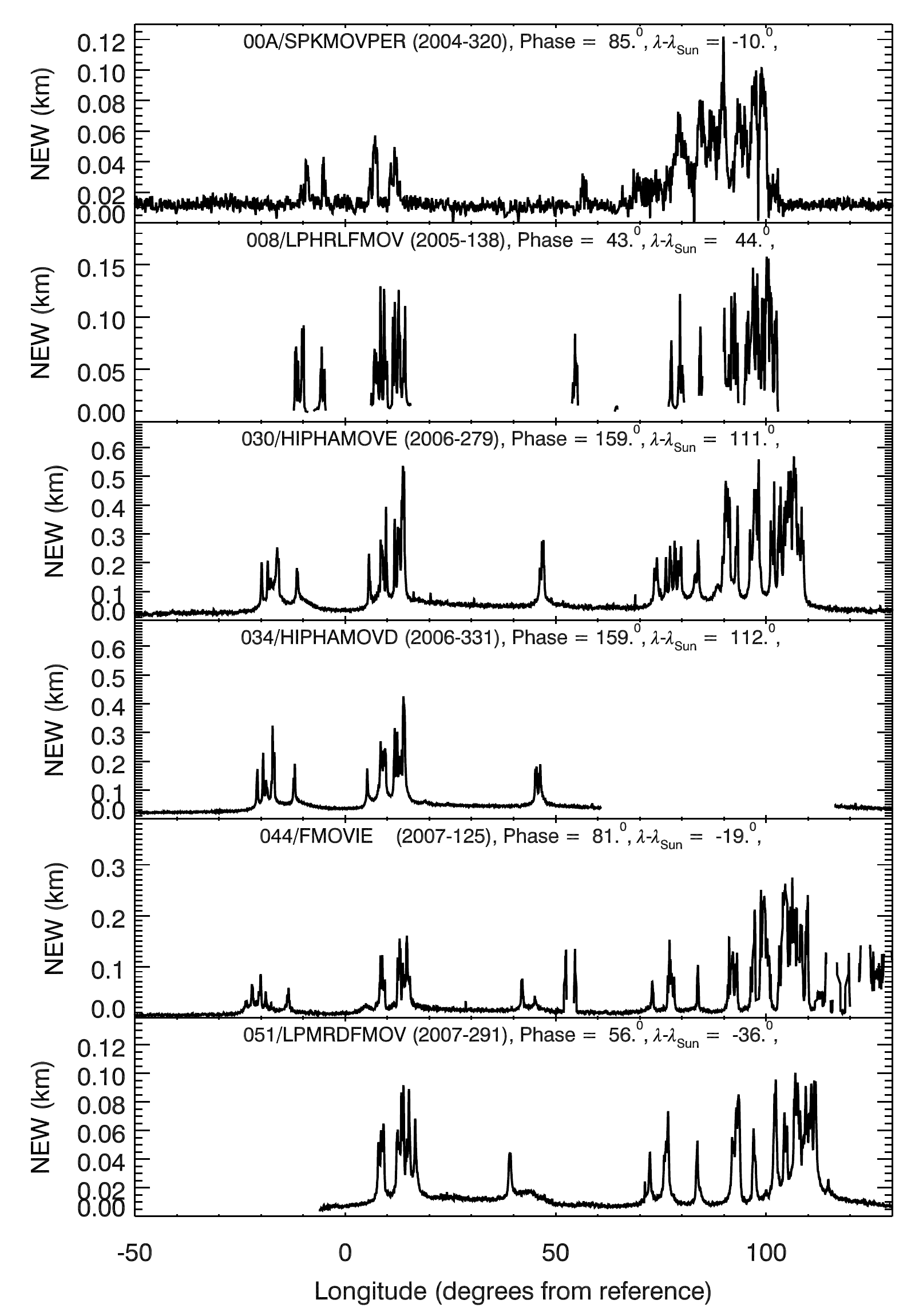}}}
{\resizebox{3in}{!}{\includegraphics{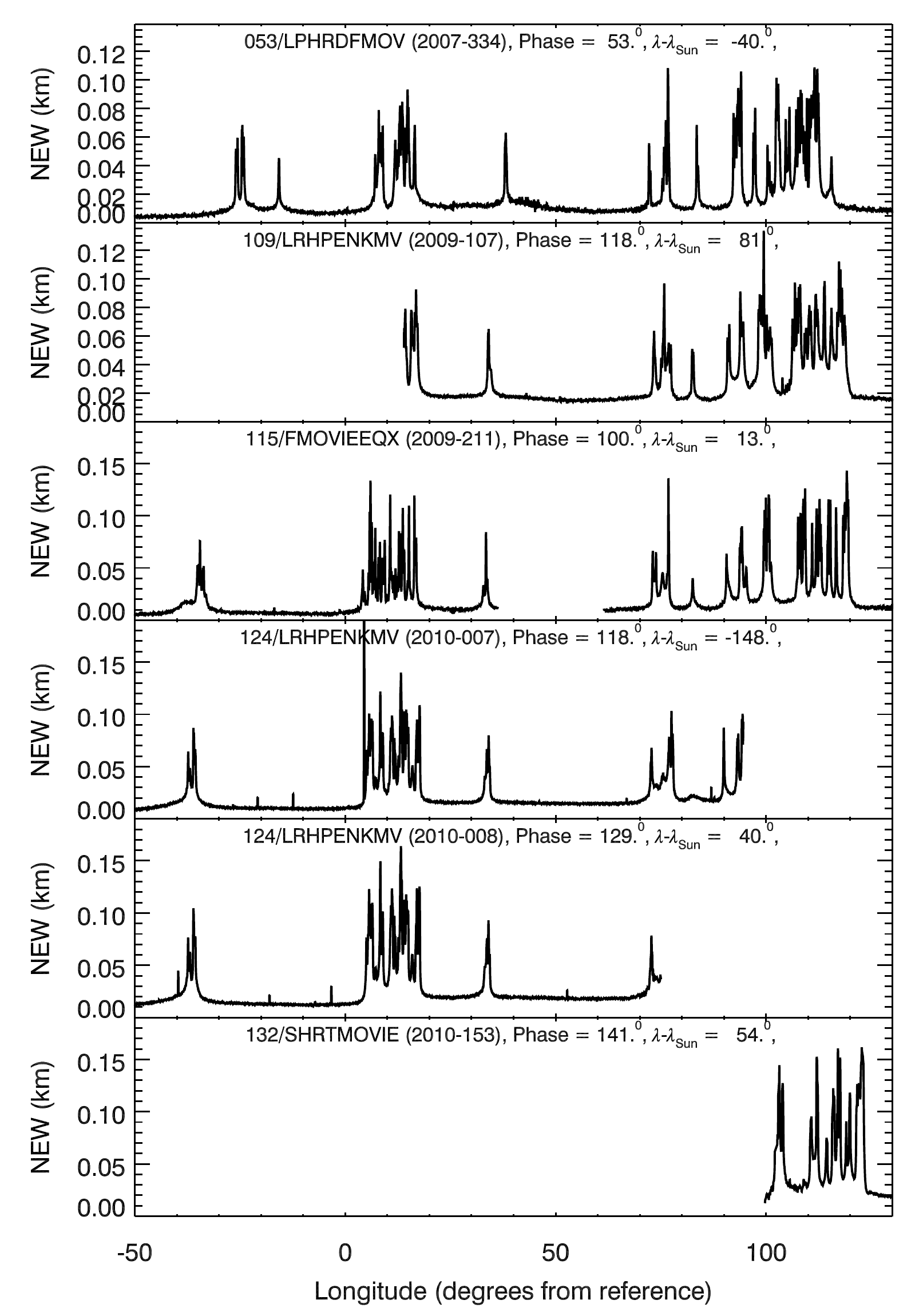}}}
\caption{Plot of the {\bf inner ringlet's} radially-integrated brightness (normal equivalent width)  versus co-moving longitude based on the data from the observations listed in Table~\ref{moslist}. 
This longitude system drifts forward relative to Pan at 0.7060$^\circ$/day with an epoch time of 170000000 ET (2005-142T02:12:15 UTC). These brightness profiles are all derived from Lorentzian fits to the ringlet, except for the 00A/SPKMOVPER observation, which is derived from direct  radial integration. Fits with peak radii more than 20 km from 133,490 km or peak widths greater than 100 km are removed and the remaining data smoothed over 5 samples to improve the display. Note the region in front of the clumps in the 044/ FMOVIE data is noisy due to nearby data gaps.}
\label{inb1}
\end{figure}

\begin{figure}
\centerline{\resizebox{5in}{!}{\includegraphics{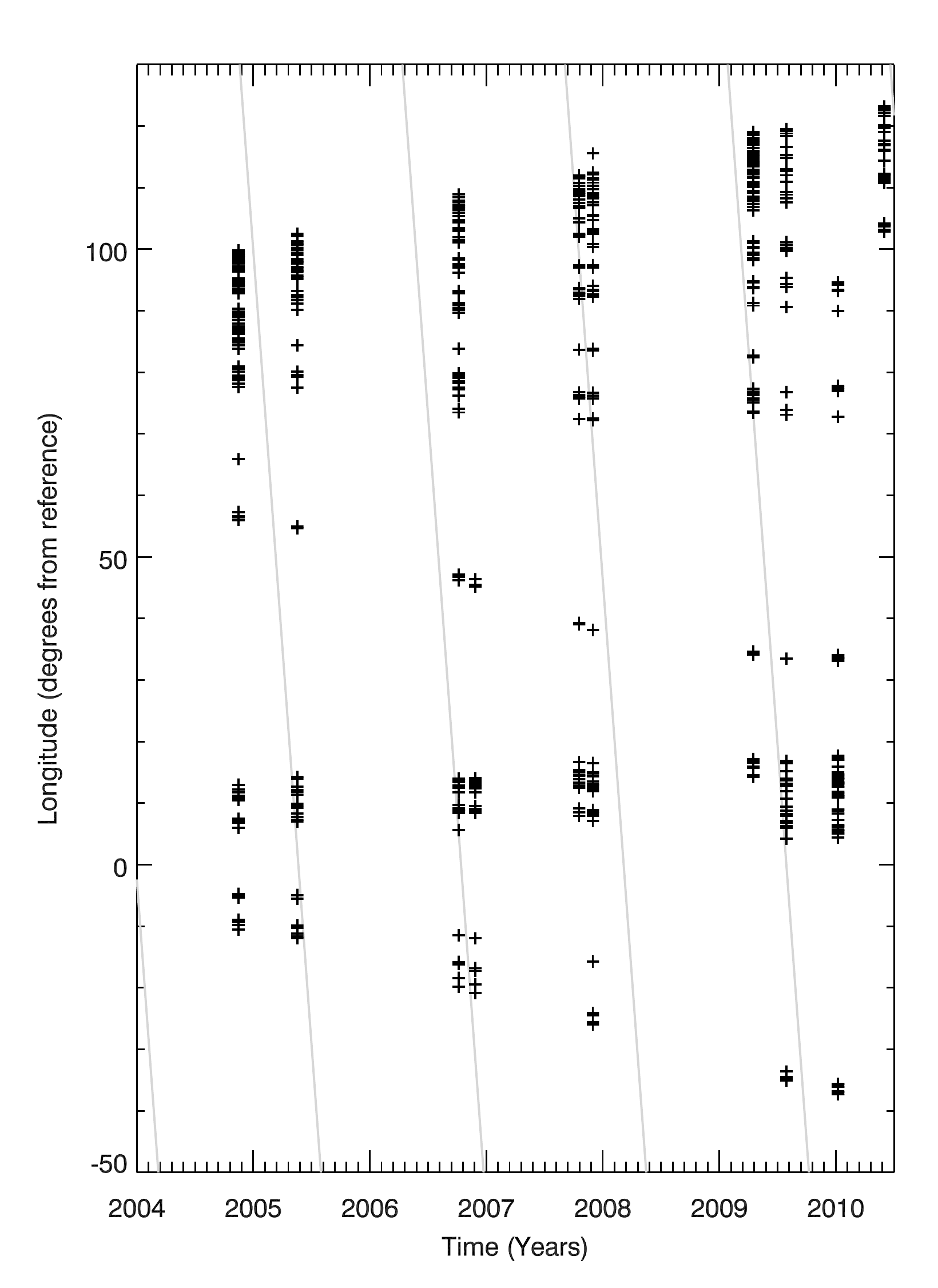}}}
\caption{Plot showing the locations of brightness peaks in the {\bf inner ringlet} as  functions of longitude and time.  The longitude system drifts forward relative to Pan at 0.7060$^\circ$/day with an epoch time of 170000000 ET (2005-142T02:12:15 UTC). The black plusses are measurements derived from the mostly complete mosaics shown in Figure~\ref{inb1}. Note that some clumps are missing at certain times owing to data gaps in the observations. The gray lines indicate Pan's longitude in this coordinate system.}
\label{clumptrackin}
\end{figure}

The inner ringlet data shown in Figures~\ref{inb1} and~\ref{clumptrackin}  are plotted in a longitude system that drifts {\em forward} relative to Pan at a rate of 0.7060$^\circ$/day, and has its origin at Pan's location at an epoch time of 170000000 ET (2005-142T02:12:15 UTC). Assuming the \citet{Jacobson06} values for Saturn's gravitational field parameters, this rate corresponds to a semi-major axis of 133,484 km, which is consistent with the observed location of this ringlet (Figure~\ref{egapprof}). When the clumps in this ringlet were first seen in 2004-2005, they could also be divided into  a few large groups. The largest cluster of clumps was located at co-rotating longitudes of about 90$^\circ$, while two smaller clusters  were found at +10$^\circ$ and -10$^\circ$.  Finally, an isolated clump could be seen around 60$^\circ$ co-rotating longitude. These clumps dispersed from a region 110$^\circ$ wide in 2004 to cover a region about 160$^\circ$ wide in 2010. This expansion is due to a combination of the steady backward drift of the most trailing set of clumps and the steady forward drift of the leading edge of the large clump cluster during this time.  However, the trailing edge of the large clump cluster remains fixed around $80^\circ$ during the same time period, so this cluster actually disperses during this time. Indeed, this group of clumps seem to split in two, with a gap forming around 85$^\circ$. The clump cluster around $10^\circ$ also does not move much in this coordinate system, but it does seem to spread and grow in complexity as time goes on. Finally, the isolated feature that was at $60^\circ$  in 2004 initially drifts backward at a steady rate, but then seems to stall sometime in 2008 or 2009 at a longitude of about 30$^\circ$.

 The fastest relative motions are between the two ends of the clump region, which drifted 0.025$^\circ$/day to 0.030$^\circ$/day relative to each other. This is comparable to the fastest drift rates observed in the Pan ringlet, indicating a basic similarity in the dynamics within these two regions. If these drift rates were simply due to differences in the particles' mean motions, this would imply that the clumps cover a semi-major axis range of about 4 km. However, as with the Pan ringlet, such an interpretation is questionable because the clumps do not always follow simple trajectories. For example, the clump initially at 60$^\circ$ went from drifting backwards at a rate of about 0.02$^\circ$/day to nearly motionless in this coordinate system, which would correspond to a semi-major axis shift of over 2 km if this clump were simply a test particle. While this clump did have conjunctions with Pan in early 2008 and 2009 (see Figure~\ref{clumptrackin}), these Pan encounters probably cannot explain the sudden deceleration of this clump. The expected semi-major axis shift experienced by a particle on a semi-major axis $a_P\pm\Delta a$ due to an encounter with Pan can be estimated by combining Equations 10.52 and 10.57 of \citet{MD99}:
\begin{equation}
{\delta a} \sim 3.3a\left(\frac{m_p}{M_S}\right)^2\left(\frac{a}{\Delta a}\right)^5.
\label{dap}
\end{equation}
For the inner ringlet,  $\Delta a \simeq 100$ km, so $\delta a$ is only 0.1-0.2 km, much smaller than the shift required to explain the change in this clump's drift rate. Again, the unusual accelerations of this clump suggest that the motions of these clumps are more complex than those of isolated particles.

\subsection{Outer ringlet}

\begin{figure}
{\resizebox{3in}{!}{\includegraphics{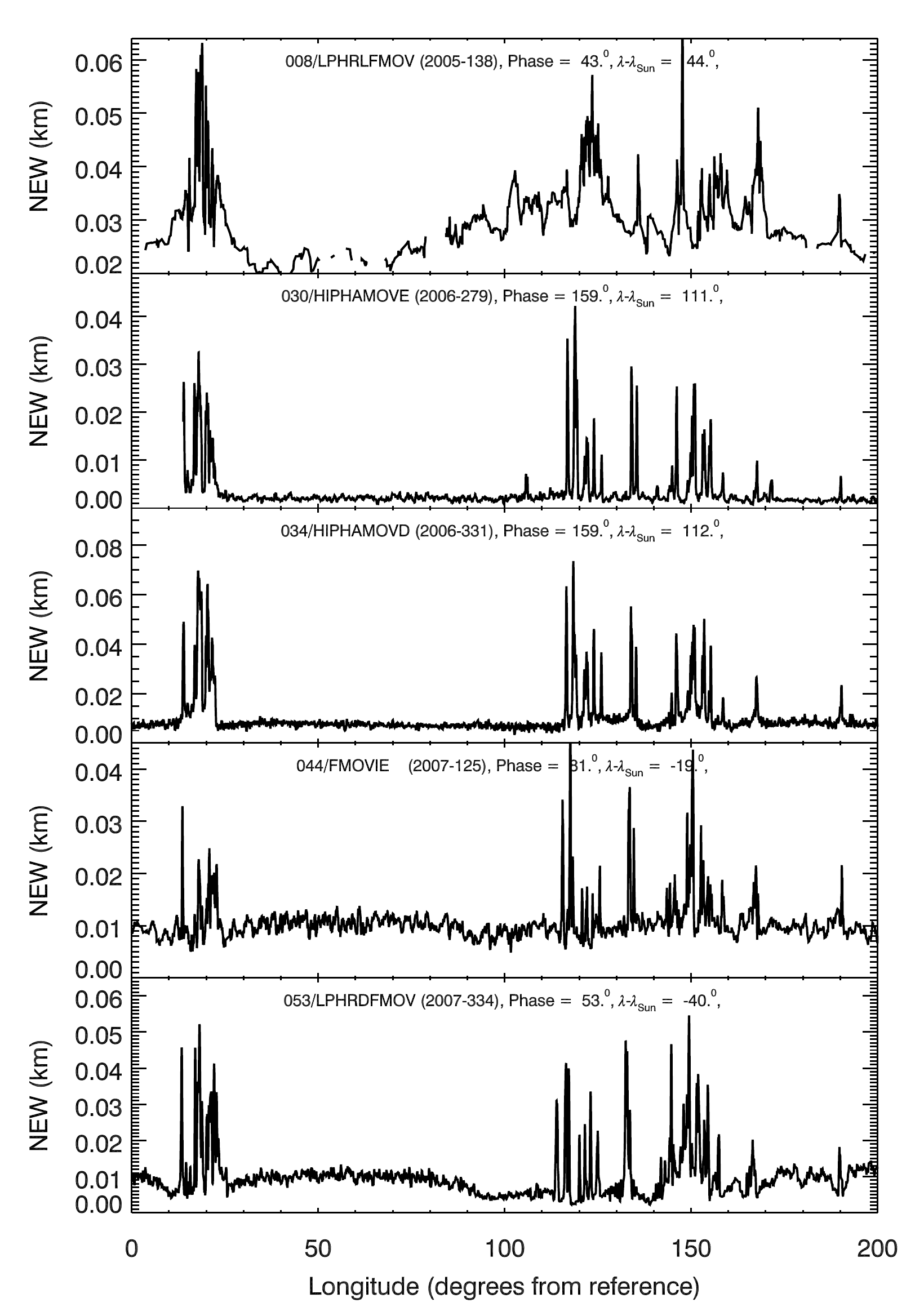}}}
{\resizebox{3in}{!}{\includegraphics{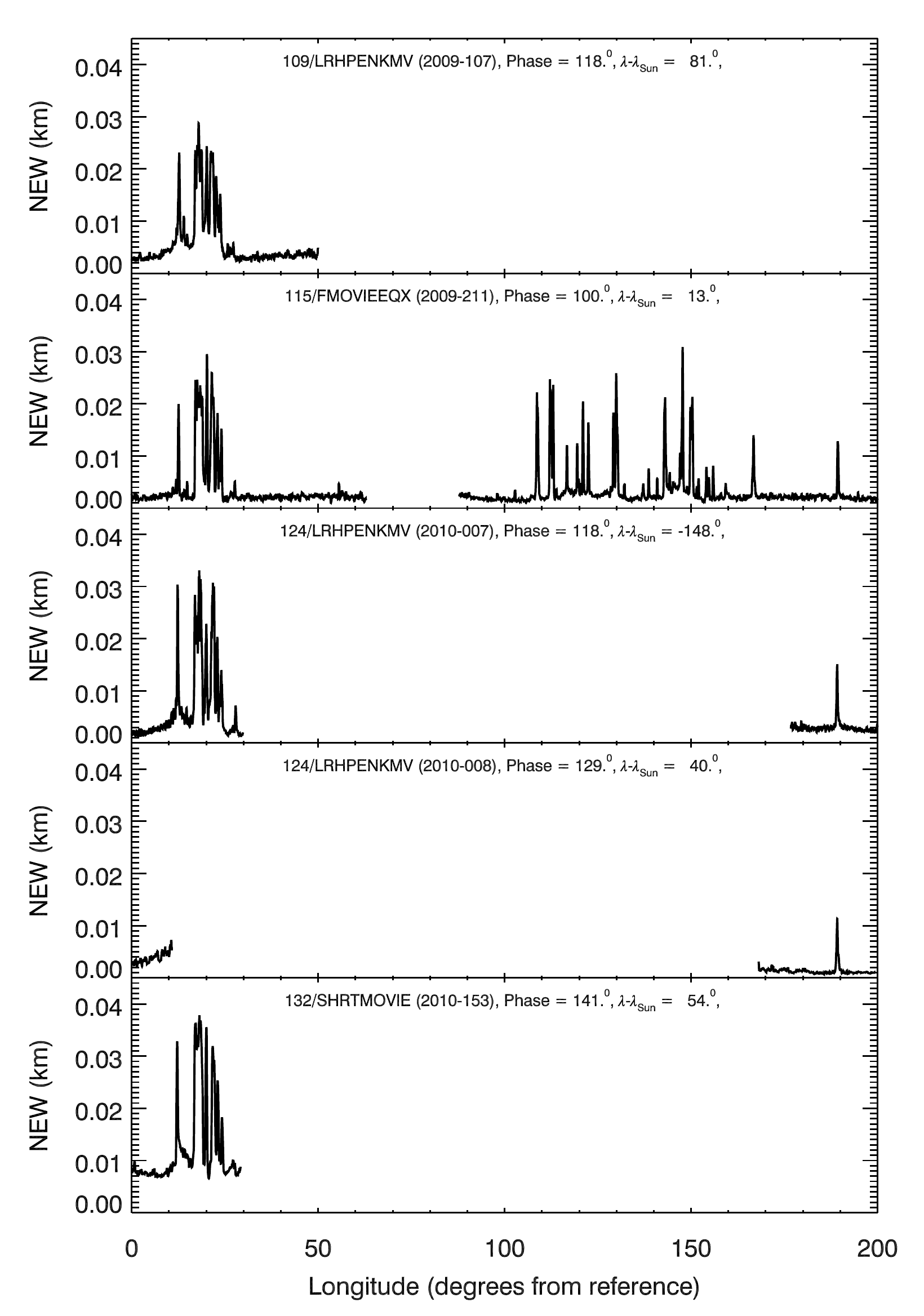}}}
\caption{Plot of the {\bf outer ringlet's} radially-integrated brightness (normal equivalent width) versus co-moving longitude based on the data from the observations listed in Table~\ref{moslist}. 
This longitude system drifts backwards relative to Pan at 0.9581$^\circ$/day with an epoch time of 170000000 ET (2005-142T02:12:15 UTC). These brightness profiles are all derived from Lorentzian fits to the ringlet, except for the 008/LPHRLFMOV observation, which is derived from direct integration. Fits with peak radii more than 20 km from 133,715 km or peak widths greater than 40 km  or less than 10 km are removed and the remaining data smoothed over 5 samples for the sake of clarity.}
\label{outb1}
\end{figure}

\begin{figure}
\centerline{\resizebox{5in}{!}{\includegraphics{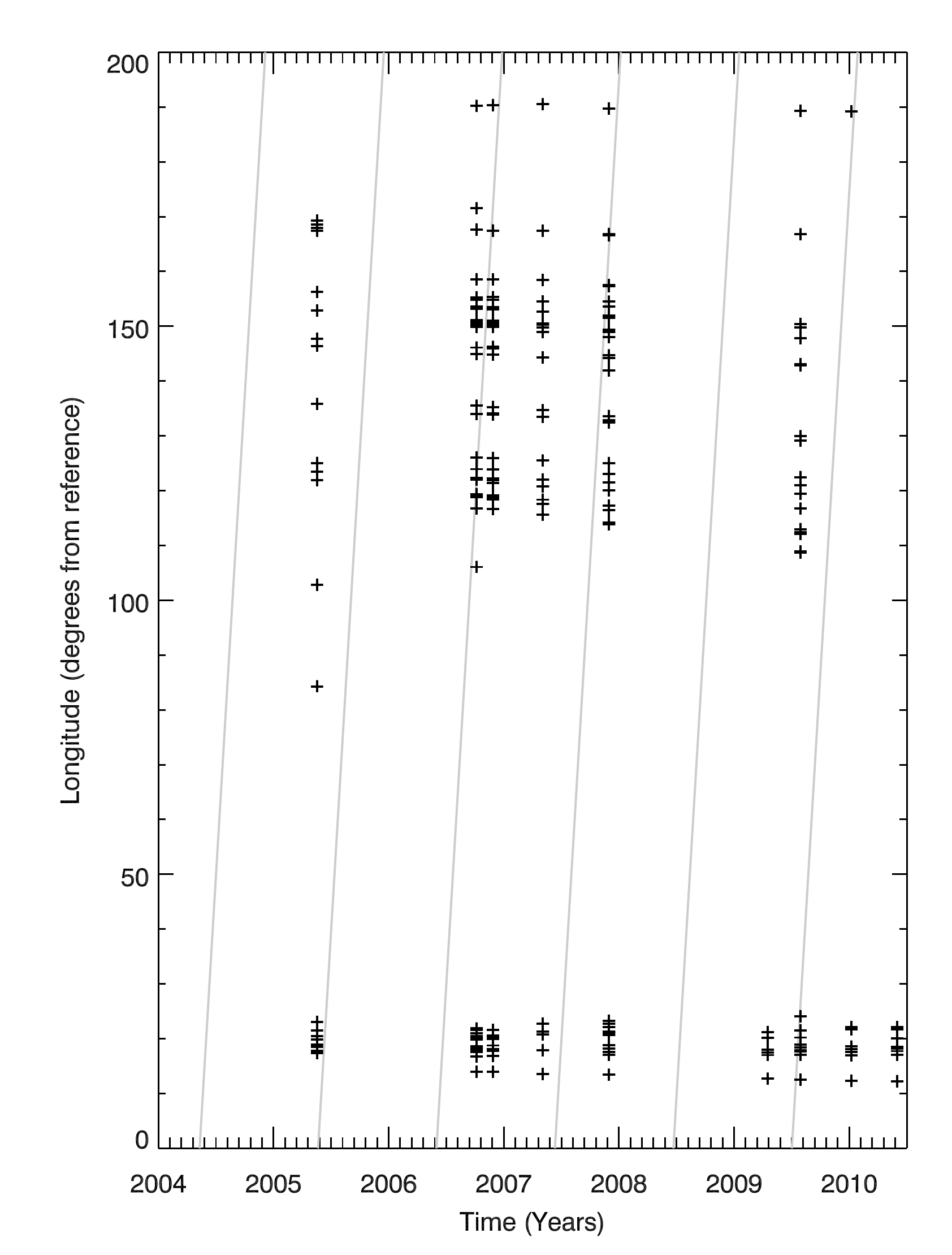}}}
\caption{Plot showing the locations of brightness peaks in the {\bf  outer ringlet}  as functions of longitude and time.  The longitude system drifts backward relative to Pan at 0.9590$^\circ$/day with an epoch time of 170000000 ET (2005-142T02:12:15 UTC). The black plusses are measurements derived from the largely complete mosaics shown in Figure~\ref{outb1}. Note that some clumps are missing in certain time periods due to data gaps in the observations. The gray lines indicate the longitude of Pan in this coordinate system. Note the data from 2005 were noisy, so the few  peaks between 80$^\circ$ and 110$^\circ$ are likely spurious.}
\label{clumptrackout}
\end{figure}

The outer ringlet data shown in Figures~\ref{outb1} and~\ref{clumptrackout} are plotted using a longitude system that drifts  {\em backwards} relative to Pan at a rate of 0.9581$^\circ$/day and has its origin at Pan's longitude at an epoch time of 170000000 ET (2005-142T02:12:15 UTC). This corresponds to a semi-major axis of 133,720 km assuming \citet{Jacobson06} values for Saturn's gravity field. Again, this semi-major axis is consistent with the observed location of the ringlet. Since this ringlet lies just 30 km interior to the Encke gap's outer edge, only 10 of the mosaics yielded useful profiles. Still, these are enough to document that the clumps in this ringlet form two well-separated groups. One tight cluster of clumps is located at a co-rotating longitude of about 20$^\circ$, while a more dispersed archipelago of peaks extends between about 110$^\circ$ and 160$^\circ$, with a  couple of outlying isolated clumps at 170$^\circ$ and 190$^\circ$.

Compared to the clumps in the Pan and inner ringlets, the clumps in the outer ringlet seem less time-variable. For example, the dense clump cluster always has a sharp isolated spike at about 12$^\circ$, a broader peak around 19$^\circ$, and a series of narrow spikes at larger longitudes. The pattern of narrow spikes between 110$^\circ$ and 200$^\circ$ is also remarkably repeatable across the observations. Indeed, the most obvious change in these clumps is a slight backwards drift of the material between 110$^\circ$ and 130$^\circ$ between 2007 and 2009. Even this drift is less than 0.01$^\circ$/day, so the relative drift rates in this ringlet are much less than those found in the other two ringlets.  If we assume the drifts are due to different particle mean motions, then these clumps would have a semi-major axis spread of only about 1.5 km, as opposed to the 4-km widths of the other two ringlets. However, given the trajectories of the clumps in the other two ringlets are inconsistent with those of test particle orbits, we caution against taking these numbers too literally. Nevertheless, the outer ringlet does appear to have a narrower radial profile than either the inner and outer ringlets (see Figure~\ref{egapprof}), so the particles in this ringlet may be more tightly confined in semi-major axis than those in the other two.

\section{Pan's perturbations on the other ringlets}

\begin{figure}
\resizebox{6in}{!}{\includegraphics{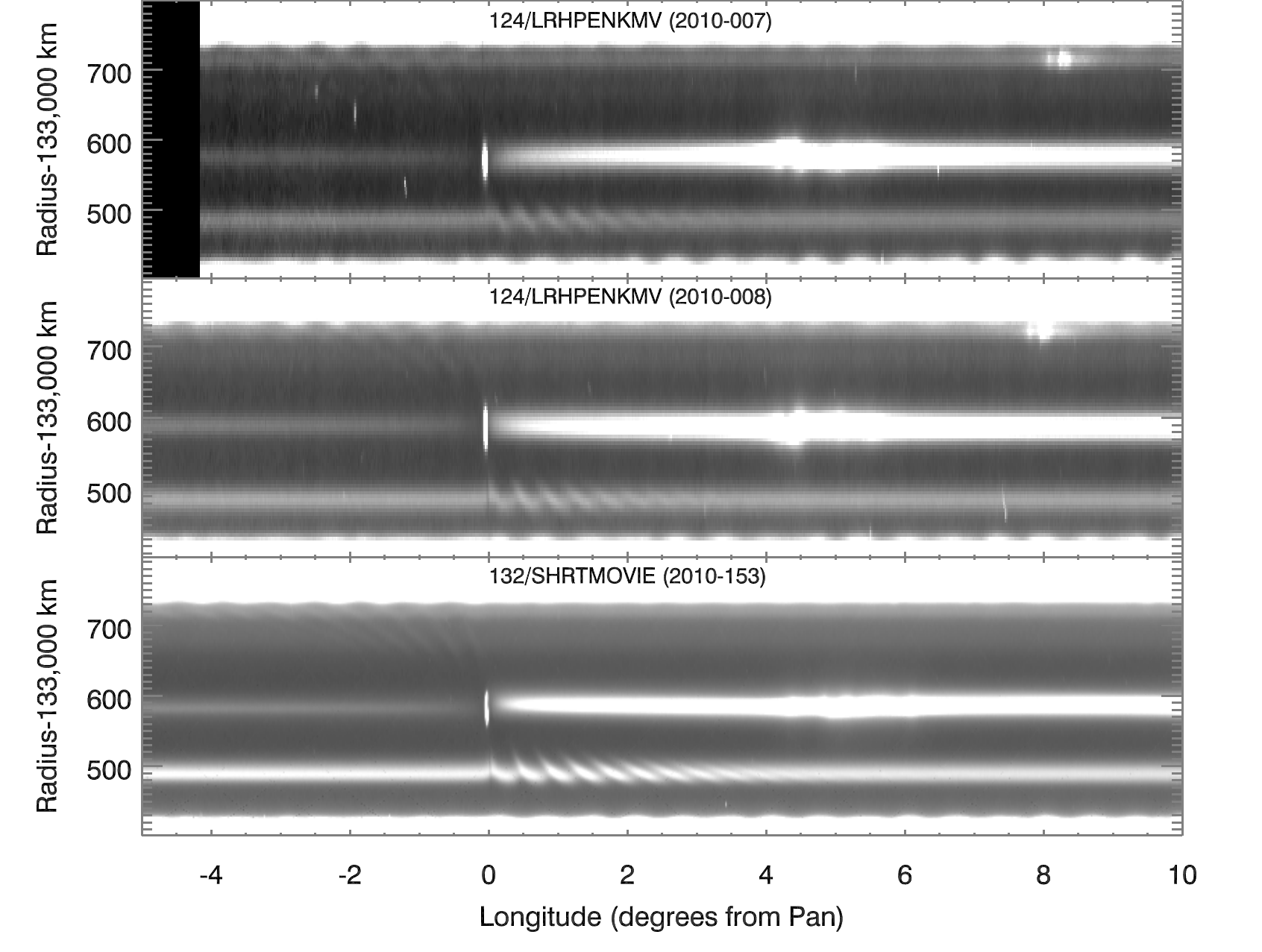}}
\caption{Images of the region around Pan in the three highest signal-to-noise mosaics. Note the Pan-induced waves and wakes in the inner, outer and fourth ringlets (as well as the gap edges). Also note the differences in the inner-ringlet's wave morphology among the observations, which are likely due to differences in the ringlet-particles' true anomalies prior to their conjunctions with Pan.}
\label{wakedispdet}
\end{figure}

\begin{figure}
\resizebox{3in}{!}{\includegraphics{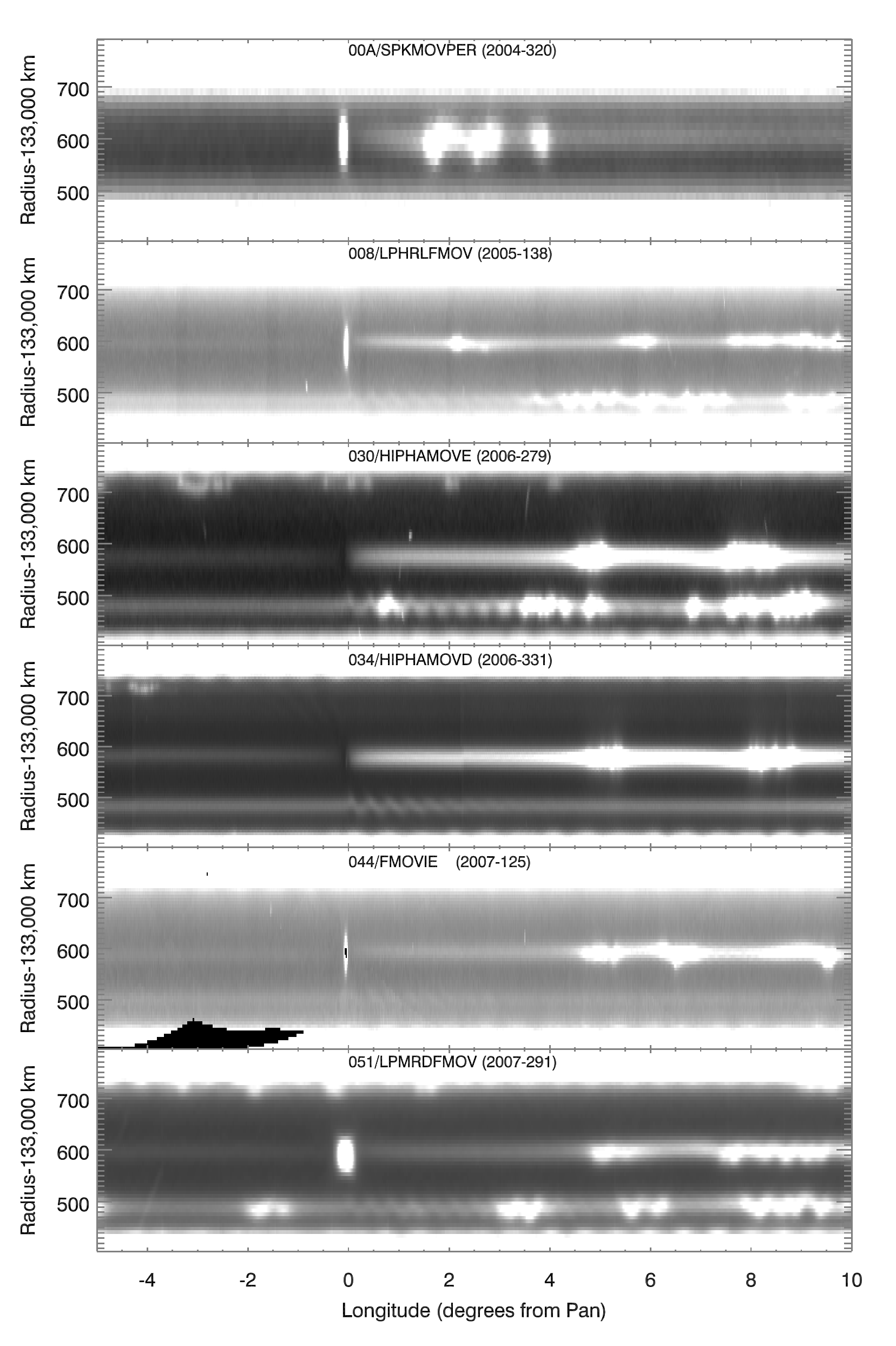}}
\resizebox{3in}{!}{\includegraphics{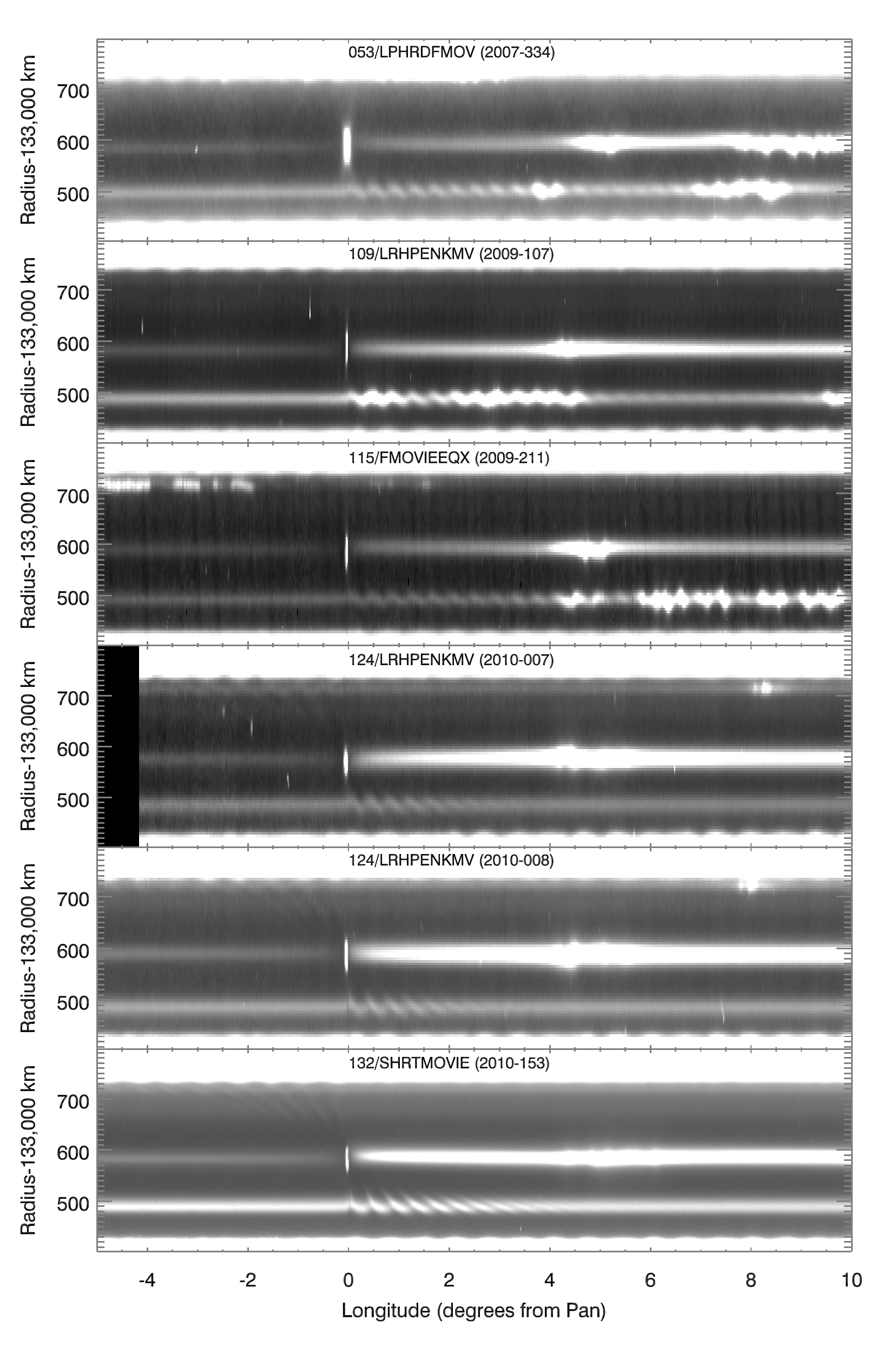}}
\caption{Images of the regions around Pan derived from most of the observations
listed in Table~\ref{moslist}. Note the waves in the inner ringlet generated by Pan's  gravitational perturbations. Whenever the disturbed part of the ringlet is clump-free, 
the wave damps within about 3$^\circ$. By contrast, the waves in the clumpy regions can persist over 10$^\circ$ downstream from the moon.}
\label{wakedispov}
\end{figure}

One way to probe the various ringlets' orbital properties is to examine how they respond to Pan's gravitational perturbations. These are most clearly seen in Figure~\ref{wakedispdet}, which shows close-ups of the region around Pan in the three highest signal-to-noise mosaics derived from the observations in Table~\ref{moslist}. In all these mosaics, the portion of the inner ringlet just in front of Pan exhibits periodic wiggles. Close inspection of these images reveals that the part of the outer ringlet immediately behind Pan also displays a series of wiggles, and  a similarly periodic brightness variation can even be seen in the fourth ringlet. All of these periodic patterns are likely due to Pan's gravitational perturbations on this ring material. 

Particles drifting past a massive object like Pan will have their orbits perturbed by the moon's gravity. If the particles were initially on circular orbits, then the moon's gravity throws the particles onto eccentric orbits with initially aligned pericenters (see Figure~\ref{hillplot}). These particles' organized epicyclic motion causes them to move in and out as they drift downstream of the moon, forming a series of ripples with a characteristic wavelength of $3\pi\Delta a$, where $\Delta a$ is the semi-major axis difference between the particles and the moon \citep{Dermott81, SB82}. The wavelengths of the ripples in both the inner and outer ringlets are consistent with this explanation. 

In reality,  the particles in these ringlets do not all have the same semi-major axis, so their epicyclic motions gradually slip out of phase, producing density variations like those seen in the fourth ringlet, and perhaps the inner ringlet as well. In dense rings, these density variations eventually lead to collisions that should cause any coherent pattern to dissipate. However, in these low optical depth ringlets, collisions are rare. Even so, as the epicyclic motions of the particles slip further and further out of phase,  any coherent pattern should eventually dissipate. The distance these patterns extend beyond Pan therefore provides information about the range of semi-major axes present in these ringlets. 

While the qualitative appearance of these structures is reasonable, a truly rigorous analysis of such structures would need to account for the fact that the particles do not approach Pan on circular orbits. For example, as we will discuss in more detail below, the inner ringlet possesses finite forced and free eccentricities. The orbital changes induced by Pan therefore depend not only on the particles' semi-major axis, but also their true anomalies during conjunction \citep{SB82, DQT89}. Indeed, if we compare the mosaics derived from the two LRHPENKMV observations from Rev 124, we can see some differences in the wave morphology in the inner ringlet that can be attributed to its finite eccentricity. In the earlier observation, the minima in radius appear to be sharper than the maxima, while in the later observation, which was obtained on the opposite side of the ring and thus viewed the same material half an orbital/epicyclic period later, the maxima appear to be sharper than the minima. Such patterns could be consistent with Pan's gravitational perturbations on an eccentric ringlet, but confirming this will require detailed simulations that are beyond the scope of this report.

While a rigorous analysis of these wavy patterns is not feasible here, we can use fairly simple arguments to obtain some useful insights into the semi-major axis dispersion in different regions of the inner ringlet. Consider Figure~\ref{wakedispov}, which shows close-ups of all the relevant mosaics. These reveal that the ripples in the inner ringlet extend different distances downstream from Pan depending on whether the disturbed region contains clumps or not. When there are no clumps in the disturbed region (the Rev 34 HIPHAMOVD, Rev 44 FMOVIE, Rev 124 LRHPENKMV and Rev 132 SHRTMOVIE observations), the ripples in the inner ringlet dissipate within a few degrees of Pan. By contrast, when the disturbed region does contain clumps, as in the Rev 008 LPHRLFMOV, Rev 030 HIPHAMOVE, Rev 51 LPMRDFMOV, Rev 053 LPHRDFMOV, Rev 109 LRHPENKMOV and Rev 115 FMOVIEEQX observations, the ripples can persist as far as 10$^\circ$ downstream of Pan. Since the distance the ripples extend downstream of Pan is set by the semi-major axis dispersion within the ringlet, this suggests that the clumps contain particles with a smaller range of semi-major axes than the rest of the ringlet.

We can make this qualitative  observation a bit more quantitative if we assume the center of the ringlet is $\Delta a$ from $a_P$, and the ringlet consists of particles with a range of semi-major axes $\delta a$.  In this case, we expect any coherent pattern produced by Pan to smear out when the epicyclic motions of particles at $\Delta a \pm \delta a$ are out of phase by 180$^\circ$. This will occur at a distance $x_d$ downstream from Pan where $x_d=3\pi(\Delta a + \delta a/2)(N-1/4)$ and $x_d=3\pi(\Delta a - \delta a/2)(N+1/4)$ for the same $N$. This condition is satisfied when $ N \simeq \Delta a/(2\delta a)$, or when  $x_d \simeq (3\pi/2) \Delta a^2/\delta a$. 

In the clump-free regions of the Pan ringlet, the wave seems to damp within $1^\circ-2^\circ$ of Pan, so $x_d$ is between 2500 and 5000 km, which corresponds to a semi-major axis spread $\delta a$ between 10 and 20 km. By contrast, in the clumpy regions the waves extend over $10^\circ-15^\circ$, implying damping lengths of order 30,000 km, and semi-major axis spreads of order 1-2 km. Both of these numbers are reasonable, given the overall width of the ringlet, the persistence of the clumps, and the slow drift rates of clumps relative to each other. 

\section{Ringlet Orbital Parameters}

The above analysis of the distribution and evolution of the clumps in these various ringlets reveals some surprising patterns. In particular, the relative motions of these features are inconsistent with those expected for clumps of material moving in the combined gravity fields of Saturn and Pan. Thus, in order to better understand the dynamics of both these features and the ringlets as a whole, we will now use the apparent radial positions of these ringlets to investigate their orbital properties. 

The following studies will focus exclusively on the Pan and inner ringlets because both these ringlets are sufficiently far from the Encke Gap edges that our fitting algorithms can yield reliable estimates of their radial positions. By contrast, for most of the observations considered here, the outer ringlet is only barely resolved from the outer gap edge. While  our ringlet-fitting procedures can still provide useful information about the morphology and distribution of the clumps in the outer ringlet, the corresponding radial position estimates are more sensitive to the background signal from the nearby gap edge. Obtaining robust estimates of the outer ringlet's position is particularly difficult outside of the clumps, where the ringlet is comparatively faint. As will become clear below, detailed comparisons among multiple observations over a broad range of longitudes are needed to make sense of the radial positions of the inner and Pan ringlets. At present, the outer ringlet data are not sufficient to do these comparisons,  so we will not examine the radial structure of the outer ringlet further here.

Determining the orbital properties of the clumpy  inner and Pan ringlets is not as straightforward as measuring the shapes of such non-circular ring  features  like the dense Huygens ringlet or even the dusty ringlet in the outer Cassini Division. The shapes of the latter ring features can be determined by simply measuring their radial positions at multiple inertial longitudes, provided we assume that the ring particles' orbital properties are the same at all co-rotating longitudes. This, however, is clearly not a valid assumption for clumpy features like the Encke Gap ringlets. Instead, we can only obtain useful information about the Encke Gap ringlets' orbital properties by comparing observations  of the same co-rotating longitudes $\lambda_c$ at different inertial longitudes $\lambda_i$. This obviously complicates the analysis, and forces us to focus our attention on a few particularly informative data sets.  Furthermore, many of the relevant observations can only provide sensible orbital information if the ringlets are assumed to exhibit ``heliotropic'' behavior similar to that previously  identified in a dusty ringlet in the Cassini Division \citep{Hedman10}. While this was not unexpected, given that both this Cassini Division ringlet and the Encke Gap ringlets are made out of comparably small particles \citep{Hedman11}, it does further complicate the analysis of the ringlets' radial structure.

After summarizing the theory and formalism for describing heliotropic ring features,  we first consider  the Rev 045 PANORBIT data, which yield complete orbit information for a small part of the Pan ringlet in the vicinity of the moon at one time.  Then we examine the Rev 124 LRHPENKMV data, where multiple  clumps in both the Pan and inner ringlets were observed at two very different  inertial longitudes. These observations clarify that the kinks associated with the clumps in both ringlets are due to variations in the particles' orbital eccentricites. Finally, we use the mosaics illustrated in Figure~\ref{mos1} to study the large-scale variations in these ringlets' orbital properties. 

\subsection{Properties of heliotropic ringlets}
\label{helio}

\citet{Hedman10} provide a detailed discussion of the dynamics of narrow heliotropic ringlets, based on observations of the dusty ``charming ringlet'' in the Cassini Division's Laplace Gap. That ringlet exhibits systematic variations in its observed radial position in a coordinate system fixed relative to the Sun, such that the geometric center of that ringlet was displaced away from Saturn's center towards the Sun. This unusual behavior is due to solar radiation pressure producing a forced eccentricity $e_f$ in the orbits of the tiny grains that form this ringlet \citep{BHS01}. However, the shape of this ringlet also varied with time. These variations could be modeled by assuming the ringlet traced out the orbit of a particle with both a forced eccentricity generated  by solar radiation pressure and a free eccentricity precessing around the planet at the local rate. While it remains unclear what process coordinates the particles' motions within the ringlet so as to maintain this free eccentricity,  this model still provides a useful way to parameterize the ringlet's morphology. As we will demonstrate below, the dusty Encke Gap ringlets also exhibit time-variable eccentricities that can be modeled as a forced component aligned with the Sun and a freely-processing component. We will therefore employ this decomposition to describe the shape of the Encke-Gap ringlets.

None of the observations to date indicates that the Encke-Gap ringlets have any detectable inclination, so (for the sake of simplicity) these ringlets will be assumed to lie exactly in Saturn's equatorial plane, In that case, the radial position of a heliotropic ringlet as a function of inertial longitude $\lambda_i$ can be expressed as:
\begin{equation}
r(\lambda_i, t)=a-ae(t)\cos[\lambda_i-\varpi(t)],
\end{equation}
where the eccentricity $e$ and pericenter $\varpi$ are slowly-varying functions of time. These quantities are given by:
\begin{equation}
e\cos(\varpi-\lambda_\Sun)=-e_f+e_l\cos(\varpi_l+\dot{\varpi}_lt)
\end{equation}
\begin{equation}
e\sin(\varpi-\lambda_\Sun)=e_l\sin(\varpi_l+\dot{\varpi}_lt),
\end{equation}
where $\lambda_\sun$ is the Sun's inertial longitude, $e_f$ is the forced eccentricity induced by solar radiation pressure,  and $e_l$, $\varpi_l$ and $\dot{\varpi}_l$ parametrize the magnitude, orientation and precession rate of the free component of the eccentricity, respectively. Note that since the alignments of the free and forced eccentricities  have different time-dependencies, these two components of the total eccentricity can be separated from one another by comparing measurements made at different times. For the purposes of this analysis, we will assume that the free eccentricity's precession rate $\dot{\varpi}_l$ is basically the precession due to Saturn's finite oblateness, $\dot{\varpi}_0$, which is $3.2^\circ$/day in the Encke Gap. Thus the orbital properties of the ringlet are specified by the parameters $a$, $e_f$, $e_l$ and $\varpi_l$, which for the Encke Gap ringlets may be functions of co-rotating longitude $\lambda_c$.

\subsection{Orbital elements of the Pan ringlet near Pan}

\begin{figure}
\centerline{\resizebox{4.5in}{!}{\includegraphics{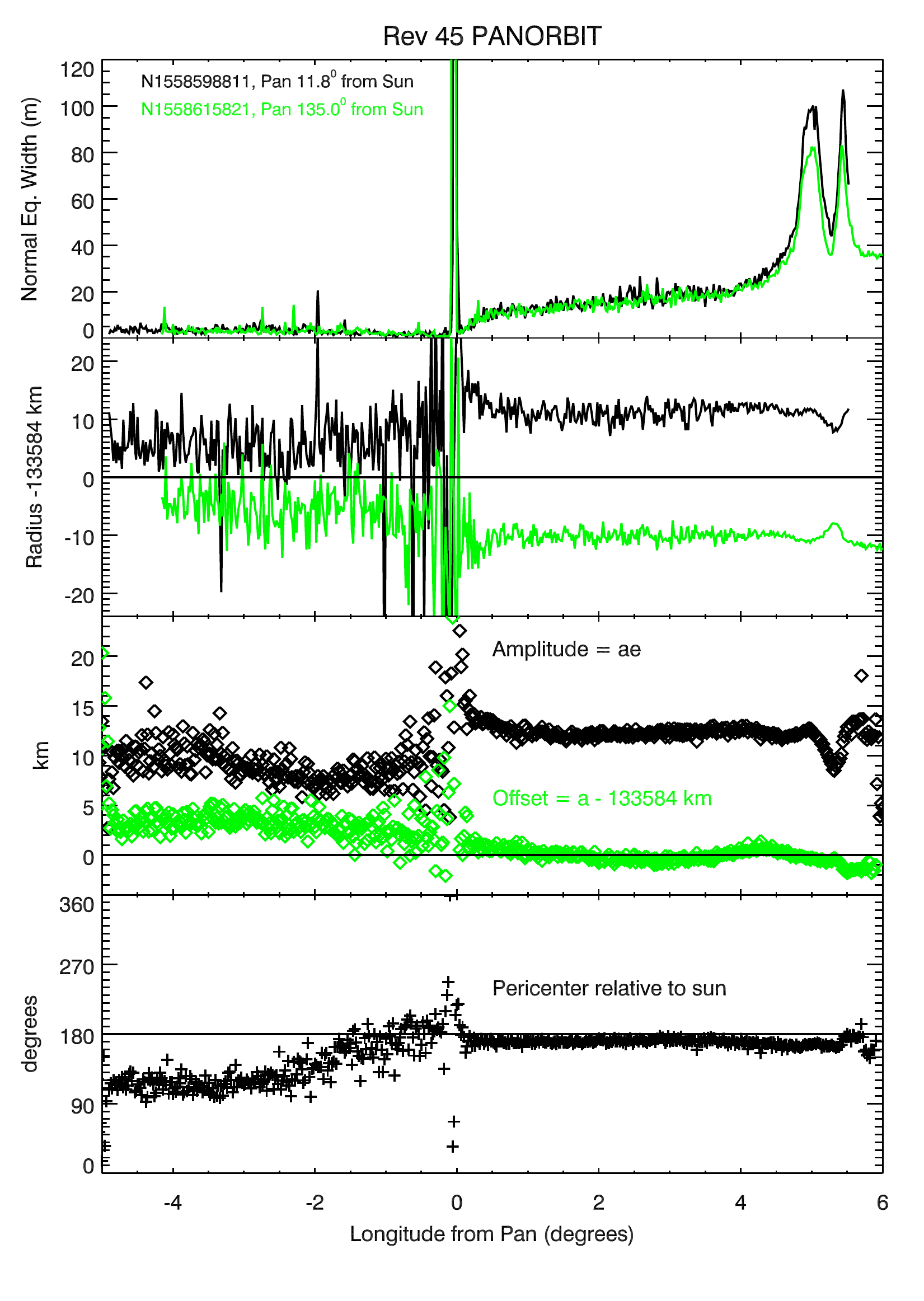}}}
\caption{Orbital elements of the Pan ringlet derived from the PANORBIT observation. The top two panels show the integrated brightness and radial position of the Pan ringlet derived from two images, one taken close to the sub-solar longitude, and the other taken near Saturn's shadow. Note that the ringlet is found displaced outward from Pan's orbit on the sunward side of the rings, and inwards on the side near Saturn's shadow. The bottom two panels show the ringlet's  semi-major axis, eccentricity and pericenter longitude derived from all the useful images in this sequence. Statistical error bars are not plotted for reasons of clarity, but are consistent with the scatter in the estimates (i.e. they are around 0.5  km in $a$ and $ae$ and 5$^\circ$ in the pericenter in front of Pan, and 1-2 km in $a$ and $ae$ and 10-20$^\circ$ in the pericenter behind Pan). In front of Pan, the ringlet has a semi-major axis close to that of Pan, a finite eccentricity, and a pericenter anti-aligned with  the Sun. Note that the eccentricity is reduced in the vicinity of the bright clumps between 5$^\circ$ and 6$^\circ$. Behind Pan, where the ringlet is fainter, the semi-major axis is systematically outside the orbit of Pan and the pericenter deviates from exactly 180$^\circ$. }
\label{panorbit}
\end{figure}

The PANORBIT observation from Rev 045 is a useful starting point for investigations of the ringlet's orbital properties because it consists of 158 images of Pan and the surrounding rings as the moon moved around the planet. The resulting images cover roughly 210$^\circ$ in true anomaly, with some gaps where the planet appeared behind the rings or when the rings themselves were in Saturn's shadow. These images were all re-projected onto a common scale in radius and longitude relative to Pan (sampling distances of 5 km and 0.02$^\circ$ respectively), and then the radial brightness profile at each longitude in each scan was fit to a Lorentzian in order to estimate the integrated brightness and radial position of the Pan ringlet. However, due to the changing viewing geometry and  resolution of the images over the course of the observation, the radial position estimates had to be refined  based on measurements of the position of the Encke-Gap's edges in each image. 

For each longitude in each image, the locations of both gap edges  were estimated as the points of maximum slope in the radial brightness profile, which were found by fitting peaks to the {\em  derivative} of the brightness profile. The edge waves generated by Pan cause the radial positions of both edges to vary by a few kilometers within each image, so we did not individually  adjust each estimate of the ringlet's radial position. Instead, we simply computed a single offset for each image based on the median deviation of both edges from their nominal positions at 133,423 km and 133,745 km. The resulting offsets varied over a range of about 6 km with an $m=2$ pattern. Such a pattern would not be confused with the $m=1$ pattern due to a real eccentricity, but removing these offsets still improves the reliability of the subsequent analysis.

The top pair of panels of Figure~\ref{panorbit} show two representative profiles of the Pan ringlets' brightness and radial position derived from two images in the PANORBIT sequence. One of these images (N1558598811) was obtained when Pan was only 12$^\circ$ from the sub-solar longitude, while the other (N1558615821) was obtained when Pan was over $130^\circ$ from the sub-solar longitude, and thus closer to Saturn's shadow. The integrated brightness profiles derived from these two images are very similar, up to an overall normalization that can probably be attributed to slight differences in the phase angles of the two observations (83$^\circ$ versus 76$^\circ$) and small uncertainties in the background subtraction. However, the radial position of the ringlet in the two images show clear systematic differences. The observation taken when Pan was near the sub-solar longitude shows the ringlet displaced exterior to Pan's semi-major axis at 133584 km, while the data taken closer to Saturn's shadow are shifted towards smaller radii. These variations in the apparent radial position of the ringlet around Pan can be most easily explained if the ringlet particles are on eccentric orbits with aligned pericenters. Furthermore, the directions of these displacements are consistent with the ringlet being heliotropic, with a forced eccentricity that tends to place the particles' orbital pericenters 180$^\circ$ from the Sun. At the same time, it is also apparent that the orbital properties of the ringlet depend upon the co-rotating longitude relative to Pan. The most obvious example of this is the distinct ``kink'' in the ringlet's radial position associated with the bright clumps around 5$^\circ$ in front of Pan.

Images from a single observing sequence (i.e. taken at a single time) do not provide sufficient information to determine all the parameters in a heliotropic model: $a$, $e_f$, $e_l$ and $\varpi_l$.  However, we can derive estimates of the instantaneous values of $a$, $e$ and $\varpi$ at each co-rotating longitude by fitting the observed radial positions $r$ from all the relevant images  to the function:
\begin{equation}
r=a-ae\cos(\lambda_i-\varpi).
\end{equation}
Note that due to variations in the viewing geometry, the range of $\lambda_i$ observed depends somewhat on $\lambda_c$. Also note that images obtained when the ring was in shadow, backlit by the planet, or yielded radial positions more than 50 km from 133584 km were excluded prior to performing these fits. Based on the residuals to these fits, we estimate the statistical uncertainties on these parameters are around 0.5  km in $a$ and $ae$ and 5$^\circ$ in $\varpi$ for longitudes in front of Pan (where the signal is stronger), and 1-2 km in $a$ and $ae$ and 10-20$^\circ$ in $\varpi$ for longitudes behind Pan

The bottom two panels of Figure~\ref{panorbit} show the estimated values of $a$, $ae$ and $\varpi$ as functions of co-rotating longitude relative to Pan.  These plots indicate that for the portion of the ringlet in front of Pan, $a$ is close to Pan's semi-major axis, $ae$ is around 12 km, and the orbital pericenter is almost exactly 180$^\circ$ from the Sun. On the  other hand, the part of the ringlet falling behind Pan displays a slightly lower eccentricity, a pericenter that gets as far as $80^\circ$ from the anti-Sun direction, and a semi-major axis that is displaced by about 3 km exterior to $a_P$.  

No single observation can prove that this ringlet is heliotropic, but $\varpi$ always being almost exactly 180$^\circ$ from the Sun at all longitudes in front of Pan is certainly consistent with what one would expect for a heliotropic ringlet with  $e_f>>e_l$. However, since the pericenter does deviate from $\lambda_\sun+180^\circ$ behind Pan, the entire ringlet cannot just have eccentricities forced by solar radiation pressure. Both these results are consistent with the analysis of the mosaics described at the end of this section, which provides separate estimates of $e_f$ and $e_l$. 

While these data do not provide strong constraints on the origin of the ringlet's eccentricity,  they do clearly demonstrate that the kink in the ringlet's radial position at $5^\circ$ corresponds to a region of reduced eccentricity.  By contrast, neither $a$ nor $\varpi$ vary noticeably within this region. 

\subsection{Orbital element variations associated with clumps}

\begin{figure}
\resizebox{6in}{!}{\includegraphics{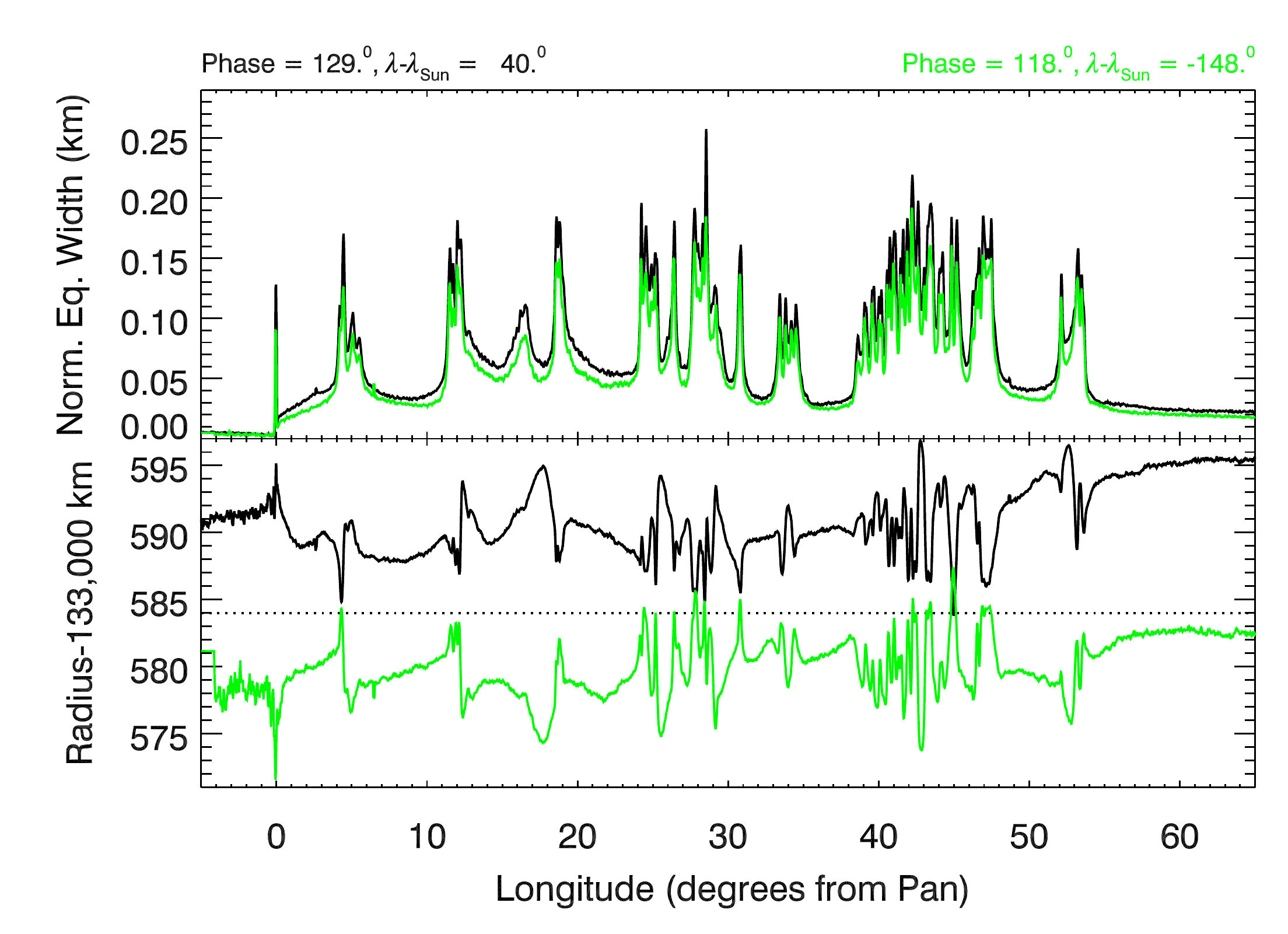}}
\caption{The integrated brightness and radial position of the clumps in the Pan ringlet obtained from the Rev 124 LRHPENKMV observations. These profiles were derived from Lorentzian fits to the radial brightness profiles whose radial scales were refined using the position of the Encke-Gap's outer edge. Fits with peak radii more than 20 km from 133,584 km are removed and the remaining data smoothed over 5 samples for the sake of clarity. These two observations imaged the same ring region at two different longitudes, one close to the sub-solar point and one close to Saturn's shadow. Note that the variations in the radial position of the ringlet are reversed at the two locations, suggesting that the observed kinks in the ringlet are due primarily to eccentricity variations.}
\label{mov124pan}
\end{figure}

\begin{figure}
\resizebox{6in}{!}{\includegraphics{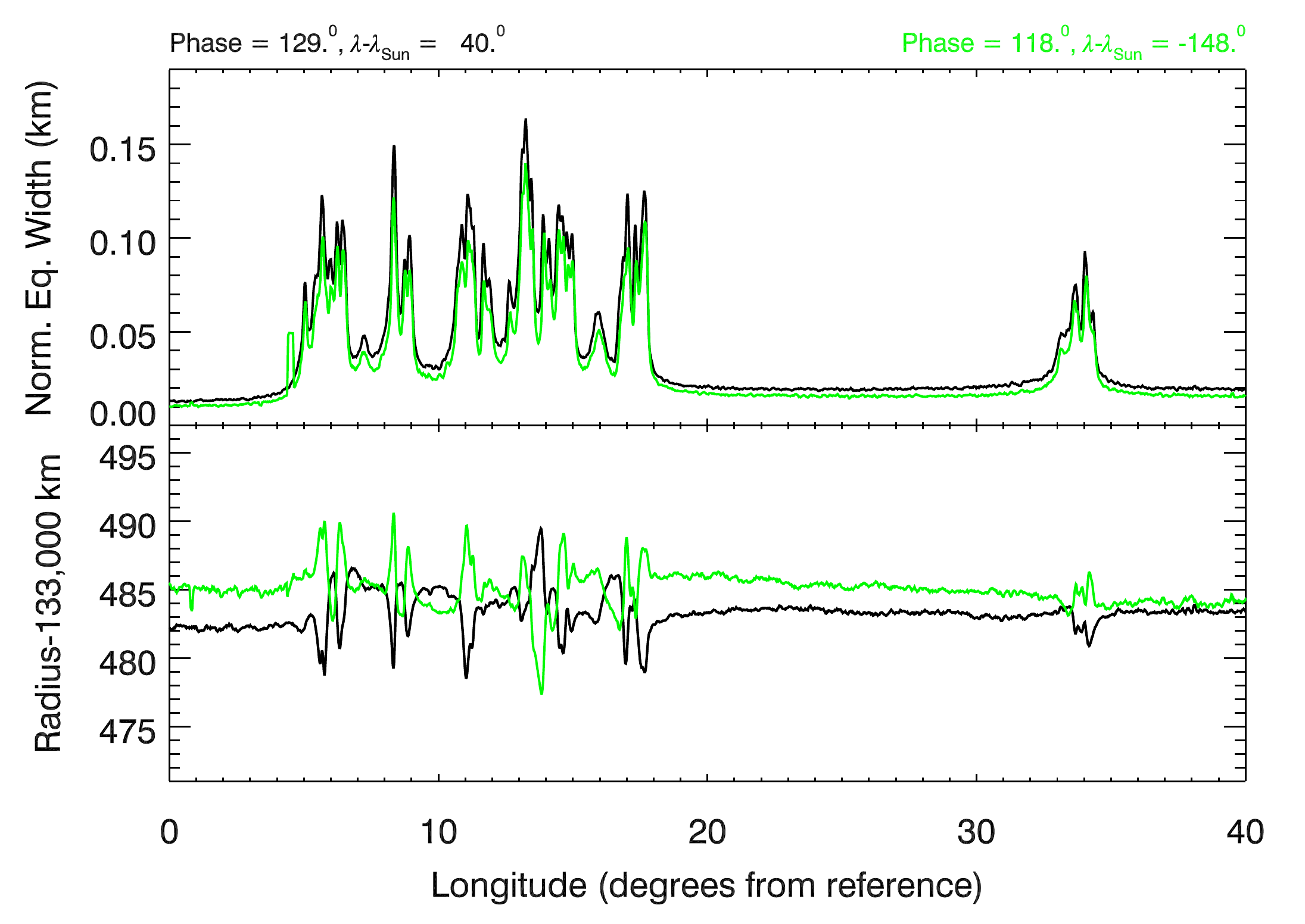}}
\caption{Brightness and radial position profiles of the clumps in the inner ringlet obtained from the Rev 124 LRHPENKMV observations. These profiles were derived from Lorentzian fits to the relevant brightness profiles whose radial scales were refined based on the observed positions of the Encke Gap's outer edge. Fits with peak radii more than 10 km from 133,484 km, widths greater than 100 km or less than 10 km, or peak brightnesses greater than 0.02 are removed and the remaining data smoothed over 5 samples for the sake of clarity. The longitude system used here drifts forward relative to Pan at 0.7060$^\circ$/day with an epoch time of 170000000 ET (2005-142T02:12:15 UTC).  These two observations imaged the same region in the ring at two different longitudes, one close to the sub-solar point and one close to Saturn's shadow. Note that the variations in the radial position of the ringlet are reversed at the two locations, suggesting that the observed kinks in the ringlet are due primarily to eccentricity variations.}
\label{mov124in}
\end{figure}

The LRHPENKMV observation sequence from Rev 124 was deliberately designed to investigate the orbital properties of the kinks in the Encke Gap ringlets. During this observation, the camera first stared at a point in the Encke Gap near the sub-solar longitude, then it looked at a point on the opposite side of the rings, near Saturn's shadow. The timing of these two pointings was chosen so that the same co-rotating longitudes would be observed at both locations. 

Figures~\ref{mov124pan} and~\ref{mov124in} show the  integrated brightness and radial position profiles for both Pan and inner ringlets derived from these observations. Again, the radial position estimates were refined based on the observed positions of the Encke Gap edges in the observed mosaics. Since we are looking at regions immediately in front of Pan, only the less-disturbed outer edge of the gap was used for this purpose. This edge position was measured by fitting a peak to the derivative of the radial brightness profiles. The edge positions were low-pass filtered using a 2$^\circ$ wide boxcar to remove fine-scale structure associated with the wavy edges, and then used to compute a correction that would place the smoothed edge at 133,745 km at all co-rotating longitudes. These corrections remove some broad-scale ripples in the ringlets' radial positions, but do not affect the fine-scale variations seen in Figures~\ref{mov124pan} and~\ref{mov124in}.

For both ringlets, the two brightness profiles are essentially the same, up to an overall normalization factor due to the slight phase-angle difference between the two observations. However, the radial positions at the two locations are quite different. Since these two data sets were obtained on opposite sides of the planet, the average of the two radial positions corresponds to the semi-major axis of the ringlet, while the difference between them is proportional to $ae$ (the constant of proportionality depending on the pericenter location).

As with the PANORBIT observations, the  Pan ringlet is displaced outwards from Pan's orbit when viewed near the sub-solar longitude and is displaced inwards when viewed near Saturn's shadow. This coincidence strongly suggests that this ringlet  exhibits heliotropic behavior. The PANORBIT and LRHPENKMV  observations were obtained 960 days apart, and the expected apsidal precession rate of this ringlet is 3.2$^\circ$/day,  so any freely-precessing eccentricity would place the pericenter on opposite sides of the planet during the two observations. Thus the ring's pericenter can only be on the anti-solar side of the planet in both observations if the eccentricity is forced by the Sun. 

On the other hand, the observed part of the inner ringlet is actually found closer to the planet on the sunward side of the rings. Thus this material  does not exhibit the same consistently heliotropic behavior as the clumps in the Pan ringlet, and it must have a finite free eccentricity. However, just as the PANORBIT observation alone could not provide solid proof that the Pan ringlet was heliotropic, these data alone cannot be used to argue that the inner ringlet has  zero forced eccentricity due to solar radiation pressure. Indeed, examinations of  the data from all the mosaics indicate that the inner ringlet does have a finite forced heliotropic eccentricity (see Section 6.4). 

For both ringlets, there is a strong anti-correlation between the radial position variations observed at the sub-solar longitude and those seen at the anti-solar longitude.  This implies that the kinks in both ringlets are primarily due to variations in the particles' orbital eccentricities, which is consistent with the analysis of the PANORBIT images described above. Furthermore, the kinks are clearly associated with the clumps in the brightness profile. In the Pan ringlet, all the locations where the separation between the two radial position curves reaches a minimum correspond to a peak in the brightness profiles. Similarly, whenever the radial position of the inner ringlet reaches a local minimum on the sunward side of the rings (and a local maximum on the anti-solar side), there is a corresponding peak in the ringlet's brightness. This implies that these brightness maxima correspond to regions with anomalous eccentricities. However,  there are also multiple brightness maxima in both ringlets that do not correspond to obvious extrema in the radial position curves. This was also the case in the PANORBIT data, where the clump closest to Pan is not associated with an obvious kink.  

Variations in the particles' semi-major axes can also be detected in these observations. For example, in the Pan ringlet  the two position profiles are roughly symmetric about $a_P=133,584$ km  along most of the region within $50^\circ$ of Pan, which requires a semi-major axis close to $a_P$. However, beyond $50^\circ$, both curves shift outwards, suggesting that the semi-major axis here is exterior to $a_P$. However, these semi-major axis variations appear to be on a broader scale than the eccentricity variations responsible from the sharp kinks in these profiles. These broad-scale trends can be clarified by comparing these data to those derived from the other mosaics.

\subsection{Large-scale orbital element variations}

\begin{figure}
\resizebox{3in}{!}{\includegraphics{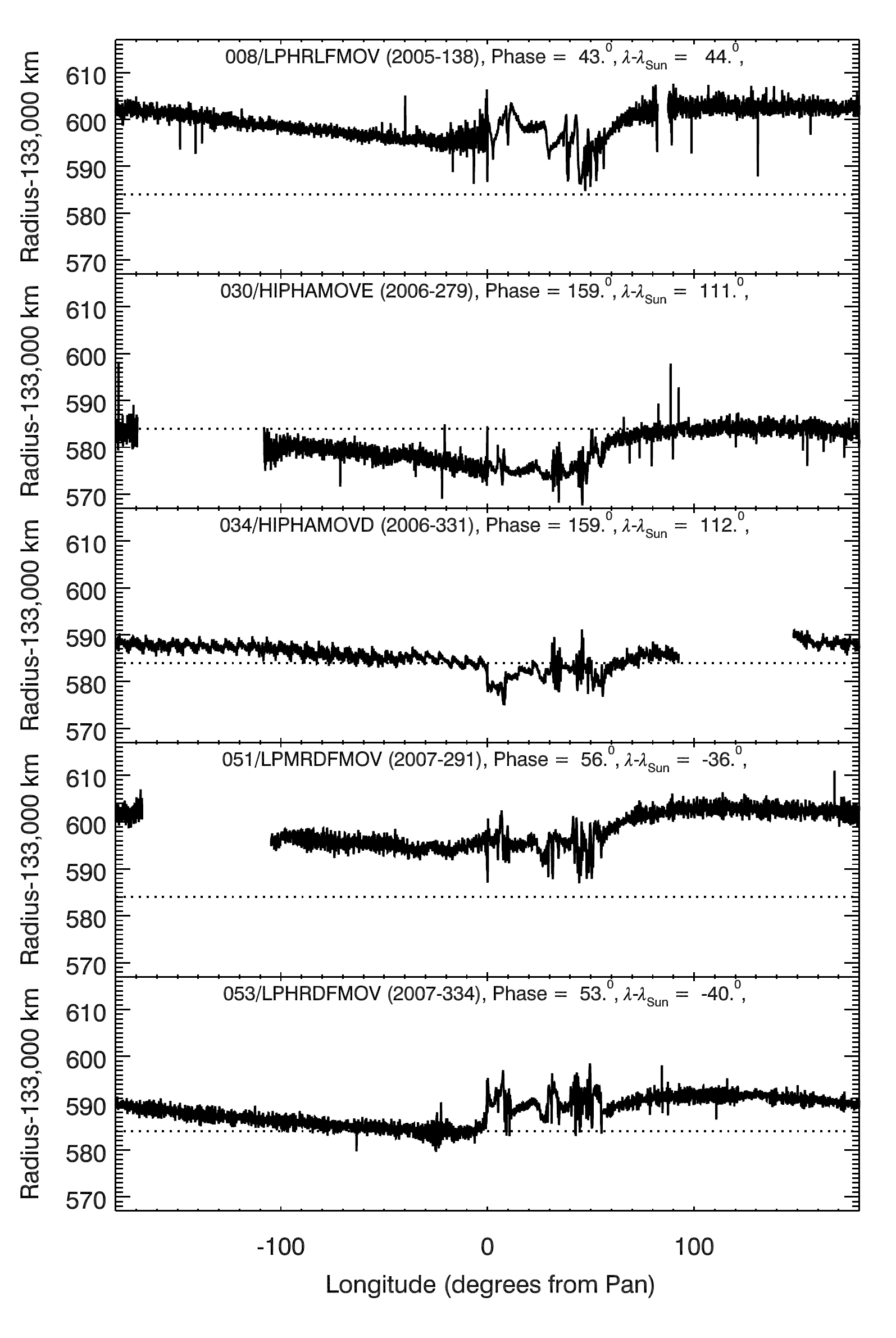}}
\resizebox{3in}{!}{\includegraphics{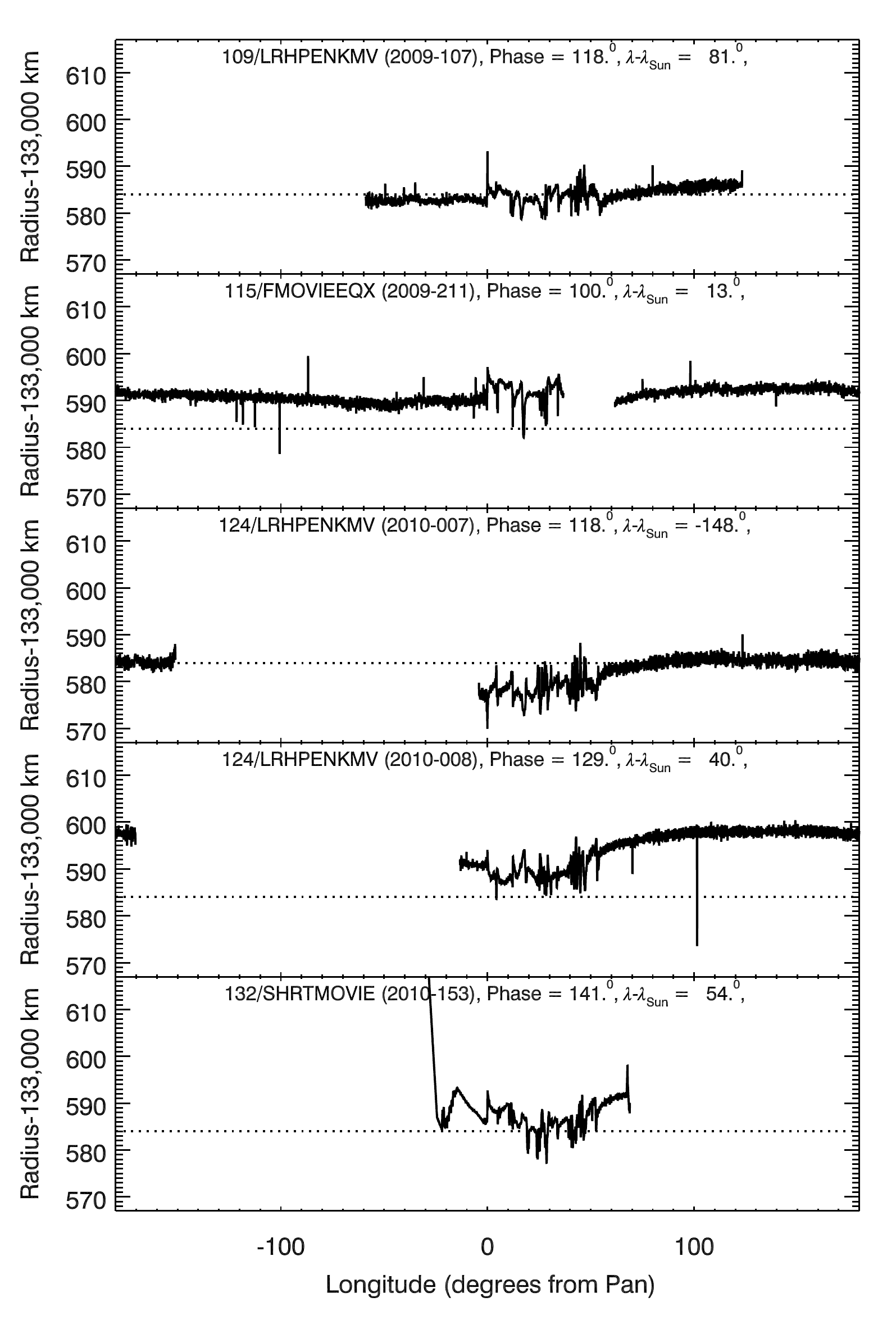}}
\caption{Plots showing the edge-corrected radial positions of the Pan ringlet as a function of co-rotating longitude. For clarity, fits with peak radii more than 30 km from 133,585 km are removed and the remaining data are smoothed over 5 samples. Still some narrow spikes corresponding to misfits can be seen in many of the profiles. The sawtooth pattern in the Rev 034 HIPHAMOVD observation is an artifact that may be associated with the finite eccentricity of this ringlet and the finite longitudinal span of the images. Also, while the Rev 132 SHRTMOVIE data are shown here, they are not used in later fits to the orbital elements due to the restricted longitudinal coverage of this data set. Nevertheless, it is clear that in all the profiles the radial position of the ringlet shifts outwards between 50$^\circ$ and 70$^\circ$ in front of Pan.}
\label{rpan}
\end{figure}

\begin{figure}
\resizebox{3in}{!}{\includegraphics{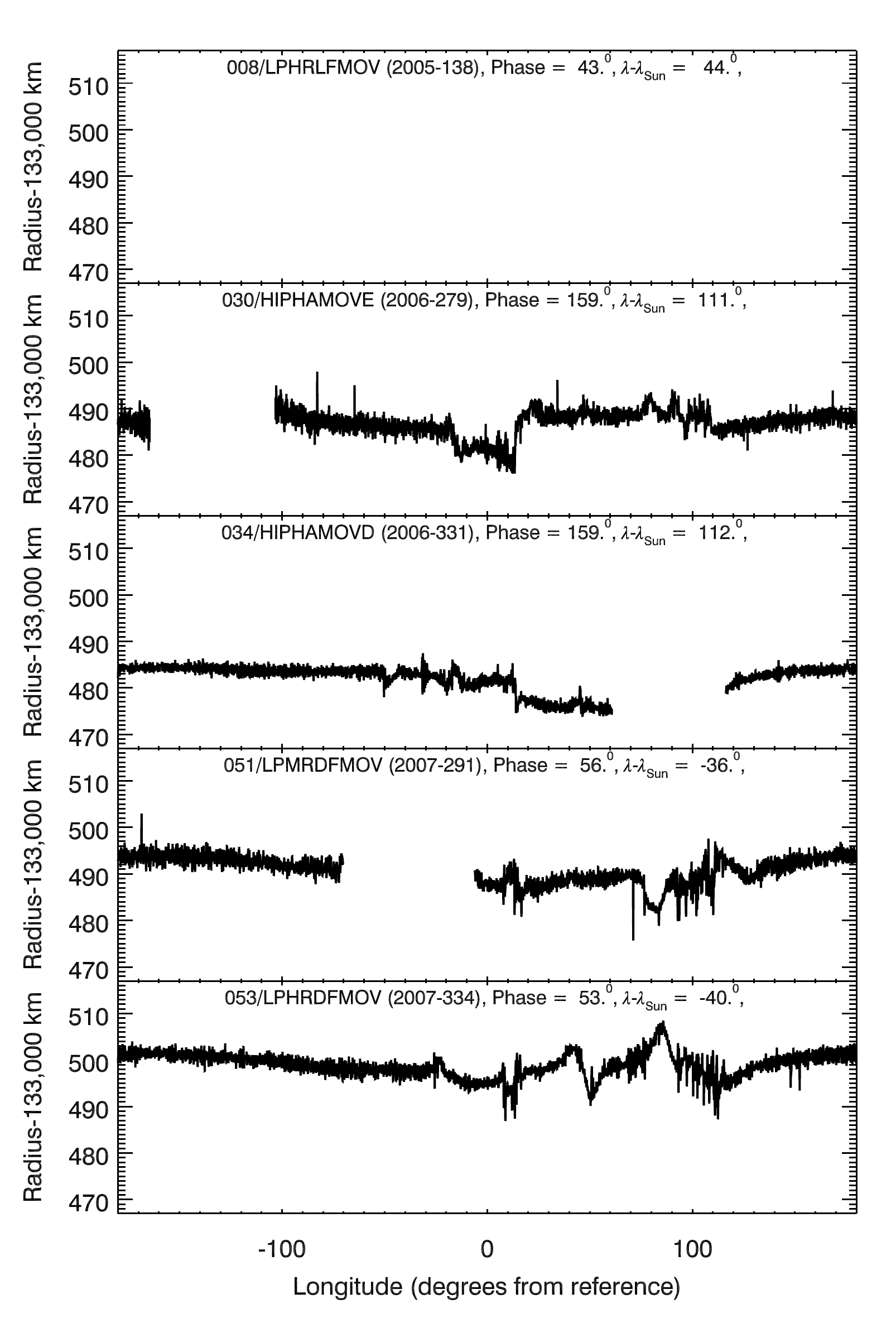}}
\resizebox{3in}{!}{\includegraphics{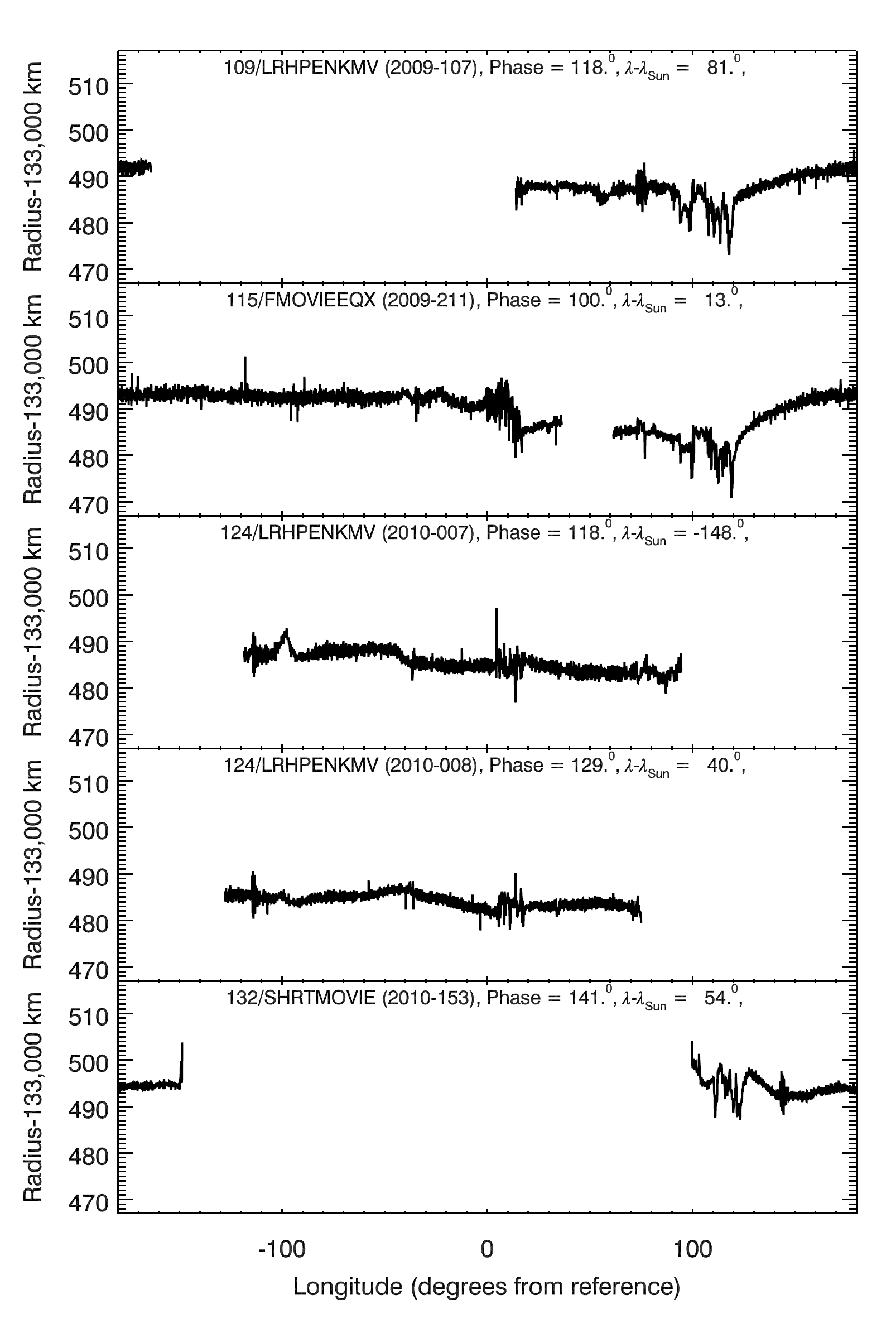}}
\caption{Plots showing the edge-corrected radial positions of the inner ringlet as a function of co-rotating longitude. This longitude system drifts forward relative to Pan at 0.7060$^\circ$/day with an epoch time of 170000000 ET (2005-142T02:12:15 UTC). These brightness profiles are all derived from Lorentzian fits to the ringlet. Fits with peak radii more than 20 km from 133,490 km or peak widths greater than 100 km are removed, and the remaining data are smoothed over 5 samples for the sake of clarity. The Rev 008 SPKMOVPER data are not shown here  because of their low quality (the panel is kept just for ease of comparison to Figure~\ref{rpan}, and the Rev 132 SHRTMOVIE data are not included in subsequent orbital fits because of their limited longitudinal extent.  In many of these profiles, there appears to be an increase in the fit radius at longitudes between 110$^\circ$  and 130$^\circ$, just in front of the clump-rich region.}
\label{rin}
\end{figure}

Both the PANORBIT and LRHPENKMV observations provide detailed but restricted information  about the variations in the ringlets'  orbital properties. In order to place these observations in context, and to better understand these ringlets'  global structure, we now turn our attention back to the large-scale mosaics. Figures~\ref{rpan} and~\ref{rin} show the edge-corrected radial positions of the ringlets as functions of co-rotating longitudes derived from the mosaics listed in Table~\ref{moslist} with sufficient resolution to obtain sensible estimates of the ringlets' radial positions. As above, these radial positions have been corrected based on the positions of the edges within the mosaic, which were measured at each longitude by fitting a peak to the derivative of the radial brightness profiles. Since we are only looking at broad-scale trends in these plots,  filtering out the edge waves was not necessary in this case. However, we avoid using either edge when it is observed between 0$^\circ$ and 40$^\circ$ downstream of Pan, due to large-scale variations in the edge position in these highly disturbed regions.

If we first consider the Pan ringlet data, we can note that the overall radial position of the ringlet depends on the observed inertial longitude relative to the Sun. The sequences taken near the sub-solar longitude (Rev 008 LPHRLFMOV, Rev 051 LPMRDFMOV, Rev 053 LPHRDFMOV, Rev 115 FMOVIEEQX, the second LRHPENKMV in Rev 124 and Rev 132 SHRTMOVIE) all show the ringlet displaced exterior to Pan's orbit, while those taken further from the sub-solar point (Rev 030 HIPHAMOVE, Rev 034 HIPHMOVD, Rev 109 LRHPENKMV, and the first LRHPENKMV in Rev 124) show the ringlet either near to, or displaced inwards from, Pan's orbit. While this suggests that this ringlet is heliotropic, there is also evidence that this ringlet's radial position is not strictly controlled by the Sun. For example, compare the Rev 008 LPHRLFMOV to the Rev 115 FMOVIEEQX data. The latter was obtained closer to the  sub-solar point, but the former shows a more extreme outward radial offset, indicating that this ringlet also has a finite free eccentricity independent of the forced heliotropic eccentricity. Furthermore, we can detect common trends among all these profiles, such as an outward shift  between $50^\circ$ and $70^\circ$ in front of Pan, that could be attributed to variations in the ringlet's semi-major axis. 

The inner ringlet profiles, by contrast, do not provide clear evidence for heliotropic behavior (The ringlets' average radial position is not obviously correlated the observed longitude relative to the Sun). Still, clear systematic variations in the ringlet's mean radial position can be found among these observations, indicating that this ringlet does have a finite eccentricity. Also, we can detect an outward shift in the region between 110$^\circ$ and  130$^\circ$ in most of the profiles. This occurs immediately in front of the clump-rich region, suggesting a change in the ringlet's semi-major axis at this location, similar to that found in the Pan ringlet.

The nature of these broad-scale variations and trends can be clarified by fitting the radial position data at each co-rotating longitude to the heliotropic model described in Section~\ref{helio} above. This model has a small number of free parameters $a$, $e_f$, $e_l$, $\varpi_l$ and possibly $\dot{\varpi}_l$; and at most co-rotating longitudes there are sufficient radial position measurements to determine this many parameters.
However, in order to keep outliers from corrupting the fits, we first down-sample the edge-corrected radial position-estimates by averaging over 1$^\circ$ wide bins in co-rotating longitude. Uncertainties is these estimates were conservatively estimated as the standard deviations of the relevant estimates, which are typically around 1 km. Furthermore, we only use a sub-set of the mosaics, which are marked with an $R$ in Table~\ref{moslist}. Specifically, we exclude the Rev 00A SPKMOVPER data (and the Rev 008 LPHRLFMOV data for the inner ringlet) due to the low spatial resolution of these images. We also exclude the Rev 044 FMOVIE data because the gaps around the inner edge corrupt the edge corrections, and  the Rev 132 SHRTMOVIE data because they only cover a small range of longitudes and at most longitudes the inner edge data are insufficient to correct the ringlets' radial positions. This leaves nine profiles for the Pan ringlet and eight profiles for the inner ringlet, which should still be enough to fit all the model parameters. However, many of these profiles do not cover all co-rotating longitudes, so at some locations the model cannot be adequately constrained.

\begin{figure}
\resizebox{6in}{!}{\includegraphics{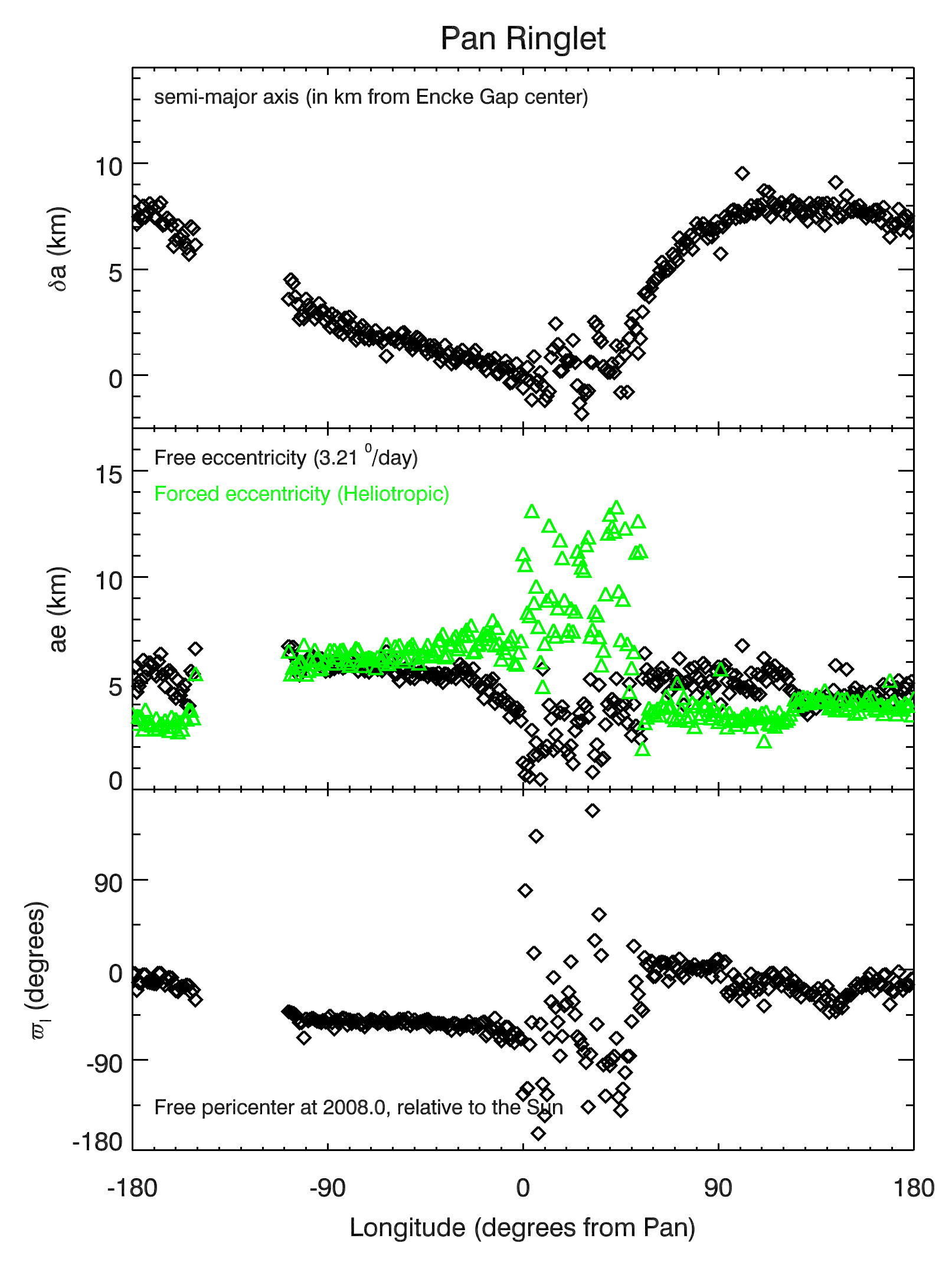}}
\caption{Plots of the Pan ringlet's orbital elements as functions of co-rotating longitude derived from the mosaics marked with an $R$ in Table~\ref{moslist}. The semi-major axis is measured from the Encke Gap center at 133,584 km. These fits assume the free precession rate was 3.21$^\circ$/day (3.18$^\circ$/day relative to the Sun), using an epoch time of 2008-001T00:00:00. Statistical error bars on these estimates are not shown for reasons of clarity, but are between 0.5 km and 1 km for $a$ and $ae$, and about 5$^\circ$ for the pericenter location.}
\label{fitpan}
\end{figure}

\begin{figure}
\resizebox{6in}{!}{\includegraphics{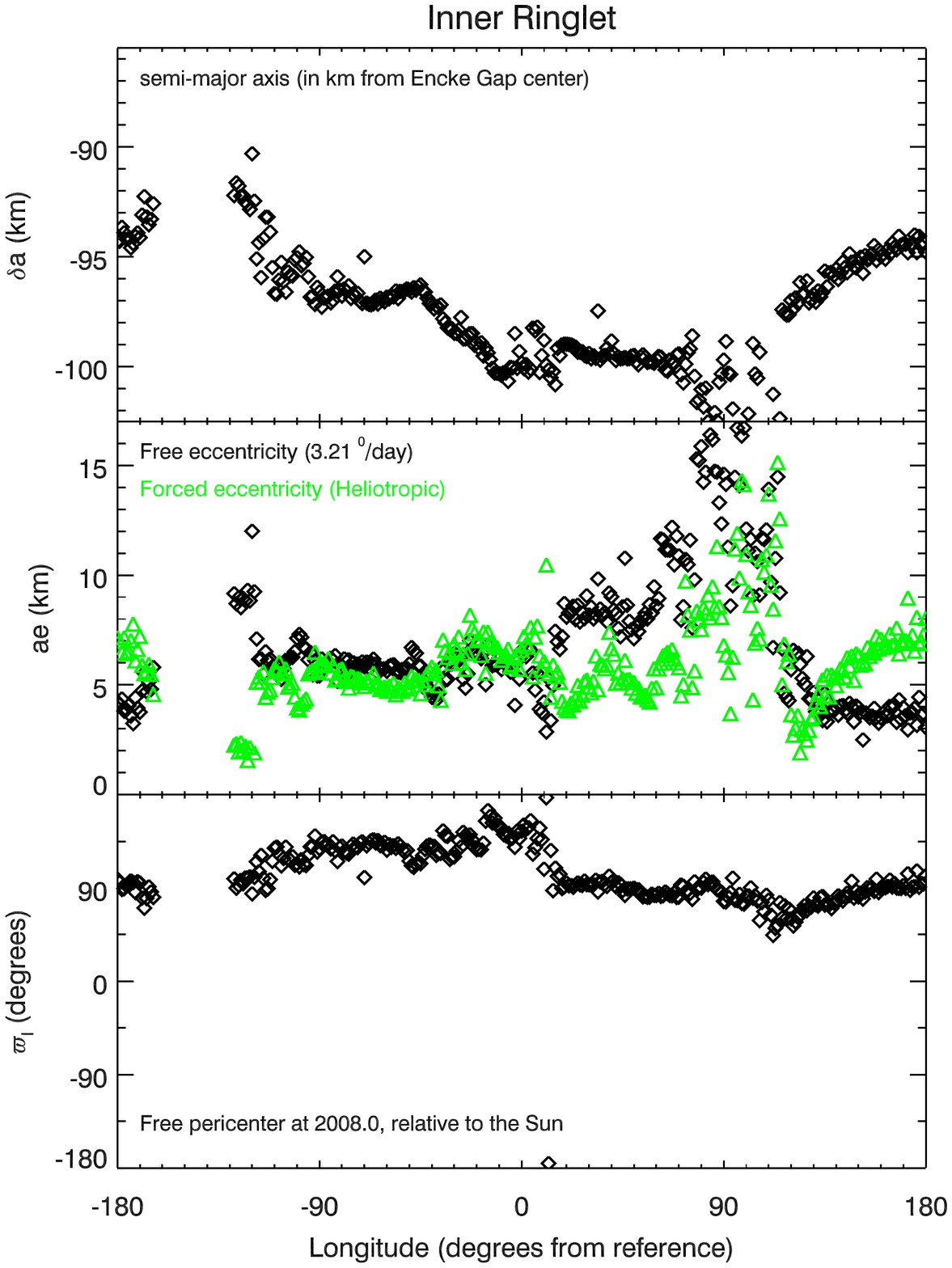}}
\caption{Plots of the inner ringlet's orbital elements as functions of co-rotating longitude derived from the  mosaics marked with $R$ in Table~\ref{moslist}. The semi-major axis is measured from the Encke Gap center at 133,584 km.  The co-rotating longitude system drifts forward relative to Pan at 0.7060$^\circ$/day with an epoch time of 170000000 ET (2005-142T02:12:15 UTC). These fits assume the free precession rate was 3.21$^\circ$/day (3.18$^\circ$/day relative to the Sun), using an epoch time of 2008-001T00:00:00.  Statistical error bars on these estimates arenot shown for reasons of clarity, but are between 0.5 km and 1 km for $a$ and $ae$, and about 5$^\circ$ for the pericenter location.}
\label{fitin}
\end{figure}

Figures~\ref{fitpan} and~\ref{fitin} show the heliotropic parameters $a$, $e_f$, $e_l$ and $\varpi_l$ as functions of co-rotating longitude in both the Pan and inner ringlets. Note that because we are mostly interested in large-scale trends, we do not attempt to account for the motions of clumps or for the waves generated by Pan in the inner ringlet in these calculations. Furthermore, in order to reduce the number of free parameters in these fits, the free precession rate was held fixed at 3.21$^\circ/$day (3.18$^\circ$/day relative to the Sun). Allowing the precession rate to float did not change the overall trends, but gave rise to  increased scatter in the parameters, especially $\varpi_l$. Varying the assumed precession rate also did not affect the trends in the fit parameters significantly. Fitted parameters are only plotted at co-rotating longitudes with more than four radial position measurements. The statistical uncertainties on these parameters are between 0.5 and 1 km for $a$, $ae_f$ and $ae_l$, and around 5$^\circ$ for $\varpi_l$. Thus the large-scale trends seen in these plots are highly significant, however we caution that smaller-scale fluctuations might reflect systematic errors in individual observations.

First, consider the fit parameters for the Pan ringlet shown in Figure~\ref{fitpan}. These parameters generally show nice, smooth trends, except in the region between 0$^\circ$ and 60$^\circ$ in front of Pan. the excess scatter in this region arises because this analysis does not account for clumps drifting through this region. Despite this, the mean orbital elements in this region are consistent with those derived from the Rev 045 PANORBIT observation. In particular, the semi-major axis scatters around $a_P$, and the forced eccentricity is much larger than the free eccentricity. Thus neglecting the motions of the clumps does not appear to prevent us from obtaining sensible orbital elements. 

Outside the clumpy region, we find that the values of $e_f$, $e_l$ and $\varpi_l$ do not vary much with co-rotating longitude. Furthermore, the forced and free components of the eccentricity are comparable to each other.  These particles' orbits therefore periodically become nearly circular, and since $\varpi_l$ varies by less than 90$^\circ$ around the ring, the eccentricity variations in the entire ringlet are synchronized somehow. This behavior is very similar to that previously observed in the dusty Cassini Division ringlet \citep{Hedman10}.

By contrast, the ringlets' semi-major axes vary systematically with co-rotating longitude outside the clump-rich region. Behind Pan, the semi-major axis seems to increase linearly with distance from Pan. This trend seems to saturate when the radial displacement reaches 8 km exterior to Pan. In front of the clump-rich region, the semi-major axis rises rapidly from $a_P$ to ($a_P+8$ km) within a  space of 60$^\circ$. The latter semi-major axis shift is responsible for the radial position shift visible in all the profiles in Figure~\ref{rpan}.

Turning to the inner ringlet's  parameters illustrated in Figure~\ref{fitin}, many of the same trends are apparent, but there are some important differences as well. In this case, the clumps extend between  co-rotating longitudes of -30$^\circ$ and 120$^\circ$, but are not common outside the regions centered around $0^\circ$ and $100^\circ$. The clump-rich region has the lowest semi-major axes of $a_P-100$ km, which corresponds to the semi-major axis required to match the clumps' mean motion. Beyond the clump-rich region, the semi-major axis is displaced outwards, following trends very similar to those seen in the Pan ringlet. Also, in the regions far from the clumps,  $e_f$, $e_l$ and $\varpi_l$ are all roughly constant, and $e_f\simeq e_l$, just like for the Pan ringlet. However, unlike the Pan ringlet,  the free eccentricity is close to, or even higher than, the forced eccentricity across the entire region covered by the clumps. This is consistent with the lack of an obvious heliotropic signature in the Rev 124 LRHPENKMV data described above (see Section 6.3). 

\section{Discussion}

The above observations reveal that the fine material in the Encke Gap is sculpted by multiple processes. The overall architecture of the dusty material and the disturbances  found near Pan demonstrate that Pan's gravity does influence the motions of particles in this region. Meanwhile, the heliotropic forced eccentricities indicate that non-gravitational forces also affect the distribution of particles within the gap. The anomalous motions of the bright clumps in the narrow ringlets suggest that interactions among the dust grains themselves probably also play a role in sculpting this material. The dynamics of the dust in the Encke gap are therefore quite complex, and a detailed theoretical analysis of this system is beyond the scope of this report. Still, we can provide some initial speculations and calculations that can provide a basis for such future modeling efforts that will be the subject of a future paper. 

First, we use the magnitude of the heliotropic forced eccentricities to estimate the typical particle sizes in the ringlets and confirm that these are broadly consistent with previous estimates based on the ringlets' light-scattering properties. Then we examine the apparent variations in the inner and Pan ringlets' semi-major axes with co-rotating longitude and explore how these could be explained by radial transport of small particles. Next, we consider the role of particle collisions and argue that they may be responsible for some of the observed longitudinal variations in these ringlets' semi-major axes, as well as the formation of bright clumps. Finally, we suggest that the locations of the clump-rich regions in the Pan and inner ringlets may be determined by the competition between non-gravitational azimuthal drag forces and Pan's gravitational perturbations. 

It is important to keep in mind that the following discussions focus primarily on dynamical phenomena that could explain some of the better documented trends in the currently-available data, and additional processes not considered below may well be important in sculpting the dusty material in the Encke Gap. For example, we are still unable to ascertain what could be exciting the ``free'' components of the ringlet's eccentricities. Also, since we have not yet been able to determine the outer ringlet's orbital properties, we cannot explore its dynamics in detail at present. Furthermore, the wide variety of processes considered in these discussions may interact and interfere with one another in very complex ways, and some of these still-unexplained features of these ringlets could reflect dynamical phenomena that will require some of the interpretations given below to be reconsidered and/or revised.  

\subsection{Heliotropic behavior and particle sizes}

Away from the bright clumps, the Pan ringlet and the inner ringlet exhibit similar combinations of forced and free eccentricities, with $ae_f \simeq ae_l \simeq 5$ km. The similar magnitudes of $e_f$ and $e_l$ imply that these particles' orbits periodically become exactly circular. One possible explanation for this is that the particles were launched from source bodies on nearly circular orbits. In this case, even though solar radiation pressure imparts a forced eccentricity to these particles' orbits, the condition  that they began on circular orbits would require that $e_f \simeq e_l$ and that the particles' orbits periodically return to a circular state. However, this simple explanation is complicated by the observation that  $\varpi_l$ doesn't vary much with longitude in either ringlet. This means that the orbits of all the particles in each ringlet  become nearly circular at the same time, which would not naturally occur if all these particles moved independently from each other and were produced at different times. Similarly coordinated motions have been observed  previously in  the so-called ``charming ringlet'' in the Laplace Gap in the outer Cassini Division \citep{Hedman10}, so this synchronization of free pericenters appears to be a common feature of narrow dusty ringlets. 

As discussed in \citet{Hedman10}, collisions among a ringlets'  particles will naturally tend to align the particles' orbital pericenters. Such inter-particle collisions could therefore produce the observed coordinated motions if the collisions are sufficiently frequent  and if the particles can maintain finite free orbital eccentricities. Even outside the clumps, the Encke Gap ringlets' optical depths are about an order of magnitude higher than that of the ``charming ringlet'' (see Hedman {\it et al.} 2011), so collisions are more likely to be sufficiently frequent to align pericenters in the Encke Gap. Maintaining a finite free eccentricity is a bigger challenge, since collisions among the ring particles would also tend to dissipate $e_l$. \citet{Hedman10} explores what sorts of terms in the particles' equations of motion could support the free eccentricity of the dusty Cassini Division ringlet. For the Encke Gap ringlets, we have the additional constraint that $e_f \simeq e_l$, which could help clarify the origin of $e_l$ in these ringlets. For example, perhaps  it becomes easier for particles with different orbital semi-major axes to maintain their aligned pericenters against differential precession when all the particles' orbits periodically become circular. A full exploration of such ideas will likely require numerical simulations of these ringlets.

Despite this lingering uncertainty regarding the free component of the ringlets' eccentricity, the magnitude of the forced eccentricities can still provide a useful estimate of the typical particle sizes in these ringlets because the value of $e_f$ can be computed using orbital perturbation theory \citep{Hedman10}:
\begin{equation}
e_f \simeq \frac{n}{\dot{\varpi}_0}\left[\frac{3}{2}(1-\epsilon+\sin(2\pi\epsilon)/6\pi)\frac{F_\sun}{F_G}\cos B_\sun\right],
\end{equation} 
where $n$ is the particles' mean motion, $\dot{\varpi}_0$ is the apsidal precession rate,   $F_\Sun/F_G$ is the ratio of the solar radiation force acting on the particle to Saturn's gravitational force, $\epsilon$ is the fraction of the particles' orbit that is in shadow, and $B_\sun$ is the solar elevation angle. For particles in the Encke Gap, $n=626^\circ$/day, $\dot{\varpi}_0=3.2^\circ$/day and $F_\Sun/F_G \simeq 1.6*10^{-5} Q_{pr}/(r_g/1\mu m)$, where $Q_{pr}$ is an efficiency factor dependent on the particle properties \citep{BLS79}, and $r_g$ is the particle's physical radius. 
For the Encke Gap ringlets, $\epsilon < 0.15$, and for the images considered here, $|B_\Sun| < 25^\circ$, so $1-\epsilon+\sin(2\pi\epsilon)/6\pi$ and $\cos B_\sun$ can both only range between 0.9 and 1. Thus the heliotropic forced eccentricity can be expressed as a function of particle size:
\begin{equation}
e_f\simeq 0.0042  \frac{Q_{pr}}{r_g/1 \mu{\rm m}}.
\label{efpred}
\end{equation}
Strictly speaking, this calculation applies to individual ring particles, and the observed radial displacements of the ringlet represent the average motions of all the particles within the ringlet. Thus the measured heliotropic components of the ringlets' eccentricities provide estimates of an effective mean particle size in these ringlets.

For both the inner and Pan ringlets,  $ae_f \sim 5$ km, implying that the particles in both ringlets have effective mean radii around 100$Q_{pr}$ microns. This estimate is plausible given previous studies of these and other dusty, heliotropic rings. For example, the ``charming ringlet" exhibits larger heliotropic radial excursions than the Encke Gap ringlets, indicating that the typical particle size is around 20$Q_{pr}$ microns \citep{Hedman10}, or a few times smaller than the particles in the Encke Gap. This is consistent with studies of the transmission spectra of all these ringlets, which contain a narrow dip that can be attributed to particles in the 10-50 micron size range \citep{Hedman11}. This spectral feature is weaker in the Encke Gap ringlets than it is in the ``charming ringlet'', implying that the Encke Gap ringlets contain a bigger fraction of larger particles.

\subsection{Radial transport in the Encke Gap}

Turning from eccentricities to semi-major axes, the longitudinal variations in the mean radial position of the inner and Pan ringlets outside of the clump-rich regions suggest that the semi-major axes of the ringlets' particles are drifting towards and away from Saturn. Since the particles in the clumps have the smallest semi-major axes, they should also have the shortest orbital periods and fastest orbital speeds. Hence we may also reasonably infer that the particles outside the clump-rich regions are drifting backwards in longitude relative to the clumps, and thus there is a steady stream of material flowing out from the trailing edge of the clump-rich region in each ringlet. If this is correct, then the observed trends in both ringlets' positions imply that the particles outside the clumps initially move outwards away from Saturn,  but then reverse course and move back inwards when they approach the leading edge of the clump-rich regions.

More quantitatively, the observed trends in the ringlets' positions can be translated into estimates of the particles' radial migration rate.  Say that at a given location in a ringlet, the particles' average semi-major axis drift rate $da/dt=v_a$. Furthermore, say the average semi-major axis of these particles $a$ is different from that of Pan or the clumps $a_0$. In that case, the particles will also drift longitudinally in a co-rotating system fixed to Pan or the clumps at a speed $v_\lambda=-1.5n(a-a_0)$, where $n$ is the mean motion of the clumps. The trajectory of these particles in the co-rotating frame therefore has the following slope:
\begin{equation}
\theta =\frac{1}{a_0}\frac{da}{d\lambda_c}=\frac{v_{a}}{v_{\lambda }}=-\frac{2}{3}\frac{v_a}{n}\frac{1}{a-a_0}.
\label{slope}
\end{equation}
Hence an observed slope $\theta$ in the ringlet implies a radial migration rate $v_a=-1.5n(a-a_0)\theta$.

Such migration rates may be compared with the rates that could be generated by various perturbation forces. Changing a particle's orbital semi-major axis also changes its orbital energy, so the most efficient way to generate a nonzero $v_a$ is to accelerate the particle along its direction of motion with an azimuthal force. If the average azimuthal force applied to the ring particle over one orbit is $F_\lambda$, then the particle's semi-major axis will drift at the following rate \citep{Burns76}:
\begin{equation}
v_{a} \simeq 2an\frac{F_\lambda}{F_G}, 
\label{drag1}
\end{equation}
where $F_G$ is Saturn's central gravitational force on the particle. Note the above equation assumes the particle's orbital eccentricity is small, which is reasonable for the Encke Gap ringlets. Combined with Equation~\ref{slope}, this expression can be used to estimate the forces required to produce an observed trend in a given ringlet. 

The following subsections will explore what processes might be responsible for the various trends observed in the ringlets. First, we examine the apparent outwards motion behind the clumps and investigate whether this can be ascribed to  interactions with the magnetospheric plasma. Then we consider the inwards motion just in front of the clumps and suggest that this may be due to collisions among different populations of ring particles. 

\subsection{Outwards migration due to drag forces}

In both the inner and Pan ringlets, the semi-major axis drops steadily by about 7 km between $-180^\circ$ and $0^\circ$ in the co-rotating frame, which implies that: $\theta  \simeq  -1.7\times10^{-5}$. Hence, Equation~\ref{slope} implies that  the particles in this particular region are drifting outwards at the following rate:
\begin{equation}
v_{aD} \sim +3\times10^{-5} {\rm m/s} \frac{(a-a_0)}{10 {\rm km}}.
\label{vadest}
\end{equation}
Similarly, Equation~\ref{drag1} implies that the magnitudes of the azimuthal force in these regions are:
\begin{equation}
\frac{F_\lambda }{F_G} \simeq 10^{-9} \frac{(a-a_0)}{10 {\rm km}}.
\label{fdfg}
\end{equation}
Note that both the migration rate and the perturbing force must increase with distance from the clump's semi-major axis in order to maintain the observed nearly constant slope. 

One possible explanation for these radial motions is an interaction with the magnetospheric plasma. The ions in the plasma co-rotate with Saturn's magnetic field and thus move around the planet faster than particles orbiting at the Keplerian rate inside the Encke Gap. Thus, when these ions collide with the charged dust grains, the resulting momentum exchange accelerates the ring particles and causes them to slowly spiral outwards, as desired. Furthermore, the variations in the migration rate with distance from $a_0$ could be explained if the  moon and/or dense clumps in these ringlets absorbed the plasma in their vicinity, sharply reducing the plasma density around the clumps' semi-major axis.  

Unfortunately,  it is not yet clear whether these sorts of interactions with plasma ions are sufficient to produce the observed trends in the ringlets' radial positions. The simplest expression for the azimuthal force experienced by a particle of radius $r_g$ due to these interactions is $F_D=\pi r_g^2\rho_i w^2$, where $\rho_i$ is the plasma ion mass density,  $w=a(n-\Omega_S)$ is the azimuthal speed of the plasma ions relative to the ring particles, and $\Omega_S \simeq 810^\circ/$day is Saturn's rotation rate. Note that this is a highly over-simplified expression for the plasma interaction force, but it is a reasonable approximation for the tenuous plasma expected to exist within the rings \citep{Grun84}. Meanwhile, Saturn's gravitational pull on the particle $F_G$ can be written as $n^2a m$, where $n$ and $a$ are the particle's orbital mean motion and semi-major axis, and  $m$ is the particle's mass, which can in turn be expressed in terms of the particle's radius $r_g$ and mass density $\rho_g$.  The ratio of these two forces then becomes:
\begin{equation}
\frac{F_D}{F_G} \simeq \frac{3}{4}\frac{\rho_i}{\rho_g} \frac{a}{r_g} \left(1-\Omega_S/n\right)^2.
\end{equation}
For the particles in the Encke gap, $a \simeq 133,500$ km and $n \simeq 626^\circ/$day. Also, since these ringlets  are composed primarily of water ice,  we may assume that  $\rho_g \simeq 1$ g/cm$^3$. Furthermore  the magnitude of the ringlets' heliotropic forced eccentricities implies that  $r_g \simeq 100 \mu$m (see above). Finally, the mass density of the plasma in the Encke gap can be estimated from data obtained by Cassini when it flew over the A ring during Saturn orbit insertion. Measurements made by various  instruments demonstrate that the plasma surrounding the rings consists primarily of O$^+$ and O$^+_2$ \citep{Tokar05, Waite05, Young05}, so the mass per ion should be between 16 and 32 amu.  Unfortunately, the number density of ions within the Encke Gap $n_i$ is not so well determined. During its passage over the rings, Cassini encountered ion densities above the rings between 0.1/cm$^3$ and 1.0/cm$^3$ \citep{Tokar05, Waite05}, but numerical models suggest that the ion number density at the ringplane could be as high as 10-100/cm$^3$\citep{Tseng10, Tseng11}. Taking $n_i=10/$cm$^3$ as a fiducial number, and assuming an equal mix of O$^+$ and O$^+_2$ in the ring's ionosphere, we can then estimate the above force ratio as:
\begin{equation}
\frac{F_D}{F_G} \simeq 3\times10^{-11}\left(\frac{n_i}{10{\rm/cm}^3}\right)\left(\frac{100 \mu{\rm m}}{r_g}\right)
\end{equation}
This is an order of magnitude less than the force required to produce the observed trends, and so simple plasma drag may be insufficient to produce the required outwards migration. However, the above calculation is very rough, and the force would be larger if the ion density in the Encke Gap is higher than 10/cm$^3$, the particles are less massive than assumed here, or the coupling between the plasma and the ring particles has been significantly underestimated by neglecting the Coloumb scattering between the charged grains and plasma ions (cf. Gr\"un {\it et al.} 1984). More detailed simulations of the plasma environment within the Encke Gap will therefore be needed in order to determine whether plasma drag could be responsible for the outward motions of these small grains. 

Thus far, we have not been able to identify any other plausible physical process that could produce the observed outward trends in the ringlets' radial positions. However, whatever is causing these motions does not appear to be a localized phenomenon. Given that the radial positions of both the inner and Pan ringlets drift steadily outwards for over 180$^\circ$ in co-rotating longitude, some process is likely causing particles to accelerate azimuthally throughout the inner and central parts of the Encke Gap (the situation in the outer part of the gap is less clear). This perturbation therefore could have some relevance to other aspects of the ringlets' structure, even if we cannot yet identify how it is generated. In the following discussions, we use the generic term ``drag force'' to describe this as-yet unidentified azimuthal acceleration.

\subsection{Inwards migration from collisions and clump formation from instabilities}

While steady azimuthal forces can potentially explain both ringlets' outward displacement with increasing distance behind the clump-rich regions, it does not explain the opposite trend found just in front of these regions. This trend would require some process that transports material back inwards towards the planet and towards the clumps' semi-major axis. We propose that collisions among the particles in each ringlet are responsible for this inward motion. Furthermore, we suggest that the clumps themselves arise from an instability associated with such inter-particle collisions.

Whatever their origin, the drag forces discussed in the previous section cause the particles to spiral away from the planet, and to drift further and further outwards and backwards relative to the clump-rich part of the ringlet. Eventually, these ``drifters'' will move sufficiently far backwards  that they will pass by the clump-rich regions. Extrapolating from the observed trends, these drifters will have semi-major axes that are only about 10-15 km exterior to the clump particles.  If all the drifting particles had the same semi-major axes and were on perfectly circular orbits, they could just pass by the clumps and continue to spiral outwards. However, these particles are not all on simple circular orbits. Besides the mean forced and free components of the eccentricity discussed above, the  finite widths of these ringlets suggest that their particles possess a finite range of eccentricities and semi-major axes. The radial widths of both the inner and central ringlets are greater than 10 km (see Figure~\ref{egapprof}), so the drifters can actually pass through the clumps and collide with that material. Furthermore, the relative velocities of the drifters and the clumps is small, so there are many opportunities for particles to collide before they drift past the clumps.  

Since the drifting particles' semi-major axes are larger than those of the typical clump particles, the drifters are most likely to experience collisions with clump material near the periapses of their own orbits, when they will be moving faster than most of the clump material. Such collisions will therefore tend to knock the drifters backwards, slowing their orbital motion and causing their semi-major axes to decay inwards towards Saturn and the clump.  The rate at which the drifting particles migrate towards the clumps due to such collisions is just the product of the semi-major axis shift induced by each collision and the collision frequency. To first order, the semi-major-axis shift per collision will be of order the semi-major axis difference between the drifter and the clumps, while the collision rate for a drifter will be the particle's mean motion times the clumps' optical depth. Hence the relevant radial drift rate should be of order:
\begin{equation}
v_{ac} \sim -\tau_c n (a-a_c),
\end{equation}
where $a_{c}$ is the semi-major axis  of the clump particles, and $\tau_c$ is the clump optical depth. When the drifting particles initially encounter the clumps, they will have $a-a_c \simeq 10$ km, and the typical clump optical depth $\tau_c \simeq 0.1$ \citep{Hedman11}, so $v_{ac} \simeq -0.1$ m/s. By comparison the outward migration rate due to the drag forces is only $v_{aD} \sim 3\times10^{-5}$ m/s (see Equation~\ref{vadest}). Hence, collisions with the clump particles should be an efficient way to halt and reverse the outward migration of the drifting material. 

It is important to note that these collisions not only affect the radial migration of  particles, but also their longitudinal motion. By forcing the particles' semi-major axes to converge towards that of the clump, these interactions reduce the rate at which these particles drift past the clumps. Thus particles initially drifting past the clumps could get stuck in the clumps, raising the clump's density  and increasing the likelihood that additional drifting particles will slow down in the clump's vicinity. This instability could potentially also explain the unusual motions of the clumps. In this scenario, the clumps would not represent a fixed set of particles. Instead, particles would be constantly entering and leaving the clump. Hence the apparent motion of the clump is controlled by how quickly particles get trapped or escape from this region, which does not necessarily correspond to the trajectory of any individual ring particle. Furthermore, as particles with different orbital elements converge on these dense regions, gradual variations in orbital eccentricities could transform into sharp features like the kinks. The dynamics of these clumps are quite complex and numerical simulations along the lines of those done by \citet{Lewis11} will likely be needed to evaluate whether the accelerations and orbital characteristics of the observed clumps are consistent with the above hypotheses. Such simulations will also probably be needed to determine whether  inter-particle collisions can cause the radial position of the ringlet to begin to fall $\sim30^\circ$ in front of the clump-rich regions.

\subsection{Pan's gravity, the distribution of clumps and the location of the ringlets}

\begin{figure}[tbp]
\resizebox{6in}{!}{\includegraphics{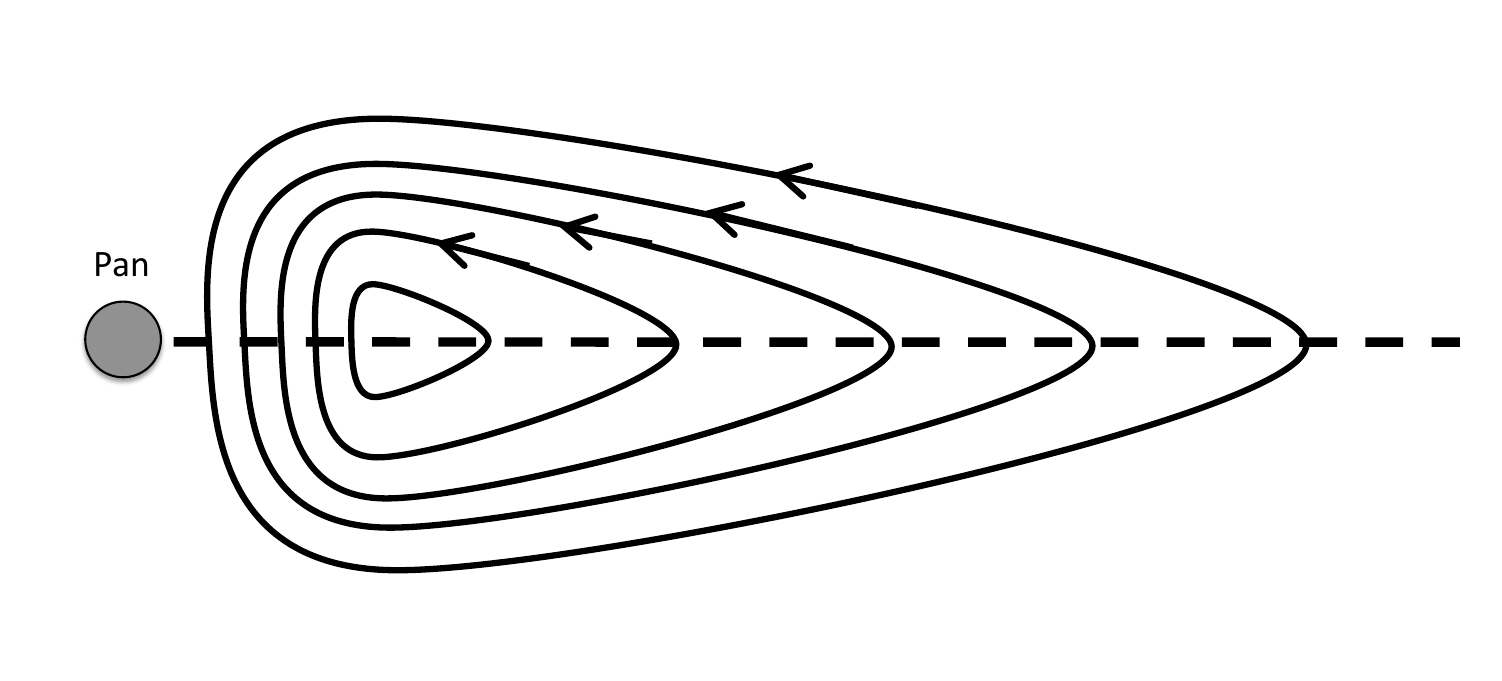}}
\caption{Schematic representation of the asymmetric trajectories of the particles in the Pan ringlet due to the combined action of drag forces and Pan's gravity in a reference frame that co-rotates with Pan. Note radius increases upwards in this diagram, longitude increases to the right, and Pan's orbit is displayed as the dashed line. Also note that in this cartoon the radial (vertical) scale is highly exaggerated relative to the longitudinal (horizontal) scale. The particles are assumed to remain on nearly circular orbits in this cartoon, and initially have a range of longitudes along Pan's semi-major axis. On the right side of the figure, the particles are drifting outwards due to drag forces, while on the left they are undergoing horseshoe motion due to Pan's gravitational perturbations. Due to the intrinsic asymmetry of these motions, these particles are more likely to be found just in front of Pan, which is also where the clumps are located.}
\label{horseshoedrag}
\end{figure}

In the previous subsection, we proposed that collisions among the ringlet particles could keep ringlet  material from drifting too far away from the semi-major axes of the relevant clumps. However, we still need to find a way to anchor the clumps  at particular semi-major axes and prevent them from slowly drifting outwards under the influence of the relevant drag forces. It turns out that for both the Pan and the inner ringlets, the gravitational perturbations from Pan are likely responsible for maintaining the clumps at nearly constant semi-major axes.

For the Pan ringlet, the importance of Pan's gravity is not surprising. As discussed above, the entire Pan ringlet occupies the horseshoe zone surrounding Pan's orbit. As demonstrated by \citet{Murray94}, particles  can be trapped in this region even in the presence of drag forces, so long as the latter do not allow a particle to escape the horseshoe region before it has a close encounter with the moon (see also Murray and Dermott 1999). In this case, we can estimate that the outwardly-drifting particles would have semi-major axes around 15 km exterior to Pan if they avoided collisions with any clump material. This lies comfortably within $\Delta a_h$ for Pan, so Pan's gravity should be able to keep the particles in such a ringlet from dispersing. 

Furthermore, the combination of Pan's gravity and the outward migration induced by the drag forces  could naturally produce the asymmetric distribution of clumps in the Pan ringlet (see Figure~\ref{horseshoedrag}). Imagine we launch fine debris on circular orbits at a range of longitudes relative to Pan, and for the sake of simplicity, let us neglect eccentricities driven by solar radiation pressure.  These particles will then remain on circular orbits but they will all migrate outwards and drift backwards relative to Pan under the influence of the drag forces. These particles will encounter Pan at various positive values of $\delta a_{\rm before}=a-a_P$, and Pan's gravity will force all of them  onto orbits with $\delta a_{\rm after}=-\delta a_{\rm before}$,  so that they will begin to move forward relative to Pan. After the encounter, the steady outward migration will resume, and barring any collisions among the ring particles, the trajectories of the particles will form closed loops with one end at their start location and the other on the leading side of Pan. The average semi-major axis of all these particles therefore equals $a_P$, and the density of particles is highest in the region just in front of Pan. Since material naturally collects in front of Pan, collisions among the ringlet particles will favor the formation of clumps in this region, consistent with the observations. (Recall that because the clumps might not follow the trajectory of any individual particles,  the clumps themselves would not necessarily follow trajectories like those shown in Figure~\ref{horseshoedrag}.)

Since no comparably massive moon has been identified in the inner ringlet, the clumps here cannot be similarly anchored by such horseshoe motion. Instead, we argue that the material in the inner ringlet is maintained by a balance between drag forces pulling particles outwards and Pan's gravitational perturbations pushing them inwards. As discussed above, some process is causing the particles far from the clumps to drift outwards at a rate of $v_{aD} \sim +3\times10^{-5} {\rm m/s}[(a-a_0)/10 {\rm km}]. $
On the other hand, each time the particles in the inner ringlet pass by  Pan, their semi-major axes will be shifted inwards by the amount stipulated in Equation~\ref{dap}. The frequency of such encounters is  $\Delta n=1.5n\Delta a/a$, so these perturbations will cause the particles to migrate inward at a rate:
\begin{equation}
v_{aP} \sim -5an\left(\frac{m_p}{M_S}\right)^2\left(\frac{a}{\Delta a}\right)^4.
\end{equation}
For the inner ringlet, $\Delta a /a \sim 0.0007$, which together with the current estimate of Pan's mass $m_p/M_S \sim 0.8*10^{-11}$ \citep{Porco07, Weiss09} yields $v_{aP} \sim -2\times10^{-5}$ m/s, which is remarkably close to the above value for $v_{aD}$. Hence the inner ringlet may well be situated in a region where the torques from drag forces and Pan's gravity balance, halting the radial motion of material. Indeed, material dispersed within the inner half of the gap will naturally collect at this location, as material  closer to the planet is pushed outwards by drag forces and material closer to Pan is driven inwards. These competing forces, coupled with collisions among the particles, could then lead to the formation of a narrow ringlet.

A similar balancing of forces could potentially explain the distribution of material in the outer part of the Encke Gap (i.e., the narrow outer ringlet and the broader ``fourth ringlet''). However, since Pan's gravitational perturbations should always cause material to move away from Pan's orbit, such a balancing act would require some process that caused material in the outer part of Encke Gap to migrate inwards. One way this could occur is if the processes that accelerate particles in the inner and Pan ringlets  decelerate the particles in the outer part of the Encke Gap, and thus cause particles to move away from both edges of the gap.  Unfortunately, the data considered here do not have sufficient resolution to provide secure information about the orbital properties of the outer ringlet. Hence we cannot evaluate such possibilities at present. Future studies using higher-resolution observations should clarify the orbital properties of this ringlet, and thus provide additional insights into the dynamics of dust within the Encke Gap. For example, any trends in the semi-major axis could reveal whether particles in the outer half of the gap are migrating radially in the same way as the other two ringlets. 

\section{Summary}

The Cassini observations of the dusty ringlets in the Encke Gap reveal a number of interesting dynamical phenomena:
\begin{itemize}
\item The bright clumps in the central Pan ringlet are confined to a longitudinal region roughly 60$^\circ$ wide just in front of Pan.
\item The bright clumps in the inner and outer ringlets cover less than 180$^\circ$ in co-rotating longitude, and the distribution of clumps is not obviously disrupted by conjunctions with Pan.
\item Within the inner and Pan ringlets, clumps drift relative to each other at rates of up 0.04$^\circ$/day, while the largest relative drift rates observed in the outer ringlet are near 0.01$^\circ$/day.
\item Clumps in the Pan and inner ringlets are observed to merge and split. They also accelerate in surprising ways and follow trajectories that are inconsistent with those expected for isolated particles moving in the combined gravitational fields of Saturn and Pan.
\item The orbital elements of the particles in both the inner and Pan ringlets vary systematically with co-rotating longitude. 
\item Both the inner and Pan ringlets exhibit some heliotropic behavior, and outside the clumps, the free eccentricity is approximately equal to the forced eccentricity that is induced by solar radiation pressure.
 \item ``Kinks" in the Pan and inner ringlets associated with the clumps appear to correspond to variations in the ring-particle's eccentricities. In the Pan ringlet, these kinks
 seem to be locations where the heliotropic forced eccentricity is reduced.
 \item The semi-major axes of both the inner and Pan ringlets vary with co-rotating longitude. They reach a minimum within the clump-rich regions and are up to 10 km larger outside of this region.
 \end{itemize}
 
 \section*{Acknowledgements}
 
 We acknowledge the support of the Cassini Imaging Team, the Cassini Project and NASA. This work was funded by NASA Cassini Data Analysis Program Grants
 NNX09AE74G and NNX12AC29G. We also wish to thank S. Charnoz and an anonymous reviewer for their helpful comments.

\begin{table}
\caption{Movie sequences used to construct mosaics}
\label{moslist}
\resizebox{6.5in}{!}{\begin{tabular}{|c|c|l|l|c|c|c|c|c|c|c|c|c|}\hline
Rev & Sequence & Date & Images  &
	Em. & Phase & Solar & Obs. & Mosaic & \multicolumn{3}{c|}{Quality Flags$^c$} \\ 
 & & & & Angle & Angle & Long.$^a$ & Long.$^a$ & Res.$^b$ & Pan & Inner & Outer \\ \hline
000 & SATSRCH       & 2004-173 & N1466448221-N1466504861 (119) &
	106$^\circ$ & 67$^\circ$  & 159$^\circ$ & 178$^\circ$ & 20 km/pix & I & X & X\\
00A & SPKMOVPER & 2004-320 & N1479201492-N1479254052 (74) & 
	102$^\circ$ & 84$^\circ$ & 165$^\circ$ & 156$^\circ$ & 14 km/pix & P & I & X\\ 
008 & LPHRLFMOV  & 2005-138 & N1495091875-N1495139739 (194) &
	109$^\circ$ & 42$^\circ$ & 172$^\circ$ & 216$^\circ$ & 5 km/pix & R & P & I  \\
030 & HIPHAMOVE & 2006-279 & N1538861755-N1538900050 (70) &
	77$^\circ$ & 159$^\circ$ & 191$^\circ$ & 302$^\circ$ & 6 km/pix & R & R & P\\
034 & HIPHAMOVD & 2006-331& N1543346569-N1543387061 (46) &
	70$^\circ$ & 158$^\circ$ & 193$^\circ$ & 305$^\circ$ & 5 km/pix & R & R & P\\
044 & FMOVIE          & 2007-125 & N1557020880-N1557071468 (134) &
	61$^\circ$ & 81$^\circ$ & 198$^\circ$ & 180$^\circ$ & 6 km/pix & P & P & P\\
051 & LPMRDFMOV & 2007-291 & N1571435192-N1571475337 (260) &
	86$^\circ$ & 56$^\circ$ & 204$^\circ$ & 170$^\circ$ & 7 km/pix & R& R & X\\
053 & LPHRDFMOV & 2007-334 & N1575141899-N1575189603 (134) &
	80$^\circ$ & 52$^\circ$ & 205$^\circ$ & 165$^\circ$ & 5 km/pix & R & R & P\\
109 & LRHPENKMV & 2009-107 & N1618663507-N1618688110 (60) &
	47$^\circ$ & 117$^\circ$ & 221$^\circ$ & 302$^\circ$ & 4 km/pix & R & R & P\\
115 & FMOVIEEQX & 2009-211& N1637609661-N1627655251 (149) &
	62$^\circ$ & 100$^\circ$ & 224$^\circ$ & 237$^\circ$ & 5 km/pix & R & R & P\\
124 & LRHPENKMV & 2010-007 & N1641576230-N1641603998 (104) &
	106$^\circ$ & 118$^\circ$ & 229$^\circ$ & 81$^\circ$ & 5 km/pix & R & R& P\\
124 & LRHRENKMV & 2010-008 & N1641604730-N1641631010 (91) &
	107$^\circ$ & 129$^\circ$ & 229$^\circ$ & 268$^\circ$ & 5 km/pix & R & R& P\\
132 & SHRTMOVIE & 2010-153 & N165413619- N1654175167 (240) & 
	78$^\circ$ & 141$^\circ$  & 233$^\circ$ & 289$^\circ$ & 2 km/pix & P & P & P\\
	\hline
\end{tabular}}

$^a$ Longitudes measured relative to ring's ascending node on the J2000 coordinate system.

$^b$ Resolution of mosaics generated from the images, which oversample
the original pixels by roughly a factor of 2.

$^c$ X=no attempt to derive brightness profiles. I=Brightness profiles derived by integration over a radial range. P=Brightness profiles derived using a peak-fitting routine. R=Radial locations derived from peak-fitting routine suitable to
determining ringlet orbital elements.

\end{table}


\begin{table}
\caption{Supplementary images containing the region around Pan}
\label{suplist}
\resizebox{6.5in}{!}{\begin{tabular}{|c c|c c|c c|c c|c c|}\hline
Image & Date & Image & Date & Image & Date & Image & Date & Image & Date \\ \hline
N1492024160 & 2005-102  & N1552731154 & 2007-075  & N1575012478 & 2007-333  & N1583628328 & 2008-068  & N1603375318 & 2008-296  \\
N1492759120 & 2005-111  & N1552731197 & 2007-075  & N1575012511 & 2007-333  & N1583758349 & 2008-069  & N1603375361 & 2008-296  \\
N1493446920 & 2005-119  & N1553898401 & 2007-088  & N1575055318 & 2007-333  & N1583758382 & 2008-069  & N1603721360 & 2008-300  \\
N1493544975 & 2005-120  & N1553898444 & 2007-088  & N1575055351 & 2007-333  & N1586079511 & 2008-096  & N1603721403 & 2008-300  \\
N1495641779 & 2005-144  & N1553936876 & 2007-089  & N1575629792 & 2007-340  & N1586079554 & 2008-096  & N1604570501 & 2008-310  \\
N1495713539 & 2005-145  & N1553936919 & 2007-089  & N1575629835 & 2007-340  & N1586106286 & 2008-096  & N1604570544 & 2008-310  \\
N1495770990 & 2005-146  & N1554110742 & 2007-091  & N1575676367 & 2007-340  & N1586106329 & 2008-096  & N1606481890 & 2008-332  \\
N1495814115 & 2005-146  & N1554110785 & 2007-091  & N1575676410 & 2007-340  & N1586166616 & 2008-097  & N1607328286 & 2008-342  \\
N1496700636 & 2005-156  & N1555229824 & 2007-104  & N1575800823 & 2007-342  & N1586166659 & 2008-097  & N1607328329 & 2008-342  \\
N1497235299 & 2005-163  & N1555229867 & 2007-104  & N1575800866 & 2007-342  & N1587821608 & 2008-116  & N1610355419 & 2009-011  \\
N1497276055 & 2005-163  & N1555508391 & 2007-107  & N1576171776 & 2007-346  & N1587821651 & 2008-116  & N1610355462 & 2009-011  \\
N1498058015 & 2005-172  & N1555508434 & 2007-107  & N1576171819 & 2007-346  & N1588751210 & 2008-127  & N1610899512 & 2009-017  \\
N1498825460 & 2005-181  & N1555556437 & 2007-108  & N1577141652 & 2007-357  & N1588751253 & 2008-127  & N1610899555 & 2009-017  \\
N1499520329 & 2005-189  & N1555556480 & 2007-108  & N1577141695 & 2007-357  & N1590835414 & 2008-151  & N1612537044 & 2009-036  \\
N1499726971 & 2005-191  & N1555615492 & 2007-108  & N1577512965 & 2007-362  & N1591525824 & 2008-159  & N1612537087 & 2009-036  \\
N1500341195 & 2005-199  & N1555615535 & 2007-108  & N1577513008 & 2007-362  & N1591525867 & 2008-159  & N1616991490 & 2009-088  \\
N1500516231 & 2005-201  & N1555708703 & 2007-109  & N1578630743 & 2008-010  & N1591997427 & 2008-164  & N1616991533 & 2009-088  \\
N1501156540 & 2005-208  & N1555708746 & 2007-109  & N1578630786 & 2008-010  & N1591997460 & 2008-164  & N1619963567 & 2009-122  \\
N1502133340 & 2005-219  & N1556520958 & 2007-119  & N1579656750 & 2008-022  & N1592072518 & 2008-165  & N1619963610 & 2009-122  \\
N1502133373 & 2005-219  & N1556520991 & 2007-119  & N1579656793 & 2008-022  & N1592072551 & 2008-165  & N1622382064 & 2009-150  \\
N1502581803 & 2005-224  & N1558417179 & 2007-141  & N1579750261 & 2008-023  & N1596292933 & 2008-214  & N1622382097 & 2009-150  \\
N1502581836 & 2005-224  & N1558417222 & 2007-141  & N1579750304 & 2008-023  & N1596292976 & 2008-214  & N1622592755 & 2009-152  \\
N1502650783 & 2005-225  & N1558547905 & 2007-142  & N1580528781 & 2008-032  & N1596720406 & 2008-219  & N1622592788 & 2009-152  \\
N1502650816 & 2005-225  & N1558547948 & 2007-142  & N1580528824 & 2008-032  & N1596720449 & 2008-219  & N1623652033 & 2009-165  \\
N1503573529 & 2005-236  & N1559285595 & 2007-151  & N1580566252 & 2008-032  & N1597462656 & 2008-228  & N1623652076 & 2009-165  \\
N1503573562 & 2005-236  & N1559285638 & 2007-151  & N1580566295 & 2008-032  & N1597462699 & 2008-228  & N1623757093 & 2009-166  \\
N1504218268 & 2005-243  & N1559710457 & 2007-156  & N1580614147 & 2008-033  & N1597488396 & 2008-228  & N1623757136 & 2009-166  \\
N1504341929 & 2005-245  & N1559710500 & 2007-156  & N1580614190 & 2008-033  & N1597488439 & 2008-228  & N1623822254 & 2009-167  \\
N1549374582 & 2007-036  & N1559841843 & 2007-157  & N1580653027 & 2008-033  & N1600167160 & 2008-259  & N1623822297 & 2009-167  \\
N1549374625 & 2007-036  & N1559841886 & 2007-157  & N1580653070 & 2008-033  & N1600167203 & 2008-259  & N1625116703 & 2009-182  \\
N1552517897 & 2007-072  & N1559885869 & 2007-158  & N1580766488 & 2008-034  & N1601291283 & 2008-272  & N1625116736 & 2009-182  \\
N1552517940 & 2007-072  & N1559885912 & 2007-158  & N1580766531 & 2008-034  & N1601291316 & 2008-272  & N1627546060 & 2009-210  \\
N1552606713 & 2007-073  & N1560054860 & 2007-160  & N1581513703 & 2008-043  & N1602109066 & 2008-281  & N1627546103 & 2009-210  \\
N1552606756 & 2007-073  & N1560054903 & 2007-160  & N1581513746 & 2008-043  & N1602109109 & 2008-281  & N1628912570 & 2009-226  \\
N1552645698 & 2007-074  & N1573672968 & 2007-317  & N1582637241 & 2008-056  & N1602501762 & 2008-286  & N1628912603 & 2009-226  \\
N1552645741 & 2007-074  & N1573673011 & 2007-317  & N1582637274 & 2008-056  & N1602501805 & 2008-286  & N1633029034 & 2009-273  \\
N1552688328 & 2007-074  & N1574856717 & 2007-331  & N1583401346 & 2008-065  & N1603175686 & 2008-294  & N1633029067 & 2009-273  \\
N1552688371 & 2007-074  & N1574856760 & 2007-331  & N1583401389 & 2008-065  & N1603175729 & 2008-294  & &  \\
\hline
\end{tabular}}
\end{table}


\end{document}